\documentclass[11pt,a4paper]{article} 
\usepackage{jheppub_kim}

\def\a{\alpha}

\def\d{\delta}
\def\e{\epsilon}           

\def\k{\kappa}                    

\def\m{\mu}
\def\n{\nu}
  
\def\t{\tau}

\def\z{\zeta}
\def\D{\Delta}

\def\L{\Lambda}

\def\S{\Sigma}

\def\del{\partial}              
\let\a=\alpha   \let\d=\delta \let\e=\epsilon
\let\z=\zeta    \let\k=\kappa
 \let\m=\mu \let\n=\nu  
 \let\t=\tau    
  \let\D=\Delta  \let\L=\Lambda

\let\la=\label  
  
\def\nn{\nonumber} \def\bd{\begin{document}} \def\ed{\end{document}}
\def\ds{\documentstyle} \let\fr=\frac \let\bl=\bigl \let\br=\bigr
\let\Br=\Bigr \let\Bl=\Bigl
\let\bm=\bibitem
\let\na=\nabla
\let\pa=\partial \let\ov=\overline
\newcommand{\be}{\begin{equation}}
\newcommand{\ee}{\end{equation}}
\def\ba{\begin{array}}
\def\ea{\end{array}}
\def\ft#1#2{{\textstyle{{\scriptstyle #1}\over {\scriptstyle #2}}}}
\def\fft#1#2{{#1 \over #2}}
\def\del{\partial}
\def\sst#1{{\scriptscriptstyle #1}}
 \def\oneone{\rlap 1\mkern4mu{\rm l}}
\def\ie{{\it i.e.\ }}
\def\via{{\it via}}
\def\semi{{\ltimes}}
\def\str{{\rm str}}
\def\Dm{{{D_{\sst{max}}}}}
\def\vac{ \left | 0 \right \rangle }
\def\kvac{ \left | k \right \rangle }

\def\sp{\; \; \;}

\def\bol{ \left | B (p^+) \right \rangle}
\def\bo1{ \left | B^0 (p^+) \right \rangle}

\def\bolt{ \left | B (p^+) \right \rangle_{\t}}

\def\boxl{ \left | B (x^-) \right \rangle}
\newcommand{\bea}{\begin{eqnarray}}
\newcommand{\eea}{\end{eqnarray}}

\def\<{ \langle }
\def\>{ \rangle }

\def\S{\Sigma}

\usepackage{amsmath,latexsym,amssymb,slashed}
\usepackage{graphicx}
\usepackage[latin1]{inputenc}
\usepackage{subfigure}

\newcommand\ca{\mathcal{A}}
\newcommand\vp{\varphi}
\newcommand\beal{\begin{align}}
\newcommand\bbone{\ensuremath{\mathbbm{1}}}

\newcommand{\eq}[1]{\begin{equation}#1\end{equation}}
\newcommand{\spl}[1]{\begin{split}#1\end{split}}
\newcommand{\al}[1]{\begin{align}#1\end{align}}
\newcommand{\subeq}[1]{\begin{subequations}#1\end{subequations}}

\newcommand{\arXividhepth}[1]{\href{http://arxiv.org/abs/#1}arXiv:{\tt #1} [hep-th]}
\newcommand{\arXividother}[2]{\href{http://arxiv.org/abs/#1}arXiv:{\tt #1} [#2]}

\newcommand{\bg}[1]{\hat{#1}}
\newcommand{\wj}{\widetilde{J}}
\newcommand{\reo}{\mathrm{Re}~\!\omega}
\newcommand{\imo}{\mathrm{Im}~\!\omega}
\newcommand{\ads}{AdS_4}
\newcommand{\mcal}{\mathcal{M}}
\newcommand{\ccal}{\mathcal{C}}
\newcommand{\ncal}{\mathcal{N}}
\newcommand{\boxedeq}[1]{
\begin{equation}
\fbox{
\rule[0.7cm]{0pt}{0pt}
$#1$
\rule[-0.45cm]{0pt}{0pt}
}
\end{equation}
}

\def\d{\text{d}}

\def\slashchar#1{\setbox0=\hbox{$#1$}           
\dimen0=\wd0                                 
\setbox1=\hbox{/} \dimen1=\wd1               
\ifdim\dimen0>\dimen1                        
\rlap{\hbox to \dimen0{\hfil/\hfil}}      
#1                                        
\else                                        
\rlap{\hbox to \dimen1{\hfil$#1$\hfil}}   
/                                         
\fi}

\def\Re           {{\rm Re\hskip0.1em}}
\def\Im           {{\rm Im\hskip0.1em}}

\newcommand{\E}{\text{\tiny E}}

\newcommand{\tV}{{\widetilde{V}}}
\newcommand{\tH}{{\tilde{h}}}
\newcommand{\tm}{{{m}}}
\newcommand{\tmu}{{\tilde{\mu}}}
\newcommand{\trho}{{\tilde{\rho}}}
\newcommand{\tv}{{\tilde{v}}}
\newcommand{\calo}{\mbox{${\cal O}$}}
\newcommand{\cala}{\mbox{${\cal A}$}}
\newcommand{\dd}{\mathrm{d}}
\newcommand{\ra}{\rightarrow}
\newcommand{\calv}{\mbox{${\cal V}$}}
\newcommand{\calh}{\mbox{${\cal H}$}}
\newcommand{\calm}{\mbox{${\cal M}$}}
\newcommand{\abs}[1]{\left| #1 \right|}
\newcommand{\zetaa}{{\psi}}
\newcommand{\tr}{{\rm tr}\,}

\newcommand{\ky}[1]{{\color{blue}{#1}}}

\newcommand{\td}{\tilde}
 \newcommand{\bc}{\begin{center}}
 \newcommand{\ec}{\end{center}}
 \newcommand{\bfr}{\begin{flushright}}
 \newcommand{\efr}{\end{flushright}}
 \newcommand{\bfl}{\begin{flushleft}}
 \newcommand{\efl}{\end{flushleft}}
 \newcommand{\bt}{\begin{tabular}}
 \newcommand{\et}{\end{tabular}}

\title{Holographic d-wave superconductors}

\author[a,b]{Keun-Young Kim}
\author[a,c]{Marika Taylor}

\emailAdd{fortoe@gist.ac.kr}
\emailAdd{m.m.taylor@soton.ac.uk}

\affiliation[a]{Institute of Theoretical Physics,
Science Park 904, Postbus 94485, \\
1090 GL Amsterdam, The Netherlands
}
\affiliation[b]{GIST College, Gwangju Institute of Science and Technology,
123 Cheomdangwagi-ro Buk-gu, Gwangju, 500-712 South Korea
}
\affiliation[c]{School of Mathematical Sciences and STAG Research Centre, University of Southampton, \\
Highfield, Southampton, SO17 1BJ, UK.
}

\abstract{
We construct top down models for holographic d-wave superfluids in which the order parameter is a charged spin two field in the bulk. Close to the transition temperature the condensed phase can be captured by a charged spin two field in an R-charged black hole background (downstairs picture) or equivalently by specific graviton perturbations of a spinning black brane (upstairs picture). We analyse the necessary conditions on the mass and the charge of the spin two field for a condensed phase to exist and we discuss the competition of the d-wave phase with other phases such as s-wave superfluids.
}

\keywords{Gauge/Gravity duality, Superconductor}

\begin{document}

\maketitle

\section{Introduction}

Over the past few years there has been growing interest in using gauge/gravity duality to model strongly interacting systems relevant for condensed matter physics, see the reviews \cite{Hartnoll:2009sz,Herzog:2009xv,McGreevy:2009xe,Horowitz:2010gk}. In particular, we have learnt that AdS black holes generically develop hair at low temperature, with the hair capturing the phenomenology of superfluid phases in the dual field theory \cite{Gubser:2008px,Hartnoll:2008kx}.

It is remarkable that rather simple and generic gravity models can capture many features of the phase structure of strongly interacting systems but to construct more sophisticated and more realistic models one clearly needs to include additional ingredients. Every essential ingredient or feature of the strongly interacting system must have a counterpart in the bulk holographic model. There has been considerable interest recently in developing bulk descriptions for key missing ingredients such as broken translation invariance (necessary to realise finite DC conductivity in the ordinary state) \cite{Horowitz:2012ky,Horowitz:2012gs}. 

Moreover one needs to take into account that there can be a rich variety of phases even within rather simple models. Isotropic superfluid phases are not necessarily the preferred phase at low temperature: spatially modulated phases arise in rather generic models, see for example \cite{Nakamura:2009tf,Ooguri:2010kt,Ooguri:2010xs,Bergman:2011rf,Donos:2011qt}. One may also find other phases, including
striped phases \cite{Donos:2011bh}, helical phases \cite{Donos:2012gg}, and simple models can exhibit competitions between such phases \cite{Donos:2012yu}. 

The focus of this paper is on realising an important missing phase, d-wave superfluidity, within a top down holographic model. The importance of modelling d-wave superfluidity is self-evident: many unconventional superconductors admit either d-wave or mixed symmetry. A natural candidate for modelling the d-wave condensate is to use a charged spin two field in the bulk, instead of a charged scalar field. This approach was taken in
\cite{Chen:2010mk,Zeng:2010vp,Chen:2011ny,Gao:2011aa,Ge:2012vp} and also in \cite{Benini:2010qc,Benini:2010pr}. The problem is that massive spin two equations have issues with consistency, causality and hyperbolicity, and they do not follow from consistent truncations of reductions of top down models.

The approach in \cite{Chen:2010mk} and subsequent related works was to write down minimal equations for the spin two field (without looking in detail at the constraint equations required to obtain the correct number of propagating degrees of freedom) while \cite{Benini:2010qc,Benini:2010pr} looked in more detail at the effective action for the spin two field and how the constraint equations could be satisfied. In practice, evaluated on a static ansatz, the effective equations of motion are the same in both approaches. The coupled equations reduce to those for the charged scalar field/gauge field system used in \cite{Hartnoll:2008kx} and therefore an analogous condensed phase exists at low temperature. Evaluating the onshell action, one can show that it reduces to the same action as for the Maxwell-scalar case, and therefore the condensed phase is thermodynamically favoured.

The starting point of this work is the observation that massive spin two modes are generic in
compactifications from ten and eleven dimensions which give rise to lower-dimensional Anti-de Sitter backgrounds. Moreover such modes are generically charged under the $U(1)$ symmetry corresponding to the R symmetry of the dual conformal field theory and thus the corresponding spin two operators
in the dual CFT are natural candidates for d-wave order parameters. (An analogous observation was made for circle reductions of geons in \cite{Hartnett:2012np}, although the charged spin-2 condensates obtained in this case were not thermodynamically preferred.)

While charged spin two fields are generic, they cannot be retained as part of a consistent truncation and therefore one has to derive their field equations directly from the higher dimensional equations of motion. The linear field equations suffice to obtain the spectrum but,
just as for the coupled system of a charged scalar and gauge field, non-linear interactions are essential in
\cite{Chen:2010mk,Benini:2010pr} for the condensate phase to exist. The problem is that as soon as one allows interactions between Kaluza-Klein modes Pandora's box has been opened and one does not expect to be able to truncate to a finite set of equations: the spin two field will source an infinite number of other Kaluza-Klein fields.

Clearly the system of equations can only be tractable if one can suppress the backreaction of the spin two field on the other Kaluza-Klein fields, including the metric. There are (at least) two distinct possibilities for suppressing the backreaction. The first is to work in a limit in which the charge $q$ of the spin two field is large, following the arguments given in \cite{Hartnoll:2008kx,Horowitz:2010gk} for a charged scalar field, see also the discussions around \eqref{tcp} and \eqref{largeq}; then the idea is that the backreaction would be suppressed by $1/q$ factors. The second possibility is to work close to the critical temperature $T_c$ when the spin two field magnitude is proportional to $\e = (1- T/T_c)^{\lambda} \ll 1$ (for a specific exponent $\lambda$); the backreaction of the spin two is then suppressed by factors of $\e^2$ or smaller.  In this paper we will consider both possibilities but the analysis is simpler and more robust in the latter case.

Large charge $q$ certainly suppresses the backreaction, at least in part. However, in a top down model one needs to take into account that the mass and the charge are not independent: there is a maximal charge for any given mass. In particular this implies that the critical temperature is not reached for spherical or simple Sasaki-Eintein reductions within the approximation that the backreaction on the metric is ignored. (The probe approximation may however suffice for other reductions giving AdS vacua, such as warped compactifications.)
The issue is that the backreaction of the gauge field on the metric is controlled by $\mu/T$ where $\mu$ is the chemical potential and is only negligible when $\mu/T \ll 1$. The mass and charge parameters in the compactifications are however such that $T_c/\mu < 1$. This implies that one must take into account the backreaction of the gauge field, which in the ordinary phase gives rise to an R charged black hole.

Suppose one considers a spin two field in the (fixed) background of an R charged black hole. Looking at the linear equation for the spin two field in such a background, one finds that a condensed phase indeed generically exists below a critical temperature. One does not need non-linear equations for the condensed phase to exist. This might sound surprising given the discussion above but in fact this is exactly what happens close to the critical temperature in the s-wave case: the non-linear coupled equations of the gauge field and charged field are solved perturbatively in the parameter $\e$ given above, and the leading order equation for the charged field becomes a decoupled linear equation in this limit, see \cite{Maeda:2008ir,Herzog:2010vz,Siopsis:2010uq} and appendix \ref{appc}.

The linear equation for the spin two field is however only a valid approximation if the backreaction of the spin two onto the metric, gauge field and other Kaluza-Klein fields is suppressed. The backreaction onto the metric and the gauge field is suppressed either for large $q$ (with the amplitude of the spin two field scaling as $1/q$) or close to the transition temperature. The backreaction on the other Kaluza-Klein fields is however more subtle because it involves looking in detail at the interactions in the Kaluza-Klein reduction. Here we will argue in detail that the other Kaluza-Klein fields can indeed be neglected close to the critical temperature, postponing a detailed analysis of the large $q$ case for future work.

Close to the critical temperature the description of the spin two condensed phase is simple: a linear perturbation around the charged black hole background. This description can straightforwardly be uplifted to ten or eleven dimensional supergravity; the spin two field then corresponds to a specific graviton perturbation in the background of decoupled spinning D3-branes or M2-branes. This upstairs picture would be the natural starting point for the construction of a fully non-linear condensed phase solution.

Although the focus of this work is on d-wave phases, Kaluza-Klein modes may be responsible for a rich variety of phases, many of which may not be visible within consistent truncations. The d-wave phases we find would have critical temperatures which are typically lower than those of s-wave phases associated with relevant scalar charged operators. In a compactification in which no such relevant operators are present, the d-wave phase may be the first superfluid phase reached as one lowers the temperature. In typical compactifications, the s-wave phase would however be reached first (unless one finds a way to lift the scalar modes). One would expect that the interactions between distinct Kaluza-Klein modes would give rise to a competition between different phases, which would be very interesting but perhaps complicated to analyse.

In the bottom up model, the onshell action for the charged spin two field system is computed, showing that the condensed phase is indeed favoured below the critical temperature. Moreover, in the probe approximation, spin two and scalar fields of the same mass and charge would have exactly the same free energy in the condensed phase, neglecting any interactions between them. Taking into account either interactions or the backreaction onto the metric would presumably break the degeneracy and determine what is the favoured phase.

Working near the critical temperature in the top down model, one cannot determine whether the condensed phase is indeed thermodynamically favoured without computing the backreaction, both on the metric and on the other fields. The most straightforward way to compute this backreaction would be to work directly with the higher dimensional description but we postpone such an analysis for future work. 

\bigskip

This paper is organised as follows. In the next section we consider the effective description of $d$-wave condensates in terms of coupled spin two and gauge field equations in black hole backgrounds.  In section \ref{kk-ap} we discuss how such equations may arise from Kaluza-Klein reductions and in section \ref{2b} we look in detail at reductions of type IIB on Sasaki-Einstein manifolds. In section \ref{uplift} we present the description of the condensed phases in terms of higher dimensional solutions and in section \ref{conc} we conclude. Various technical issues are contained in the appendices: appendix \ref{calc-spin2} relates to corrections to the spin two equations in Kaluza-Klein reductions while appendices \ref{schr-eq},
\ref{appc} and \ref{number} explore numerical and analytical solutions of the condensate equations.

\section{Effective description of $d$-wave condensates}

We will consider holographic superconductors in $(D+1)$ spacetime dimensions, using an action involving the metric $g_{\mu \nu}$ together with a spin two field $\phi_{\mu \nu}$ and a gauge field $F_{\mu \nu}$.

There has been considerable literature discussing actions for spin two fields coupled to electromagnetic fields.
Working from a bottom up perspective, \cite{Buchbinder:1999be,Buchbinder:1999ar,Buchbinder:2000fy,Buchbinder:2012iz} discuss the conditions under which such fields can
have consistent and causal equations. Inconsistencies and acausalities are analysed in
\cite{Deser:2001us,Deser:2001dt,Porrati:2011uu}. Here we will use the equations of motion following from the action given in \cite{Benini:2010qc,Benini:2010pr}\footnote{Note that there is a typo in the Lagrangian given in equation (3) of \cite{Benini:2010pr}; the term involving $R_{\mu \nu} \phi^{\ast \mu \lambda} \phi^{\nu}_{\lambda}$ should be absent, following the discussions in appendix A of \cite{Benini:2010pr}. The equations of motion in equation (4) of \cite{Benini:2010pr} indeed drop this term.}
\bea
S &=& \frac{1}{2 \kappa^2} \int d^{D+1} x \sqrt{-g} \left( R - 2 \Lambda - \frac{1}{4} F_{\mu \nu} F^{\mu \nu} \right) \label{action-Herzog}  \\
&& +
\frac{1}{2 \kappa^2} \int d^{D+1} x \sqrt{-g} \bigg [ - | D_{\rho} \phi_{\mu \nu} |^2
+ 2 | D_{\mu} \phi^{\mu \nu} |^2 + | D_{\mu} \phi |^2 - \left[ D_{\mu} \phi^{\ast \mu \nu} D_{\nu} \phi + {\rm c.c.}\right] 
\nn \\
&& \qquad  - m^2 \left( |\phi_{\mu \nu}|^2 - |\phi|^2 \right) + 2 R_{\mu \nu \rho \lambda} \phi^{\ast \mu \rho} \phi^{\nu \lambda} - \frac{1}{D+1} R | \phi |^2  - 2 i g q F_{\mu \nu} \phi^{\ast \mu \lambda} \phi^{\nu}_{\lambda}  \bigg]. \nn
\eea
Here $\phi \equiv \phi^{\mu}_{\mu}$ and $\phi_{\rho} \equiv D^{\mu} \phi_{\mu \rho}$, with the gauge covariant derivative being defined as $D_{\mu} = \nabla_{\mu} - i q A_{\mu}$. The parameter $g$ is the gyromagnetic ratio which will be discussed further below. The Chern-Simons terms involving the gauge field which are present in odd dimensions do not play a role in what follows, as we consider only static electrically charged solutions.

This action admits Einstein solutions in which the spin two and gauge field vanish, and the metric satisfies
\be
R_{\mu \nu} = \frac{2 \Lambda}{(D-1)} g_{\mu \nu}.
\ee
We will fix $\Lambda= - D (D-1)/2$ so that the Einstein metric has unit radius. For much of this paper we will treat the spin two and gauge field in the probe approximation, in which their equations of motion are solved in a fixed background Einstein metric.
The equations of motion then describe a massive charged spin two particle of mass $m$ and charge $q$ in the curved Einstein background:
\bea
(\Box - m^2) \phi_{\mu \nu} &=&  2 D_{<\mu} \phi_{\nu>} - D_{< \mu} D_{\nu>} \phi
+ g_{\mu \nu} \left[ (\Box - m^2)
\phi - D^{\rho} \phi_{\rho} \right] \label{uueq} \\
&& \qquad + 2 R_{\mu \rho \lambda \nu} \phi^{\rho \lambda} + g_{\mu \nu} \frac{R}{(D+1)} \phi
+ i g q \left(F_{\mu \rho} \phi^{\rho}_{\nu} + F_{\nu \rho} \phi^{\rho}_{\mu} \right) ; \nn \\
D_{\mu} F^{\mu \nu} &=& i q \phi^{\ast}_{\lambda \rho} \left(D^{\nu} \phi^{\lambda \rho} - 2 g D^{\lambda} \phi^{\nu \rho} \right)
\nn \\ 
&& \qquad - i q \Big(2 (1 -g) \phi^{\ast}_{\rho} \phi^{\nu \rho}  - D_{\rho} \phi^{\ast} \phi^{\nu \rho} + \left(D^{\nu} \phi^{\ast} - \phi^{\ast \nu}\right)  \phi\Big) + {\rm h.c.} \nn
\eea
(Anti)-symmetrization (without subtracting the trace) is denoted by angled parentheses and is defined with unit weight.
These equations of motion are supplemented by constraints, which when $F_{\mu \nu} = 0$ take the simple form
\be
D^{\mu} \phi_{\mu \nu} = \phi = 0.
\ee
In this case the equation of motion reduces to
\be
(\Box - m^2) \phi_{\mu \nu} + 2 R_{\mu \lambda \nu \rho} \phi^{\lambda \rho} = 0, \label{uu2eq}
\ee
subject to these constraints. These constraints arise as follows: the components of the equations of motion in the time direction do not involve second
order time derivatives and are therefore constraints. The tensor $\phi_{\mu \nu}$ has more degrees of freedom than a spin two particle,
and additional constraint equations are needed to eliminate the extra degrees of freedom. These further constraints arise
from taking the divergence of the equations of motion twice.

In the case where $F_{\mu \nu } \neq 0$ the constraints are considerably more complicated.
Only the case of $g= 1/2$ is discussed in
\cite{Benini:2010pr} but we will need the equations for general $g$ which were given in  \cite{Deser:2001dt}.
The first divergence of the field equation gives
\bea
m^2 (\phi_{\mu} - D_{\mu} \phi) &=& - i q \left ( \frac{1}{2} F_{\mu} \phi - (1 - g) F^{\nu} \phi_{\nu \mu} + g D_{\rho} F_{\mu \nu} \phi^{\nu \rho} \right . \label{constr3} \\
&& \left . (1 + g) F_{\mu \nu} \phi^{\nu} - \frac{3}{2} F_{\mu \nu} D^{\nu} \phi + (2 -g) F^{\nu \rho} D_{\rho} \phi_{\mu \nu} \right ). \nn
\eea
where $F_{\mu} = D^{\nu} F_{\nu \mu}$. The trace of the field equation gives
\be
\left(D m^2 - \frac{D-1}{D+1} R \right) \phi - (D-1) (\Box \phi - D_{\mu } \phi^{\mu}) = 0. \label{const-2}
\ee
The second divergence of the field equation combined with the trace equation gives
\bea
- m^2 \left [ \frac{D m^2}{(D-1)} - \frac{R}{(D-1)} \right ] \phi
&=& i q \left [ 2 (1 - g) (D^{\mu} F^{\nu \rho})(D_{\nu} \phi_{\rho \mu} ) - 2 g F^{\mu} \phi_{\mu} + F^{\mu} D_{\mu} \phi \right ] \nn \\
&& \qquad + (g-2) q^2 F^{\mu \rho} F_{\nu \rho} \phi^{\nu}_{\mu} + \frac{3}{4} q^2 F_{\mu \nu} F^{\mu \nu} \phi \label{const-1}  \\
&& \qquad + i q (1 - 2 g) F^{\mu \nu} D_{\mu} \phi_{\nu}  + i q (1 -2 g) D_{\mu} F_{\nu} \phi^{\mu \nu}. \nn
\eea
From these equations it is apparent that when $A_{\mu} = 0$ the constraints reduce to $\phi = \phi_{\mu} = 0$.

When $A_{\mu} \neq 0$, the problem is that these constraints are no longer algebraic constraints on $(\phi, \phi_{\mu})$; they involve first derivatives of
$\phi_{\mu \nu}$
multiplied by first derivatives of $F_{\mu \nu}$ and for generic values of $F_{\mu \nu}$ this leads to the
equations being either non-hyperbolic or non-causal  \cite{Deser:2001dt}. From the form of \eqref{const-1} it is apparent that for generic values of $g$ there is
a term multiplying the field strength with double time derivatives of the spin two field. Such a term implies a breakdown in the number of
degrees of freedom, as a constraint of the free model becomes a propagating degree of freedom. This argument has been given in the literature, see \cite{Deser:2001dt}, \cite{Benini:2010pr},
to fix $g = 1/2$, in which case the action describes $(D+1)(D-1)/2$ propagating spin two degrees of freedom.

As we see below, the equations and constraints can be consistently solved for specific static field configurations, such
as those discussed below, for any value of $g$. If one then analyzes perturbations around such a solution, generically one will have a number of degrees of freedom which does not match that of a spin two field.

Even for $g=1/2$ in a fixed Einstein background, the equations of motion for general values of the gauge field are still non-hyperbolic or lead to non-causal propagation. To suppress these effects one can work in a limit in which
\be
\frac{q |F_{\hat{\mu} \hat{\nu}}|}{m^2}, \qquad
\frac{q |F_{\hat{\mu} \hat{\nu}; \hat{\rho}}|}{m^3} \label{sup}
\ee
are small, where $\hat{\mu}$ denotes a tangent space index.
It was suggested in \cite{Benini:2010qc,Benini:2010pr}
that the problems with causality and hyperbolicity could be corrected by adding
additional higher order terms to the action.

Most of the discussions of spin two and gauge field actions in the literature follow from bottom up models.
Here we are working in the context of Kaluza-Klein reductions of string theory/M theory backgrounds compactified on Sasaki-Einstein spaces. Kaluza-Klein gravitons in such reductions correspond to massive spin two fields from the lower dimensional perspective. In the following sections we will show
that there is indeed a systematic way of expanding
the spin two/gauge field equations when they are obtained from a Kaluza-Klein reduction. In anticipation of this discussion, it is useful to note
the following point. Suppose that one treats both the gauge field and the spin two field perturbatively, i.e. one lets
\be
\phi_{\mu \nu} = \delta \phi^{(1)}_{\mu \nu} + \delta^2 \phi^{(2)}_{\mu \nu} + \cdots; \qquad
A_{\mu} = \delta A^{(1)}_{\mu} + \delta^2 A^{(2)}_{\mu} + \cdots,
\ee
with $\delta$ small. Working perturbatively in $\delta$, the constraint \eqref{const-1} remains algebraic. To leading order we obtain that the trace $\phi^{(1)} = 0$ and to next
order we obtain a constraint on $\phi^{(2)}$ in terms of the (known) fields $\phi^{(1)}_{\mu \nu}$ and $A^{(1)}_{\mu}$, i.e. we obtain from \eqref{const-1}
\be
- m^2 \frac{D}{(D-1)} \left ( m^2 + (D+1) \right ) \phi^{(2)} = 2 i q (1-g) (D^{\mu} F^{(1) \nu \rho} D_{\nu} \phi^{(1)}_{\rho \mu}).
\ee
Thus if the equations are evaluated perturbatively the restriction to $g=1/2$ is not necessary. From a top down perspective the parameter $g$ is uniquely determined by reducing the higher dimensional equations of motion over a compact space. Constraints on such couplings may be obtained by imposing causality in the dual CFT, see \cite{Kulaxizi:2012xp}. The top down will automatically respect such constraints but furthermore determine such couplings uniquely.

\subsection{The $d$-wave condensate}

Let us now turn to superfluid solutions of these equations of motion.
The Einstein condition on the metric restricts us to considering a probe limit in which the matter fields do not backreact on the metric. The relevant background metric is therefore that of a neutral black brane
\be
ds^2 = \frac{1}{z^2} \left ( - f(z) dt^2 + d \vec{x}^2 + \frac{dz^2}{f(z)} \right ), \label{metric0}
\ee
where
\be
f(z) = 1 - \left ( \frac{z}{z_h} \right )^D. \label{neutr}
\ee
The black hole horizon is located at $z = z_h$ whilst the conformal boundary of the spacetime is at $z=0$ and the temperature
of the black hole is
\be
T = \frac{D}{4 \pi z_h}.
\ee
We now consider solutions to the coupled equations of motion for the spin two field and gauge field, assuming an ansatz in which
$\phi_{\mu \nu}$ and $A_{\mu}$ depend only on the radial coordinate $z$, with only spatial components of the spin two field switched on.
Furthermore, we consider the case in which only  $\phi_{xy} \neq 0$ is non-zero, so the ansatz is
\be
A = V(z) dt; \qquad
\phi_{xy} = \frac{1}{2 z^2} h(z), \label{ans-spin2}
\ee
with the functions $(V(z),h(z))$ being real. Note that this is clearly not the most general ansatz consistent with the symmetries: one could obtain related solutions by imposing a complex ansatz for the spin two field and by breaking the rotational symmetry in a different way. Put differently, there is an arbitrariness in the choice of the phase characterising the $U(1)$ symmetry breaking and in the choice of the plane in which the rotational symmetry is broken. With this ansatz $\phi = \phi_{\mu} = F_{\mu \rho} \phi^{\rho}_{\nu} = 0$ and
the equations of motion simplify considerably - the only terms which contribute are the following
\bea
(\Box - m^2) \phi_{\mu \nu} &=& 2 R_{\mu \lambda \rho \nu} \phi^{\rho \lambda}; \label{reduced} \\
D_{\mu} F^{\mu \nu} &=& i q \phi^{\ast}_{\alpha \beta} (D^{\nu} \phi^{\alpha \beta}) + {\rm h.c.}, \nn
\eea
where $\Box = D^{\mu} D_{\mu}$. Note that the gyromagnetic parameter $g$ drops out of the gauge field equation, because of the symmetries of the background and of the ansatz.
One also needs to check that the constraints are satisfied. The conditions
$\phi = \phi_{\mu} = F_{\mu \nu} \phi^{\rho}_{\nu} = 0$ are already sufficient for \eqref{const-1} to be satisfied but the remaining constraints \eqref{constr3} and \eqref{const-2} require the stronger conditions of \eqref{ans-spin2}.

Before moving on to discuss these solutions of the equations, we should note that the coupled system could of course admit many more types of solutions and hence different phases. In what follows we will concentrate on the simplest d-wave superfluid phase, and we will postpone discussions of other phases and indeed the related question of the stability of this superfluid phase for later work.

\subsubsection*{Equations of motion}

Using the ansatz \eqref{ans-spin2} the coupled equations of motion become
\bea
V'' + \frac{3 - D}{z} V' - \frac{q^2}{z^2} h^2 V &=& 0; \label{higgs-eq} \\
h'' + \left ( \frac{f'}{f} - \frac{D-1}{z} \right ) h' + \left ( \frac{q^2 V^2}{f^2} - \frac{m^2}{z^2 f} \right ) h &=& 0, \nn
\eea
where a prime denotes a $z$ derivative. These equations of motion admit a normal phase
solution in which the spin two field is zero
\be
A = \mu \left (1 - \left ( \frac{z}{z_h} \right )^{D-2} \right ) dt; \qquad \phi_{xy} = 0,
\ee
with $\mu$ being the chemical potential.

In \cite{Benini:2010qc,Benini:2010pr} it was noted that these equations of motion are identical to those of the Abelian Higgs model
in $AdS_{D+1}$ discussed in \cite{Hartnoll:2008kx}, implying that there also exists a condensed
solution below a critical temperature $T_c$.
In the condensed solution the spin two field is non-zero, and its asymptotic expansion
is
\be
h(z) = z^{D -\Delta} \left ( h_s + {\cal O}(z) \right ) + z^{\Delta} \left ( \frac{\kappa^2}{2 \Delta - D} \< {\cal O}_{xy} \>
+ {\cal O}(z) \right ),
\ee
where $\Delta$ is the conformal dimension of the dual spin two operator and the source term $h_s$ vanishes. The conformal
dimension of the spin two field is related to the mass parameter $m^2$ as $m^2 = \Delta (\Delta - D)$. We will derive this relation between the normalisable mode and the expectation value of the dual operator $\< {\cal O}_{xy} \>$ in section \ref{therm}. Moreover, in the condensed
phase regularity conditions are imposed  in which $h(z)$ is finite everywhere and the gauge field $A_t$ vanishes on the
black hole horizon. Numerical solutions for the condensed solution in $D=3$ were presented in \cite{Benini:2010qc,Benini:2010pr}.
Note that the non-linear interactions in the equations \eqref{higgs-eq} are necessary for a condensate solution to exist.

For later use, let us notice that the ansatz is such that only two terms non-linear in $(\phi_{xy},A_{t})$ appear in the equations of motion in \eqref{reduced}. These non-linear terms follow from the covariant derivatives implicit in the action \eqref{action-Herzog}.

We now introduce a dimensionless radial
coordinate $\zeta = z/z_h$ such that $\zeta = 1$ at the horizon and $\zeta \rightarrow 0$ at the boundary. Then let the gauge field
and spin two field be rescaled as
\be
\tilde{V} (\zeta) = q z_h V (\zeta); \qquad
\tilde{h} (\zeta) = q h(\zeta).
\ee
The coupled equations of motion then become
\bea
\ddot{\tilde{V}} + \frac{3 - D}{\z} \dot{\tilde{V}} - \frac{1}{\z^2 f} \tilde{h}^2 \tilde{V} &=& 0; \label{dimless} \\
\ddot{\tilde{h}} + \left ( \frac{\dot{f}}{f} - \frac{D-1}{\z} \right ) \dot{\tilde{h}} + \left ( \frac{\tilde{V}^2}{f^2}
- \frac{m^2}{\z^2 f} \right ) \tilde{h} &=& 0, \nn
\eea
with $f(\zeta) = (1 - \zeta^D)$ and $\dot{a} \equiv \pa_{\zeta} a$.
The rescaled equations depend on only one parameter, the mass of the spin two field, and are independent
of both the charge $q$ and the gyromagnetic parameter $g$. At first sight this implies that the existence of a non-trivial condensate
solution depends only on the mass; however, the rescaled equations need to be solved subject to the boundary conditions
\be
\tilde{V}(\zeta = 0) \rightarrow \frac{q \mu D}{4 \pi T} \equiv m_o; \qquad
\tilde{V}(\zeta = 1) = 0,
\ee
with the spin two field $\tilde{h}$ having no source term as $\zeta \rightarrow 0$ and being finite at the horizon $\zeta = 1$. This implies
that the existence of a condensate solution does indeed depend implicitly on the temperature, charge and chemical potential in the combination
$m_o$. For
\be \label{alphac}
m_{o} \ge \alpha_c m
\ee
with $\alpha_c$ a (numerically) determined constant the condensate solution will exist and be preferred. The parameter $\alpha_c$ corresponds
to a critical temperature
\be
T_c = \frac{D q \mu}{4 \pi \alpha_c \Delta}. \label{critT}
\ee
with $\Delta$ the dimension. Note that  $\alpha_c \sim 3.68(2.41)$ for $D=3(4)$ for large $\Delta \approx m$. For small values of $m$ in $D=3$ see Table \ref{T:alphac}. For $D=4$ the parameter $\alpha_c$ also increases monotonically with dimension and is bounded as $\alpha_c > 1.4$, with the lowest value obtained at the lowest dimension, $\Delta = 2$.

In the next subsections we consider how this critical temperature may be estimated for small and large values of $m$. The reason for analysing the temperature is that
in top down models a diverse range of the parameters $(q,m)$ may be realised but they are not independent of each other. It is therefore useful to study in detail how the critical temperature varies as these parameters are adjusted, without solving the equations numerically, case by case. Note that numerical solutions become increasingly difficult as one increases $m$.

\begin{table}[]
\begin{center}
\begin{tabular}{|c||c|c|c|c|c||c|}
\hline
$\Delta_+ (m^2)$ & 2(-2) & 3(0)& 3.5(7/4) & 4(4) & 5.25(28) & $\infty$ \\
\hline
$\alpha_c$ & 2.07 & 2.52 & 2.69 & 2.806 & 3.28 & 3.68\\
\hline
\end{tabular}
\end{center}
\caption{$\alpha_c$ as a function of $\Delta_+$ from numerical analysis ($D=3$). We substitute the dual operator dimension $\Delta_+$ for $m$ in \eqref{alphac}; for relevant operators we consider the quantisation leading to a higher dimension, hence the notation $\Delta_+$.}
\label{T:alphac}
\end{table}%

\subsubsection*{Estimation of critical temperature}

In the case in which the dimension of the spin two operator is large one can estimate the critical temperature following the
arguments given in \cite{Benini:2010qc,Benini:2010pr}. The idea is to consider a massive charged point particle in the black
brane background in the ordinary phase, in which
there is a background gauge field. For mass $m$ and charge $q$ the point particle action is
\be
S = - m \int d \tau - q \int A.
\ee
Then the effective potential for a static particle is given by
\be
V_{\rm eff}(\zeta) = \frac{m}{\zeta z_h} (1 - \zeta^D)^{1/2} + q \mu (1 - \zeta^{D-2}),
\ee
with $\zeta = z/z_h$.
The extremum of the potential is at $V'_{\rm eff}(\zeta_o) = 0$; this will be a global minimum
only if $q \mu$ is large enough such that $V_{\rm eff}(\zeta_o) \le 0$, since the potential is zero at the horizon $\zeta = 1$. In $D=3$
both conditions are satisfied when \cite{Benini:2010qc,Benini:2010pr}
\be
| q \mu z_h | \sim 3.68 m,
\ee
which agrees well with numerical results (but slightly overestimates the critical temperature).
In $D=4$ the resulting condition is rather similar:
\be
| q \mu z_{h} | \sim 2.41 m,
\ee
which corresponds to a critical temperature
\be
\frac{q \mu}{T_c} \sim 7.57 \Delta,
\ee
where we use the fact that $\Delta \sim m$ for large dimension. The corresponding value of $\zeta_{o}$ is found to be $0.64$,
which is outside the horizon. Condensation thus occurs at temperatures which are
small compared to $q \mu$; there is no condensation when the chemical potential is switched off.

\begin{figure}[]
\centering
  {
  {\includegraphics[width=7cm]{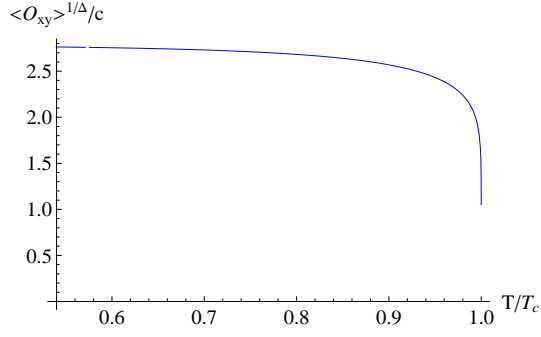}}
  }
  \caption{Condensate:  $D=4, m^2 = 5, \Delta = 5, q = 1,
  c\equiv  \kappa^{-2/\Delta} (\rho/\mu)^{1/(D-2)}$.
} \label{Fig:condensate}
\end{figure}

The actual value of the condensate can be obtained by numerical solution of the field equations, with the charge density $\rho$
and the chemical potential $\mu$ read off from the asymptotic values of the gauge field potential
\be
V(z) = \mu - \rho z^{D-2} + \cdots
\ee
For example, in the case of $D=4$ with $ m^2 = 5, \Delta = 5, \ q = 1 $, figure \ref{Fig:condensate} shows numerical condensates for which $\kappa^2 \< {\cal O}_{xy} \> \sim (\rho/\mu)^{\Delta/(D-2)}$ with a coefficient of order one below $T_c$. (Note however that in this case $T_c/\mu$ is not small, and therefore the probe approximation is not justified. We will return to this issue below.)

\subsection{Analytic approximation}

Consider the dimensionless equations \eqref{dimless}. In the near boundary region, $\zeta \rightarrow 0$, and in the near horizon region, around $\zeta = 1$, the solutions take the form
\begin{equation} \label{VA2}
\begin{split}
\tV(\zeta)& \sim
\begin{cases}
\tV_B \equiv \tmu - \trho \zeta^{D-2} & \text{if } \zeta \rightarrow 0 \\
\tV_H \equiv  \tv_0 -\tv_1\left[(1-\zeta) + \frac{1}{2}\left(-3+D-\frac{\tH_0^2}{D}\right)(1-\zeta)^2  \right] & \text{if } \zeta \rightarrow 1
\end{cases}  \,, \\
\tH(\zeta)& \sim
\begin{cases}
\tH_B \equiv \tH_+ \zeta^\D  + \tH_- \zeta^{d-\D}& \text{if } \zeta \rightarrow 0 \\
\tH_H \equiv \tH_0\left[1 + \frac{ \tm^2}{D}(1-\zeta) + \frac{(2D\tm^2 + \tm^4 - v_1^2)}{4D^2}(1-\zeta)^2 \right]  & \text{if } \zeta \rightarrow 1
\end{cases} \,.
\end{split}
\end{equation}
The near boundary data, $\tmu$, $\trho$ and $\tH_+$,  are proportional to the chemical potential, the charge density, and the one point function of the spin two operator respectively:
\begin{equation}
    \mu = \tmu  \frac{4\pi T}{q D} \,, \quad
    \rho = \trho \frac{1}{q} \left(\frac{4\pi T}{D}\right)^{(D-1)} \,, \quad
     \langle \calo_{xy} \rangle = \tH_+ \frac{(2\Delta - D)}{\kappa^2 \sqrt{2g} q} \left(\frac{4\pi T}{D}\right)^\D  \,. \label{tdmudef}
\end{equation}
The coefficient $\tH_-$ corresponds to the source of the operator and thus we set $\tH_- = 0$.
Imposing regularity at the horizon, we choose $\tv_0 = 0$ for $\tV$ and have dropped the other independent singular solution of $\tH$.

Now let us match the solutions at an intermediate value $\zeta_m$, requiring continuity of the functions $(\td{V},\td{h})$ and their first
derivatives. This gives four conditions
\begin{align}
  & \tV_B(\zeta_m) = \tV_H(\zeta_m) \,, \quad \dot{\tV}_B(\zeta_m) = \dot{\tV}_H(\zeta_m) \label{Vcond} \,, \\
    & \tH_B(\zeta_m) = \tH_H(\zeta_m) \,, \quad \dot{\tH}_B(\zeta_m) = \dot{\tH}_H(\zeta_m) \label{hcond} \,,
\end{align}
with six unknowns: $\tmu,\trho,\tv_1,\tH_+, \tH_0, \zeta_m$.
We will express four variables $\tmu,\tv_1,\tH_+, \tH_0$ in terms of $\trho,\zeta_m$.
First, from
\begin{equation}
  \frac{\tH_B(\zeta_m)}{\tH_B'(\zeta_m)} = \frac{\tH_H(\zeta_m)}{\tH_H'(\zeta_m)} \,,
\end{equation}
we have $\tv_1$ as a function of $\zeta_m$ only.
\begin{equation}
  \tv_1^2 = 2D\tm^2 + \tm^4 + \frac{2D\Delta(\tm^2 + D(2-\Delta ))}{\Delta(1-\zeta_m) + 2\zeta_m} + \frac{2D(\tm^2+D\Delta)}{1-\zeta_m} \,. \label{v12}
\end{equation}
The second equation of \eqref{Vcond} yields
\begin{equation}
  \tH_0^2 = \frac{(D-2)\zeta_m^{D-3} \trho - \tv_1\left[4-D +(D-3)\zeta_m\right]
          }{\tv_1(1-\zeta_m)/D} \,, \label{h02}
\end{equation}
with which, the first equation of \eqref{Vcond} gives
\begin{equation}
  \tmu = \frac{\tv_1(1-\zeta_m)}{2} + \frac{\trho\zeta_m^{D-3}\left(D(1-\zeta_m)+4\zeta_m-2\right)}{2} \,.
\end{equation}
Before discussing $\tH_+$, let us fix the sign of $\tv_1$ and $\tH_0$.
The expression of $\tH_0$ \eqref{h02} gives some constraints. First of all, $ [4-D +(D-3)\zeta_m] $ is mostly negative for $D > 5$. However,
if one chooses $\eta_m$ such that $ [4-D +(D-3)\zeta_m] < 0 $, we cannot see the condensed phase, whatever the sign of $\tv_1$ is.
So, this physical consideration restricts ourselves to $D=3,4$ and $[4-D +(D-3)\zeta_m] > 0$. This restriction may however be due to the limitations of our crude approximation.  The second constraint is that only $\tv_1 >0$ is allowed for $\tH_0^2 >0$. Thirdly, the sign of $\tH_0$ is arbitrary and does not matter by symmetry so we choose a positive sign.

Finally, $\tH_+ (\sim \langle \calo_{xy} \rangle)$ is solved by \eqref{hcond} with \eqref{v12} and \eqref{h02}. It is written in terms of physical variables as
\begin{equation}
  \frac{\langle \calo_{xy} \rangle}{T_c^\D} = \cala(D,\tm,\D,\k,\zeta_m,q)
  \sqrt{\left(\frac{T}{T_c}\right)^{2\D-D+1}}\sqrt{1-\left(\frac{T}{T_c}\right)^{D-1}}\,, \label{Acondensate}
\end{equation}
where
\begin{align}
  \cala &= \frac{2\D - D}{\kappa^2 \sqrt{2g} q} \left(\frac{4\pi}{D}\right)^\D
  \frac{(2D+\tm^2(1-\zeta_m))\zeta_m^{1-\D}}{D\D - D(\D-2)\zeta_m}
  \sqrt{\frac{D(4-D+(D-3)\zeta_m)}{1-\zeta_m}} \,, \\
    T_c &= \frac{D}{4\pi} \left(\frac{(D-2) \, q \, \zeta_m^{D-3}}{v_1(4-D + (D-3)\zeta_m)}  \right)^{\frac{1}{D-1}}  \rho^{\frac{1}{D-1}} \,, \\
  v_1 &= \sqrt{2 D \tm^2 + \tm^4 - \frac{2D(-\tm^2 + D(\D-2))\D}{\D(1-\zeta_m) + 2\zeta_m} + \frac{2 D \D^2}{1- \zeta_m}} \,.
\end{align}
A real solution for $\langle\calo_{xy}\rangle$ exists only when $T < T_c$ and the critical exponent is $1/2$ for all parameter set ($\calo_{xy}  \sim \sqrt{1-T/T_c}$).

With a matching point $\zeta_m=0.5$, the formula is simplified as
\begin{align}
  \cala &= \frac{2\D - D}{\kappa^2 \sqrt{2g} q} \left(\frac{\pi}{D}\right)^\D
   \frac{2^{3\D-1} (4D + \tm^2)\sqrt{-1+5/D}}{\D + 2}\,, \\
    T_c &= \frac{D}{4\pi} \left( \frac{2^{4-D}(D-2)q}{(5-D)v_1} \right)^{\frac{1}{D-1}}  \rho^{\frac{1}{D-1}} \,, \\
  v_1 &= \sqrt{\frac{16 D^2 \D + \tm^4(2+\D) + 2 D m^2 (6+5\D)}{2+\D}} \,.
\end{align}
For example, with the parameters ($D=3, \tm^2 = -2, \D = 2 , \sqrt{2g} =1, q=1$), \eqref{Acondensate} is reduced to
\begin{align}
  \frac{\langle \calo_{xy} \rangle}{T_c^2} &=  \frac{16 \pi^2}{9 \kappa^2}\frac{2+\zeta_m}{3\zeta_m} \sqrt{\frac{3}{1-\zeta_m}} \frac{T}{T_c} \sqrt{1-\left(\frac{T}{T_c}\right)^{2}}\,,  \label{condA} \\
  T_c &= \frac{3}{4\pi} \left(\frac{1-\zeta_m}{4(1+5\zeta_m)}\right)^{1/4}
  \rho^{1/2} \,,
\end{align}
which agrees with (A.30) and (A.31) in \cite{Gregory:2009fj}. (Note that such a parameter choice is valid for scalar fields but is invalid for spin 2 fields, as $m^2 \ge 0$ for unitarity.)
One has to treat this approximation with some caution since the matching point was arbitrary. However, the simple analytic calculation is in good agreement with the numerical results, particularly close to the transition temperature. Further discussion of the structure of the coupled equations is given in the appendices and will be used in the later discussions.

\subsection{Validity of approximations}

Now let us turn to the conditions under which the use of the probe equations of motion can be valid. There are two main issues. First of all, the backreaction of the matter fields on the metric has been neglected. Secondly, the action is not causal, although to suppress acausal effects one can require that all components of the gauge field strength in a tangent frame are small relative to $m^2/q$.

Looking at the first issue, notice that in the ordinary state (in which the spin two field is switched off) this condition is actually unnecessary as Einstein gravity with negative cosmological constant and gauge field is a well posed theory, which can moreover be obtained as a consistent truncation of supergravity reduced on a Sasaki-Einstein manifold. If nonetheless we consider the limit in
which the backreaction of the gauge field strength on the metric can be neglected in the ordinary state, then necessarily
\be
T \gg \mu. \label{tcp}
\ee
Noting that the critical temperature $T_c$ is smaller than $q \mu$
the neglect of the backreaction is only consistent with reaching the condensate phase if  $\mu \ll T_{c} <  q \mu$, which requires large $q$. 
This is the standard limit in which the probe approximation applies, following \cite{Hartnoll:2008kx,Horowitz:2010gk}. 
More precisely, however, looking at the condition for the critical temperature \eqref{critT} one sees that the condition required is
\be
q \gg \frac{4 \pi \a_c m}{D}. \label{largeq}
\ee
In a bottom up model in which $q$ and $m$ are independent, one can always choose $q$ such that this condition holds. In a top down model, or indeed 
taking into account the bounds on R charge for operators of a given dimension, $q$ cannot be arbitrarily large for a given $m$. In the following section we 
will consider the case where this large charge condition is dropped. 

Let us turn now to the second issue of causality and hyperbolicity violations.
It was pointed out in \cite{Benini:2010qc,Benini:2010pr} that these could be suppressed by imposing \eqref{sup}, which in turn requires that the dimension of the spin two operator is such that $\Delta \gg 1$.
One should note however that such a condition is not needed for the static condensate solution to exist; it would be needed in analysing the dynamics of fluctuations around the condensate. In other words, for the effective action to be valid for discussing the stability of the condensate, perturbations around the condensate should behave causally. If they do not, then the action is inadequate for describing the stability of the condensate. However, in the later sections of the paper we will argue
that in top down models one will in any case need to take into account other fields when discussing the stability, and the stability is best analysed directly from the higher dimensional equations of motion. 

\subsection{Near critical temperature} \label{subct}

We now consider the case where the parameter choices $(q,m)$ are such that the critical temperature obtained is not large compared to the chemical potential. In this case the probe approximation is insufficient to analyse the existence of a condensed phase. However, one can analyse such cases as follows.
In the ordinary phase the backreacted (Reissner-Nordstrom) metric can be written as
\bea
ds^2 &=& \frac{1}{z^2} \left ( - F(z) dt^2 + d \vec{x}^2 + \frac{dz^2}{F(z)} \right ); \label{rn} \\
F(z) &=& f(z) + \frac{\mu^2 z_h^2}{\gamma^2} \frac{z^D}{z_h^D} \left (\frac{z^{D-2}}{z_h^{D-2}} - 1 \right ); \nn \\
A &=& \mu \left (1 - \frac{z^{D-2}}{z_h^{D-2}} \right ) dt \equiv V^o(z) dt , \nn
\eea
where $f(z)$ was given in \eqref{neutr} and
\be
\gamma^2 = \frac{2 (D-1)}{(D-2)}.
\ee
The temperature is now given by
\be
T = \frac{D}{4 \pi z_h} \left (1 -  \frac{ (D-2) \mu^2 z_h^2}{\gamma^2 D} \right ) \equiv \frac{D}{4 \pi z_h} \left ( 1 - \frac{(D-2)}{D} \eta \right )
\ee
with
\be
\eta = \frac{\mu^2 z_h^2}{\gamma^2},
\ee
where $z_h$ still denotes the horizon location and the parameter $0 \le \eta \le D/(D-2)$. One can also write
\be
\frac{T}{\mu} = \frac{D}{4 \pi \gamma \sqrt{\eta}} \left (1 - \frac{(D-2)}{D} \eta \right ). \label{up}
\ee
The extremal limit of the black hole is obtained when $\eta \rightarrow D/(D-2)$ with the uncharged limit being $\eta \rightarrow 0$.

We now consider the static ansatz for the spin two field, $\phi_{xy} = 1/(2z^2) h(z)$, in this background.
Close to a (continuous) phase transition the condensate field $h(z)$ would scale as $\epsilon = (1 - T/T_c)^{\lambda} \ll 1$ where $\lambda$ determines the critical exponent, see the discussions in appendix \ref{appc}. The exponent obtained from the non-linear equations is $\lambda = 1/2$.
In this limit the backreaction of the spin two field on both the metric and on the gauge field would be of order $\epsilon^2$ and thus negligible. So to analyse the existence of a condensate phase we need to look at the linearized equation of motion for the spin two field in the charged black brane background. Using the spin two equation given in \eqref{uueq} for a transverse and traceless field we obtain
\be
\left(\Box - m^2 \right) \phi_{\mu \nu} = 2 R_{\mu \rho \lambda \nu} \phi^{\rho \lambda}. \label{Eff1}
\ee
The resulting equation is very similar to that obtained in \eqref{higgs-eq},
\be
h'' + \left ( \frac{F'}{F} - \frac{D-1}{z} \right ) h' + \left ( \frac{q^2 (V^o)^2}{F^2} - \frac{m^2}{z^2 F} \right ) h = 0.
\ee
Rewriting in terms of the dimensionless variable $\zeta$ one obtains the linear equation
\be
\ddot{{h}} + \left ( \frac{\dot{F}}{F} - \frac{D-1}{\z} \right ) \dot{{h}} + \left ( \frac{\tilde{V}^2}{F^2}
- \frac{m^2}{\z^2 F}   \right ) {h} = 0,
\ee
with $F(\zeta) = (1 - \zeta^D + \eta \zeta^D (\zeta^{D-2} - 1)))$ and
\be
\tilde{V} = q \mu z_h \left(1 - \zeta^{D-2}\right).
\ee
The linearised equation needs to be solved subject to the condition that the field is regular at the horizon and has no source behaviour at infinity. From the rescaling it is manifest that the condensation condition can involve only the parameters $(m, m_o = q \mu z_h,z_h)$ together with the dimensionless number $\eta$ which characterises the non-extremality of the black hole.

The critical temperature $T_c$ can be parameterised as
\be
T_c = \frac{D}{4 \pi \alpha_c (\eta,m)} \frac{q \mu}{m} \left (1 - (D-2) \frac{\eta}{D} \right ). \label{crit-2}
\ee
where the parameter $\alpha_c(\eta,m)$ now depends on $\eta$ as well as on $m$. For fixed chemical potential one can combine this equation with \eqref{up} to obtain a bound on the parameter $\eta$ for a condensate to form:
\be
\eta \ge \frac{\alpha_c(\eta,m)^2}{\gamma^2} \frac{m^2}{q^2}. \label{eta}
\ee
Repeating the estimates of the critical temperature and the analytic approximations, one finds that $\alpha_c(\eta)$ is actually only weakly dependent on $\eta$. For example, if one uses the effective potential technique for high masses, the relevant potential is now
\be
V_{\rm eff}  \propto \frac{1}{\zeta} \left(1 - \zeta^D + \eta \zeta^D (\zeta^{D-2} - 1)\right)^{1/2} - \frac{|m_o|}{|m|} \left(1  - \zeta^{D-2}\right)
\ee
with $\alpha_c (\eta)$ being the value of $|m_o/m|$ for which the potential admits a minimum at a negative value. The parameter $\alpha_c$ decreases with increasing $\eta$ but remains of the same order. In $D=3$ at large $m$ we find $\alpha_c (0) \sim 3.68$ and $\alpha_c(1.5) \sim 3.35$ while $\alpha_c (3) \sim 2.46$ (with $\eta \rightarrow 3$ being the extremal limit). In $D=4$ at large $m$ we find $\alpha_c(0) \sim 2.41$ and $\alpha_c(1) \sim 2.24$ with $\alpha_c(2) \sim 1.74$ (where $\eta \rightarrow 2$ is the extremal limit).

Returning to \eqref{eta}, and recalling that $\eta \le D/(D-2)$, we see that a condensate only forms at finite temperature if the mass and charge parameters of the condensate satisfy
\be
\frac{m^2}{q^2}\le \frac{2 D (D-1)}{(D-2)^2 \alpha_c^2} \label{mc-rel}
\ee
For $D=3$ a conservative estimate (at large $m$) would be $m^2/q^2 < 4$ and in $D=4$ at large $m$ we find that $m^2/q^2 < 2$.

For smaller values of the mass we can use the methods discussed in appendix \ref{smallm-meth} to estimate $\alpha_c$. As one would anticipate from the previous results, $\alpha_c$ decreases with decreasing operator dimension and decreasing temperature, i.e. closer to the extremal limit. Looking at the lowest operator dimension in $D=2$, $\Delta = 2$, one can estimate that $\alpha_c(0) \sim 1.4$ while $\alpha_c(1.75) \sim 1.1$. (For larger values of $\eta$ the approximation becomes increasingly sensitive to the details of the stability criteria.)  Therefore in the low dimension limit in $D=4$ we see that
\be
\frac{|m^2|}{q^2} \le 6,
\ee
where the magnitude is used since for relevant dimension operators $m^2 < 0$. A similar bound was discussed in \cite{Horowitz:2010gk}.
We will use these facts later when discussing the modes which arise in Kaluza-Klein reductions.

\subsection{Thermodynamics} \label{therm}

The existence of a condensate solution does not guarantee that the condensate phase is thermodynamically favoured over the ordinary phase. In this section we will analyse the thermodynamics and show by computing the onshell action that the condensate phase is indeed preferred.

Naively one might think that one cannot compute the onshell action to the required order without analysing the backreaction of the metric. In the probe approximation, we work with a fixed background which solves the field equations and solve the gauge field/charged field equations in this background, for small chemical potential
$\mu \ll T$ and fixed $(m,q)$. If one looks at the contribution to the onshell action from the gauge field, it would appear to be of the same order ($\mu^2/T^2$) as the backreaction of the gauge field and charged field on the metric, which suggests that one needs to know the backreaction to compute the onshell action.

This naive reasoning is not correct, however. To understand this point let us consider a more general situation where one has an action $S[g,\Phi^M]$  involving a metric $g_{\mu \nu}$  and a collection of matter fields $\Phi^M$ (which are not necessarily scalar fields). Let $g^o_{\mu \nu}$ and $(\Phi^o)^M$ be a solution of the equations of motion following from the action and consider perturbations around this solution such that
\be
g_{\mu \nu} = g^{o}_{\mu \nu} + \Delta g_{\mu \nu}; \qquad
\Phi^M = (\Phi^o)^M + \Delta \Phi^M.
\ee
The onshell action for the perturbed solution is then
\be
S_{\rm onshell} (g,\Phi^M)  = S_{\rm onshell} [g^o, (\Phi^o)^m] + \delta S_{\rm onshell}
\ee
with the change in the action being computed using
\be
\delta S = \int d^{d+1} x \left ( \frac{\delta \cal L }{\delta g^{\mu \nu}} \delta g_{\mu \nu} +
\frac{\delta \cal L}{\delta \Phi^M} \delta \Phi^M \right ),
\ee
with ${\cal L}$ being the Lagrangian.
By definition the action is stationary on any onshell solution. In particular, this implies that the first
variation of the action with respect to the background solution is zero (for any variation) since
\be
\left.\frac{\delta \cal L }{\delta g^{\mu \nu}}\right|_{g^o, (\Phi^o)^m}  = 0; \qquad
\left. \frac{\delta \cal L}{\delta \Phi^M}\right|_{g^o,(\Phi^o)^m}  = 0.
\ee
Therefore the change in the onshell action is quadratic in the perturbations of the metric and matter fields,
\be
\delta S_{\rm onshell}  = \int d^{d+1} x \left ( \left.\frac{\delta \cal L }{\delta g^{\mu \nu}} \right|_{\Delta g, \Delta \phi^M} \Delta g_{\mu \nu} +  \left.\frac{\delta \cal L}{\delta \Phi^M} \right|_{\Delta g, \Delta \Phi^M}   \Delta \Phi^M \right ).
\ee
For example, in the case of a purely gravitational action, the change in the onshell action would be expressed in terms of the Einstein operator acting on the metric perturbation, and would be quadratic in the metric perturbation.

Going back to the case of interest, one wishes to compare the onshell actions for the ordinary and superfluid phases, at fixed chemical potential $\mu$ and temperature $T$. The metric perturbation is quadratic in the dimensionless ratio $\mu/T$, which in turn implies that the metric contribution to the onshell action is quartic in $\mu/T$. The latter is therefore subleading compared to the matter terms: these perturbations are linear in the chemical potential, and thus give rise to corrections to the action which are quadratic in $\mu/T$.

We can illustrate this point by the ordinary phase solution, for which the backreacted solution is Reissner-Nordstrom \eqref{rn}. As anticipated in the limit of small chemical potential the backreaction on the metric is suppressed by factors of $\mu^2 z_h^2 \sim \mu^2/T^2$. The exact expression for the (holographically renormalized) onshell Euclidean\footnote{The Euclidean continuation of the action is denoted by a superscript. In the static case the Euclidean action is obtained by $iS \rightarrow - S^E$.} action in the grand canonical ensemble with $\mu$ fixed is
\be
S^{E}_{\rm onshell} = \frac{\beta}{2 \kappa^2 z_h^D} \left (1 + \frac{ \mu^2 z_h^2}{\gamma^2}  \right ) V_{D-1},
\ee
with $\beta$ the inverse temperature and $V_{D-1}$ the regulated volume in the spatial directions. The Euclidean onshell action is related to the free energy $F$ as $S^E = - \beta F$. Expanding this expression for small $\mu z_h$ one obtains
\be
S^{E}_{\rm onshell}
= \frac{2 \pi V_{D-1}} {D \kappa^2 z_h^{D-1} }  \left (1 + (D-2) \mu^2 z_h^2  + \cdots \right  ).
\ee
By the logic above we should get the same answer, to this order, by evaluating the action for the gauge field in the Schwarzschild background. In other words, the change in the action relative to that of the neutral black hole is
\be
\delta S = - \frac{1}{8 \kappa^2} \int d^{D+1} x \sqrt{-g^o} F^{\mu \nu} F_{\mu \nu},
\ee
where $g^o$ again denotes the background metric. Evaluating this one indeed obtains
\be
\delta S^{E}_{\rm onshell}  = \frac{2 \pi V_{D-1}}{D \kappa^2 z_{h}^{D-3}}  (D-2) \mu^2,
\ee
in agreement with the first correction given above. One could similarly analyse the case with fixed charge density $\rho$, i.e. the canonical ensemble, by adding the boundary term
\be
\frac{1}{4 \kappa^2} \int d \Sigma^{\mu} F_{\mu\nu} A^{\nu}
\ee
which enforces a boundary condition on $n^{\mu} F_{\mu \nu}$ rather than $A_{\mu}$. In this case the onshell action is shifted by $\mu \rho V_{D-1}$, with $\rho$ being the charge density.

Let us now turn to condensed phases, beginning with a scalar model in which the matter part of the action is
\bea
S = \frac{1}{2 \kappa^2} \int d^{D+1} x \sqrt{-g} \left [ - \frac{1}{4} F^{\mu \nu} F_{\mu \nu} - | \nabla \psi - i q A \psi |^2 - m^2 |\psi|^2 \right ].
\eea
Here the complex scalar $\psi$ has mass $m^2$ and charge $q$. For a static solution in which
$A_t = V (z)$ and $\psi = h(z)$ with $h(z)$ real, the coupled equations of motion reduce to \eqref{higgs-eq}. The onshell action for the matter fields in the fixed gravitational background $g^o$ is
\be
\delta S_{\rm onshell} = \frac{1}{2 \kappa^2} \int d^{D+1} x \sqrt{-g^o} \left [ - \frac{1}{4} F^{\mu \nu} F_{\mu \nu} - \nabla_{\mu} (\psi \nabla^{\mu} \psi) \right ].
\ee
The scalar term vanishes because
\be
\int d \Sigma^{\mu} \psi \nabla_{\mu} \psi
\ee
vanishes both on the horizon (due to $\psi$ remaining finite there) and at infinity (since $\psi$ falls off fast enough). Therefore the only term contributing in the onshell action is
\be
\delta S_{\rm onshell} = \frac{1}{2 \kappa^2} \int d^{D+1} x \sqrt{-g^o} \left [ - \frac{1}{4} F^{\mu \nu} F_{\mu \nu}  \right ] \label{kinetic}
\ee
(together with the appropriate boundary term, if one works instead in the canonical ensemble). Comparison of the free energy for the ordinary and condensed phases therefore involves only the kinetic energy of the gauge field. The explicit numerical solutions, see appendix \ref{number}, show that the kinetic energy of the gauge field in the condensed field is such that the condensed phase is thermodynamically favoured.

For the spin-2 model, substituting \eqref{ans-spin2} into the action \eqref{action-Herzog} and using the field equations, one obtains
\bea
\delta S_{\rm onshell} &=& \frac{1}{2 \kappa^2} \int d^{D+1} x \sqrt{-g^o} \left [ - \frac{1}{4} F_{\mu \nu} F^{\mu \nu} \right ] \\
&& - \frac{1}{\kappa^2} \int d \Sigma^z ( \phi_{xy} \pa_z \phi^{xy} ). \nn
\eea
Working out the boundary term in the second line one finds
\be
- \frac{1}{2 \kappa^2} \int d^{D}x \frac{ h(z) f(z)}{z^D} \left ( \pa_z h - \frac{2 h}{z} \right ).
\ee
However, this boundary term vanishes on the horizon $z= z_h$ since $f(z_h) = 0$ and $h(z_h)$ is finite (condensed phase) or zero (ordinary phase). The boundary term also vanishes at infinity because of the falloff behaviour of the spin-two field: $h \sim z^{\Delta}$ as $z \rightarrow 0$ while $f(z) \rightarrow 1$ as $z \rightarrow 0$ and thus the boundary term scales as a positive power of $z$ as $z \rightarrow 0$. Therefore the onshell action for the spin two model is also given by \eqref{kinetic}, as in the case of the scalar condensate.  On the one hand this result is not entirely surprising, since the equations of motion are the same in the two cases, but the latter did not guarantee that the onshell actions would be the same.

Close to the critical temperature we argued that the condensate scales as $\e$ and the backreaction on the metric and gauge fields scales as $\e^2$. Since the spin two contribution to the action vanishes because of the boundary conditions, the change of the action in the condensed phase scales as $\e^4$. Recalling that
$\e = (1 - T/T_c)^{1/2}$ this gives a change in the free energy of order $(1 - T/T_c)^2$, which is what one would expect for a second order phase transition. To compute this change in the free energy, the backreaction on the gauge field is needed for small chemical potential $\mu/T \ll 1$, while the backreaction on both the gauge field and on the metric is needed at finite chemical potential, in which the ordinary phase is the charged black hole.

\subsection{Competing phases}

The analysis in the previous sections has an interesting interpretation in terms of competing phases. In any top down model, such as those obtained by Kaluza-Klein reductions, there will be a multitude of charged fields. Our analysis shows that, for fixed chemical potential, scalar and spin two fields having the same mass and charge parameters would condense from the ordinary phase at the same temperature.

If there are no (direct) interactions between the scalar and spin two, then one would expect a phase of mixed symmetry, in which both an s-wave condensate and a d-wave condensate are present. If one includes direct interactions, and indeed indirect interactions via the metric and gauge field, then the two phases would presumably compete with each other and one would need to determine which is indeed the preferred phase.

In general the phase structure will become very complicated even in a bottom up model involving only metric, gauge field, charged scalar and charged spin two field.  For example, to determine the preferred phase one also needs to take into account that there may be qualitatively different solutions of the field equations (as well as the mixed symmetry but isotropic phases).

In a top down model containing towers of charged fields, the phase structure is potentially very rich but would also be very difficult to analyse. From \eqref{crit-2}, one can infer that the critical temperature behaves
as
\be
\frac{T_{c}}{\mu} \sim \frac{q}{m},
\ee
which implies that, for fixed chemical potential, the critical temperature is increased by higher charge and lower mass (dimension). This is unsurprising: the lowest dimension operators become important before operators of higher dimension, and increasing the charge corresponds to increasing the coupling to the gauge field. It is interesting to note that, in a reduction in which the spin two field is the lightest charged field, it would be the first to condense into a superfluid phase.

\section{Kaluza-Klein approach} \label{kk-ap}

In this section we turn to the question of whether and under which conditions the spin two model can be obtained from the reduction of a higher dimensional supergravity theory.

To begin our discussion let us note that massive spin two fields do indeed generically arise as Kaluza-Klein modes in reductions leading to anti-de Sitter solutions. To see this let us denote the higher dimensional metric as $G_{mn}$ and consider the most common situation, in which the reduction to anti-de Sitter is diagonal over a Sasaki-Einstein manifold. Then the higher dimensional metric for the AdS case can be expressed as
\be
ds^2 = G_{mn} dx^{m} dx^{n} = \bar{g}_{\mu \nu} dx^{\mu} dx^{\nu} + g^o_{ab} dy^a dy^b,
\ee
with $\bar{g}_{\mu \nu}$ the anti-de Sitter metric and $g^o_{ab}$ the Sasaki-Einstein metric, and this metric is supported by appropriate fluxes. Such a solution can immediately be generalized to
\be
ds^2 = {g}^o_{\mu \nu} dx^{\mu}dx^{ \nu} + g^{o}_{ab} dy^{a} dy^{b}
\ee
in which ${g}^o_{\mu \nu}$ is the metric on any negative-curvature Einstein manifold $M$, with the fluxes being proportional to the volume forms on the compact and non-compact manifolds. (This follows from the fact that Einstein gravity with a negative cosmological constant is always a consistent truncation of the higher dimensional theory.)

The spectrum of the theory is then obtained by considering linearised fluctuations around such a background, and then diagonalising the linearised field equations. Since the metric fluctuations are coupled to the flux fluctuations, this diagonalization is rather non trivial and will be discussed in detail below for the specific case of  compactifications of type IIB to five dimensions. It turns out, however, that in the case of the massive spin two fields, the diagonalization is in fact immediate:  the spin two field arises from the metric fluctuation
\be
g_{\mu \nu} = g^{o}_{\mu \nu} + h_{\mu \nu}.
\ee
This fluctuation $h_{\mu\nu}$ depends on all coordinates, both $x^{\mu}$ and $y^a$, and can be expressed in terms of the complete basis formed by the harmonics of the Sasaki-Einstein manifold. Since it is a scalar from the perspective of the latter manifold, it must be decomposable in terms of  the scalar harmonics $Y^{I}(y^a)$ on the compact space as
\be
h_{\mu \nu} = \sum_{I} h_{\mu \nu}^{I}(x^{\rho}) Y^{I}(y^a).
\ee
Each such mode results in a spin two field in the lower dimensional space, whose mass is given in terms of the eigenvalue of the harmonic on the compact space. (The explicit formulae in the case of type IIB reductions will be given below.) The case in which the scalar harmonic is trivial corresponds to the massless graviton. Massive spin two fields corresponding to Kaluza-Klein gravitons are also charged with respect to the gauge fields in the lower dimensional space, since the corresponding scalar harmonics transform in non-trivial representations of the isometry group of the compact space.

Thus the Kaluza-Klein spectrum indeed contains charged massive spin two fields. As highlighted earlier, the equations of motion at the linearised level do not however suffice to analyse condensate formation. If one works in the neutral black brane background, the higher order terms in \eqref{action-Herzog}, in particular the covariant derivative terms, are necessary to generate a condensate solution - without the cubic interaction terms in \eqref{higgs-eq} no condensed solution exists. 
However, once one allows for interactions involving the spin two field, then one does not expect that the effective action describing the spin two, metric and gauge field can involve only a finite number of terms. The question is thus whether from a top down perspective one can obtain effective equations of motion equivalent to those obtained from \eqref{action-Herzog}. If one works in the Reissner-Nordstrom background, as in section \ref{subct}, the spin two equation is linear, but the gauge field corresponds to a non-diagonal reduction and one needs to take into account non-linearities to probe the structure away from the critical temperature.

In this section we show that, at least close to the superfluid transition temperature, the bottom up equations can indeed capture the leading terms in the equations of motion. We also show that the gyromagnetic ratio $g$ is computable in terms of certain cubic overlaps of harmonics on the compact manifold and therefore takes a specific value. Although the gyromagnetic ratio does not play any role in the superfluid condensate solution, it is useful to derive the value that arises from top down considerations.

\subsection{Consistent truncations}

A consistent truncation consists of (i) a finite set of lower dimensional fields including the metric $g_{\mu \nu}$ together with
other matter fields, denoted schematically by $\phi^I$, which satisfy certain equations of motion and (ii) a (non-linear) map
relating these lower dimensional fields to ten-dimensional supergravity fields $(G_{mn}, F_{mnpqr}, \cdots)$. The truncation is
said to be consistent iff the lower dimensional field equations together with the uplift map give solutions of the ten-dimensional
equations of motion.

In the context of sphere, or more generally Sasaki-Einstein, reductions, the consistent truncation has traditionally been considered to be a gauged supergravity theory, or a consistent subsector of such a theory. More recently novel consistent truncations have been found in which certain higher mass fields are retained \cite{Maldacena:2008wh,Cassani:2010uw,Skenderis:2010vz,Gauntlett:2010vu,Liu:2010sa}; the corresponding operators in the dual CFT are
no longer in the stress energy tensor supermultiplet. As discussed in \cite{Skenderis:2010vz} the existence of such novel consistent truncations follows from the OPE of the dual operators.

In the case of interest here, however, one does not expect that the lower dimensional action can be part of a consistent truncation in which a finite number of fields is retained. Indeed, the problems with causality and hyperbolicity of the action \eqref{action-Herzog} are indicative that it cannot be a consistent truncation.

On the other hand, the lower dimensional metric and gauge field can by themselves form a consistent truncation, with explicit formulas for the reduction being given in the case of spherical reductions in \cite{Cvetic:1999xp}, see section \ref{uplift}. 
This fact immediately implies restrictions on the couplings between the gauge field and all other Kaluza-Klein fields. Suppose we denote the lower dimensional fields by $\Phi^M$, which consists not just of fields in the lower
dimensional consistent truncation $\phi^I$, but also all other Kaluza-Klein excitations on the compactification manifold. Let us denote the latter by $\phi^A$, so that $\Phi^M = \{ \phi^I, \phi^A \}$. Consistent truncation requires that the fields $\phi^I$ do not source $\phi^A$, which in terms of the effective lower dimensional
action implies that there are no couplings involving only one $\phi^A$
\be
\int d^{d+1} x \sqrt{-g} (\pa^n \phi^A) \prod_{I} \pa^p \phi^I,
\ee
where $\pa$ denotes schematically derivatives, and $(n,p)$ denote integers. 

Conversely any field $\phi^A$ will generically act as a source for many other fields in the reduction. Thus for the massive spin two field of interest we would expect that we cannot decouple it from the other Kaluza-Klein fields. i.e. a solution of the set of equations following from \eqref{action-Herzog} would not solve the higher dimensional field equations,  as the spin two field would source other Kaluza-Klein excitations. 

\subsection{Systematic expansion of equations}

Next we review what is known about the interactions between generic Kaluza-Klein fields. This question might at first seem intractable since one has an infinite tower of such Kaluza-Klein fields interacting with one another. However, 
in the literature the interactions have been derived using an expansion in the number of fields, for two distinct but related purposes.

The first application is in computing correlation functions around a fixed holographic background, dual to a given state in the field theory. One then considers all fluctuations to be small and in the equations of
motion terms involving $(\Phi^M)^{n+1}$ are by construction always smaller than terms involving $(\Phi^M)^{n}$, regardless of the field $\Phi^M$ under consideration. Following the holographic dictionary of \cite{Gubser:1998bc,Witten:1998qj} quadratic terms give the free action and two point functions in the dual theory, with cubic terms needed to compute three point functions and so on. In practice only a subset of cubic terms and a small number of quartic terms have been computed even in maximally supersymmetric compactifications, since considerable computation is needed.

A second related use of equations of motion expanded in the number of fields is the following. Suppose one is given an exact solution of the higher dimensional supergravity which is asymptotically $AdS \times SE$. Even if this cannot
be reduced to a lower-dimensional solution of a consistently truncated theory, one can systematically
extract holographic data from the higher-dimensional solution using the method of Kaluza-Klein holography \cite{Skenderis:2006uy}. The basic idea is to express
the asymptotic solution near the $AdS$ conformal boundary as a perturbation of $AdS \times SE$.
Let us focus on a specific example, that of IIB supergravity, for which the asymptotic form of the metric and five form should then be
\be
G_{mn} = g^{o}_{mn} + h_{mn}; \qquad F_{mnpqr} = F^{o}_{mnpqr} + f_{mnpqr},
\ee
where $g^{o}_{mn}$ is the $AdS \times SE$ metric and $F^{o}_{mnpqr}$ is the flux in that background, with $h_{mn}$ being the metric fluctuation and $f_{mnpqr}$ being the five form fluctuation. The dilaton $\Phi = \phi^o + \phi$ and the axion $C = C^o + c$ similarly approach constant values $(\phi^o,C^o)$ near the AdS boundary and the asymptotic behaviour of the complex three-form $G_{mnp} \equiv g_{mnp}$ is fixed similarly.

Kaluza-Klein holography provides a precise map between the coefficients in the asymptotic expansion of the
fluctuations $(h_{mn}, f_{mnpqr}, \phi,c,g_{mnp})$ and the expectation values of chiral primary operators in the dual CFT. Operationally this is achieved by expanding these fluctuations in the basis of Kaluza-Klein harmonics, and then carrying out a (rather complicted, non-linear) reduction to obtain the effective five-dimensional action. The crucial point is that in obtaining the one point of an operator of dimension $\Delta$ one needs only know the non-linear interactions of operators whose dimensions sum to $\Delta$. For the lowest dimension operators, which control the leading behaviour in the IR, this implies that one only needs to calculate the lowest order interactions iteratively: cubic, quartic and so on.

\subsection{Spin two equations}

The action \eqref{action-Herzog} cannot arise through a consistent truncation but as we discussed above that such terms do arise in the expansion of the effective lower dimensional action in terms of the number of fields. In writing down the equations of motion (\ref{higgs-eq}, \ref{dimless}), we are using a probe approximation and thus the fields are implicitly assumed to be small. This would suggest that the expansion of the lower dimensional action in terms of the number of fields may be valid but this is subtle since the cubic terms in (\ref{higgs-eq}, \ref{dimless}) cannot be viewed as negligible relative to the linear terms.
If however cubic terms are not subleading to the linear terms, then it is not obvious why higher point interactions should be suppressed and why other fields in the Kaluza-Klein tower would not be induced.
 
Letting $\Phi$ be a generic Kaluza-Klein field, the issue is that the equations (\ref{higgs-eq}, \ref{dimless}) should in general receive additional contributions of the schematic form
\bea
V'' + \frac{3 - D}{z} V' -  \frac{q^2}{z^2 f} h^2 V \sim \sum_{n > 1, p > 1, r > 0}(h^n V^{p} \Phi^r)  \label{q1l} \\
h'' + \left ( \frac{f'}{f} - \frac{D-1}{z} \right ) h' + \left ( \frac{q^2 V^2}{f^2} - \frac{m^2}{z^2 f} \right ) h \sim \sum_{n >1, p > 1, r > 0}(h^n V^{p} \Phi^r), \\
{\cal L} \Phi \sim \sum_{n > 1, p > 1, r > 0} (h^n V^p \Phi^r) \nn
\eea
where ${\cal L}$ is a linear differential operator, describing the dynamics of a free non-interacting field $\Phi$. The terms on the right hand side must be at least quadratic in fields and (by the argument above) the gauge field cannot act as a source for any of the Kaluza-Klein fields. Since the spin two field $h$ is charged, the specific combinations which appear should respect charge conservation. Interaction terms will be computed explicitly in the next section; we will show that as well
as the required cubic terms there are indeed additional terms of the above type. 

In the field equation for the generic Kaluza-Klein field $\Phi$, the $r=0$ terms source $\Phi$, so that setting $\Phi = 0$ is no longer a consistent solution. (Terms with $r > 0$ are less important because $\Phi = 0$ would always be a consistent solution, although it would not necessarily be stable.) In the equations for the spin two field and gauge field, higher point interaction terms are not forbidden by any symmetry (apart from charge conversation) and the induced Kaluza-Klein fields $\Phi$ backreact on these equations also.

Despite the complicated structure of these coupled equations, the terms on the left hand side are sufficient to capture the behaviour near the transition temperature.
The argument can be summarised as follows: following the discussions in appendix \ref{appc} and section \ref{subct}, close to the transition temperature the condensed solution admits an expansion
\be
V = V_o + \e^2 V_2 + \cdots; \qquad
h = \e h_1 + \e^3 h_3 + \cdots,
\ee
with $\e = (1 - T/T_c)^{\lambda}$ for exponent $\lambda$. We do not assume that $V_0$ is small compared to the temperature; we can take into account the backreaction of the ordinary phase gauge field $V_o$ by using the appropriate charged black brane solution; by choosing the $U(1)$ appropriately we obtain the Reissner-Nordstrom brane, with no other scalar fields induced.
In the following section we will show that any other Kaluza-Klein field induced by the condensate is of order $\e^2$ or smaller. This in turn implies that the terms on the right hand side of the equation for the spin two are of order $\e^2$ or smaller, and thus are subleading to those on the left hand side. In other words, even though other Kaluza-Klein fields are indeed generically induced, they do not affect the size of the condensate close to the transition temperature. The d-wave phase is not necessarily the most stable phase even close to the transition temperature and one would need to analyse the backreaction on other fields to explore stability.

\bigskip

Below the transition temperature, two factors could still suppress contributions to the spin two/gauge field equations from other Kaluza-Klein fields. As discussed earlier, large charge $q$ can suppress the backreaction of the condensate field. In such a limit, when the condensate field is small, but $q$ times the condensate is of order one, only terms involving appropriate numbers of powers of $q$ on the righthandside of \eqref{q1l} can compete with the terms on the left hand side. However, as we will see in the next section, factors of $q$ arise from taking derivatives on the compact space. The higher dimensional theory is only two derivative and so generically one does not expect higher point interactions to appear with more than two powers of $q$. Making this argument precise, however, would involve the complex details of the reduction formulae. Another factor which could suppress the effects of the higher point interactions is their finite near horizon behavior. As discussed in the appendices, the emergence of a condensed solution at low temperature is rather robust. Adding to the spin two equation additional terms
which are finite everywhere would be unlikely to change the condensed solution qualitatively.

Analyzing these possibilities in detail, and working out the backreaction of the condensate, using the equations \eqref{q1l} would be complicated. In section \ref{uplift} we will discuss the condensed phase directly from the higher-dimensional perspective. The backreaction of the condensate can  be computed directly in the higher dimensional theory and such an approach is likely to be easier than working with the dimensionally reduced equations \eqref{q1l}. For example, the Coulomb branch solutions
expressed in ten dimensional language are considerably simpler than the solutions of the corresponding  infinity set of five dimensional equations, analyzed in \cite{Skenderis:2006uy}. In that case the complete non-linear solution is best found in ten-dimensional language, although note that the reduction to five dimensions is still necessary to extract the holographic field theory information, one point functions and so on, see \cite{Skenderis:2006uy}.

\section{Reduction of type IIB supergravity} \label{2b}

In this section we will derive the relevant cubic couplings in the low energy action for
Sasaki-Einstein compactifications of type IIB supergravity to five dimensions. We will determine the gyromagnetic ratio in
\eqref{action-Herzog} from the top down model and we will show that this action indeed captures the leading terms close to the temperature at which the spin two condensate forms.

One might wonder why we choose to work with a five dimensional example, rather than one in four dimensions. One reason is for computational convenience as there
is more literature on couplings of Kaluza-Klein fields in reductions to five dimensions. However, there are conceptual reasons for working with five bulk dimensions rather than four: the Coleman-Mermin-Wagner theorem indicates that symmetry breaking in the latter case is a relic of the infinite $N$ limit. The five dimensional case is stable to finite $N$ corrections and some d-wave superfluids can in any case be best viewed as having three spatial dimensions, for example, certain pnictides.

An $AdS_{5} \times SE_5$ metric solves the IIB equations supported by a background 5-form flux proportional
to the volume forms:
\bea
ds^2 &=& \bar{g}_{\mu \nu} dx^{\mu \nu} + g^{o}_{ab} dy^{a} dy^{b}; \\
F &=& 4 ( \eta_{AdS_5}  + \eta_{SE_5}), \nn
\eea
where $(\bar{g}_{\mu \nu},g^o_{a b})$ are the metrics on $AdS_5$ and $SE_5$ respectively, whilst $\eta$ denotes the
volume form. Our conventions for the type IIB field equations are
\be
R_{mn} = \frac{1}{96} F_{mpqrs} F_{n}^{\; pqrs},
\ee
with $F = \ast F$.  As mentioned in the previous section, this solution can immediately be generalized to
\bea
ds^2 &=& {g}^o_{\mu \nu} dx^{\mu \nu} + g^{o}_{ab} dx^{a} dx^{b}; \\
F &=& 4 ( \eta_{M_5}  + \eta_{SE_5}), \nn
\eea
in which ${g}^o_{\mu \nu}$ is the metric on a negative-curvature Einstein manifold $M_5$.

\subsection{Linearized analysis}

In this section we will discuss the relevant parts of the spectrum of the theory in such a background.
Let us first consider linearized fluctuations about this background:
\bea
g_{\mu \nu} &=& {g}^o_{\mu \nu} + h_{\mu \nu}; \qquad
g_{\mu a} = h_{\mu a}; \\
g_{ab} &=& g^{o}_{ab} + h_{ab}; \qquad
H_{\mu \nu} = h_{\mu \nu} + \frac{1}{3} {g}^o_{\mu \nu} h^a_a \nn \\
C_{mnpq} &=& c^{o}_{mnpq} + c_{mnpq}, \nn
\eea
where $C_{mnpq}$ is the 4-form potential and $x^m$ runs over all ten indices $(x^\mu,y^a)$. The redefined
quantity $H_{\mu \nu}$ takes into account the linearized 5-dimensional Weyl shift needed to obtain an Einstein frame metric.

Here we will not need to consider the fluctuations of other type IIB fields.
The latter are zero (or in the case of the dilaton and axion, constant)
in the background and their fluctuations would only couple to those of the metric and five form via quadratic terms in the equations.
In other words, the other fields are at least quadratic in the fluctuations of interest, $(h_{ab}, c_{mnpq})$ and the backreaction of
these fields into the metric and five form equations is then at least cubic order in the fluctuations.

Every fluctuation can be expanded in terms of the complete set of harmonics on the compact space, for example
\be
H_{\mu \nu} = \sum H_{\mu \nu}^{I} (x^{\mu}) Y^{I} (y^a),
\ee
where $Y^I(y^a)$ are scalar harmonics. Inserting the linearized fluctuations into the field equations, and using
orthogonality properties of the harmonics, results in linearized equations for the five-dimensional fields.
In the context of the $AdS_5 \times S^5$ reduction of type IIB the linearized equations were found in \cite{Kim:1985ez},
and the linear equations given here for $AdS_5$ replaced by a negative curvature Einstein manifold are closely analogous.

There is a gauge dependence, which can be handled by working with gauge invariant fluctuations or by fixing a gauge.
For computing the spectrum at the linearized level it is convenient to impose a de Donder-Maxwell gauge in which
\bea
\nabla^a h_{(a b)} &=& \nabla^{a} h_{a \mu} = 0; \\
\nabla^{a} c_{abcd} &=& \nabla^{a} c_{a b c \mu} = \nabla^{a} c_{ab \mu \nu} = \nabla^a c_{a \mu \nu \rho} = 0. \nn
\eea
This implies that the complete expansion for the metric and four-form potential fluctuations is
\bea
H_{\mu \nu} &=& \sum H_{\mu \nu}^{I} (x) Y^{I} (y); \qquad
h_{\mu a} = \sum B_{\mu}^{I_v} (x) Y^{I_v}_{a} (y); \nn \\
h_{(ab)} &=& \sum \phi^{I_t}(x) Y^{I_t}_{(a b)}(y); \qquad
h^{a}_{a} = \sum \pi^{I}(x) Y^{I}(y); \\
c_{\mu \nu \rho \sigma} &=& \sum b^{I}_{\mu \nu \rho \sigma} (x) Y^{I}(y); \qquad
c_{\mu \nu \rho a} = \sum b^{I_v}_{\mu \nu \rho} (x) Y^{I_v}_a(y); \nn \\
c_{\mu \nu a b} &=& \sum b^{I_a}_{\mu \nu} (x) Y^{I_a}_{[ab]}(y); \qquad
c_{\mu a b c} = \sum b_{\mu}^{I_v} (x) \eta_{a b c}^{\; \; d e} \nabla_{d} Y_{e}^{I_v} (y); \nn \\
c_{a b c d} &=& \sum b^{I} (x) \eta_{a b c d}^{\; \; \; \; e} \nabla_e Y^{I} (y). \nn
\eea
Here the harmonics are the scalar harmonics $Y^{I}(y)$, the vector harmonics $Y^{I_v}_a(y)$, the
traceless symmetric tensor harmonics $Y^{I_t}_{(ab)} (y)$ and the antisymmetric tensor harmonics
$Y^{I_a}_{[ab]}(y)$. Note that on the 5-sphere these harmonics satisfy
\bea
\Delta Y^{I} &=& \Box Y^{I} = - k (k+4) Y^{I}; \\
\Delta Y^{I_v}_{a} &=& (\Box - 4) Y^{I_v}_{a} = - (k+1)(k+3) Y^{I_v}_a; \nn \\
\Delta Y^{I_t}_{(ab)} &=& (\Box - 10) Y^{I_t}_{(ab)} = - (k^2 + 4 k + 8) Y^{I_t}_{(ab)}; \nn \\
\Delta Y^{I_a}_{[ab]} &=& (\Box - 6) Y^{I_a}_{[ab]} = - (k+2)^2 Y^{I_a}_{[ab]}; \nn \\
\nabla^{a} Y^{I_v}_{a} &=& \nabla^{a} Y^{I_t}_{(ab)} = \nabla^{a} Y^{I_a}_{[ab]} = 0.
\eea
where the symbol $\Delta$ denotes the Hodge-de Rham operator. The indices $I$, $I_v$, $I_t$, $I_a$ label the quantum numbers of scalar, vector, symmetric tensor and antisymmetric tensor harmonics, respectively, of given degree $k$.

\bigskip

One can expand the ten-dimensional Ricci tensor $R_{mn}$ to linear order as
\bea
R_{mn} &=& R^{o}_{mn} - \frac{1}{2} (\Box_x + \Box_y) h_{mn} - \frac{1}{2} \nabla_{m} \nabla_{n} h_{p}^{p}  \label{lina} \\
&& + \frac{1}{2} \nabla_{m} \nabla^{p} h_{pn} + \frac{1}{2} \nabla_{n} \nabla^{p} h_{pm} + R^{o}_{mpqn} h^{pq} \nn \\
&& + \frac{1}{2} R_{m}^{o \; p} h_{np} + \frac{1}{2} R_{n}^{o \; p} h_{mp}, \nn
\eea
where $R^{o}_{mnpq}$ is the background value of the Riemann tensor, and $\nabla_m$ is the background covariant derivative.
$\Box_x$ and $\Box_y$ denote the Laplacians on the non-compact space and compact space respectively. The background
values of the Ricci tensor are $R^o_{\mu \nu} = - 4 g^{o}_{\mu \nu}$ and $R^o_{ab} = 4 g^{o}_{ab}$.
The linearized equations can be expressed as
\bea
E_{mn}^{(1)} &\equiv&  R_{mn}^{(1)}
+ \frac{1}{24} h^{kl} F^o_{mk m_1 m_2 m_3} F^o_{nl}{}^{m_1 m_2 m_3} \label{linb} \\
&-&\frac{1}{96}(f_{m m_1 m_2 m_3 m_4} F^o_n{}^{m_1m_2m_3m_4}
+ f_{nm_1m_2m_3m_4} F^o_m{}^{m_1m_2m_3m_4}) = 0 \nn \\
E^{(1)}_{m_1\ldots m_5} &\equiv&
(f-f^*)_{m_1\ldots m_5} + \frac{1}{2} h^l_l F^o_{m_1\ldots m_5}
-5 h^k_{[m_1} F^o_{m_2\ldots m_5]k} = 0
\eea
where the linearized correction to the Ricci tensor can conveniently be expressed as
\be
R_{mn}^{(1)} = \nabla_k h^k_{mn} - \frac{1}{2} \nabla_m \nabla_n h^l_l, \qquad
h^k_{mn} = \frac{1}{2} (\nabla_m h^k_n + \nabla_n h_m^k - \nabla^k h_{mn}), \label{Def}
\ee
rewriting \eqref{lina}.

Then the linearized field equations can be written down as follows. The linearized Einstein equations are
\bea
E_{\mu \nu}: \qquad && \frac{1}{2} (\Box_x + \Box_y) H_{\mu \nu} - \frac{1}{2} (\nabla_{\mu} \nabla^{\rho} H_{\nu \rho}
+ \nabla_{\nu} \nabla^{\rho} H_{\mu \rho}) + \frac{1}{2} \nabla_{\mu} \nabla_{\nu} H^{\sigma}_{\sigma}
-  R^{o}_{\mu \rho \sigma \nu} h^{\rho \sigma}
\\
&& \qquad - \frac{1}{2} (\nabla{}_{\mu} \nabla{}^{a} h_{\nu a} + \nabla{}_{\nu} \nabla{}^{a} h_{\mu a})
- \frac{1}{6} g^{o}_{\mu \nu}  (\Box_x + \Box_y) h^{c}_{c} =
- 4 H^{\sigma}_{\sigma} g^{o}_{\mu \nu} + \frac{20}{3} h^{c}_{c} g^o_{\mu \nu} \nn \\
&&
- \frac{1}{12} g^{o}_{\mu\nu} \eta^{\rho \sigma \tau \mu' \nu'} \pa_{\rho} c_{\sigma \tau \mu' \nu'}. \nn \\
E_{\mu a}: \qquad && \frac{1}{2} (\Box_x + \Box_y) h_{\mu a}  - \frac{1}{2} (\nabla_{\mu} \nabla^{\rho} h_{a \rho}
+ \nabla_{a} \nabla^{\rho} H_{\mu \rho}) - \frac{4}{15} \nabla_{\mu} \nabla_{a} h^{c}_{c} + \frac{1}{2} \nabla_{\mu} \nabla_{a}
H^{\sigma}_{\sigma} \nn \\
&&
- \frac{1}{2} \nabla_{\mu} \nabla^{b} h_{(ab)} - \frac{1}{24} \nabla_a \nabla{}^{b} h_{\mu b} \nn \\
&& \qquad
=  - \frac{1}{24} \eta_{\mu}^{\; \nu \rho \sigma \tau} (\pa_{a} c_{\nu \rho \sigma \tau}
- 4 \pa_{\nu} c_{a \rho \sigma \tau}) - \frac{1}{2} \eta_{a}^{\; bcde} (\pa_{\mu} c_{bcde}
- 4 \pa_b c_{\mu c d e}). \nn \\
E_{ab}: \qquad
&& \frac{1}{2} (\Box_x + \Box_y) h_{(ab)} - \frac{1}{2} (\nabla{}_{a} \nabla{}^{\rho} h_{b \rho}
+ \nabla{}_{b} \nabla{}^{\rho} h_{a \rho}) - \frac{8}{15} \nabla{}_{a} \nabla{}_{b} h^{c}_{c}
+ \frac{1}{2} \nabla{}_{a} \nabla{}_{b} H^{\sigma}_{\sigma} \nn \\
&& \qquad - \frac{1}{2} (\nabla{}_{a} \nabla{}^{c} h_{(bc)} + \nabla_{b} \nabla^{c} h_{(ac)})
+ \frac{1}{10} g^{o}_{ab} (\Box_x + \Box_y) h^{c}_{c} -  R^{o}_{a c d b} h^{cd} = \nn \\
&& \qquad
4 h^{c}_{c} g^o_{ab} - \frac{1}{12} g^{o}_{ab} \eta^{cdefg} \pa_{c} c_{defg}. \nn
\eea
Note that $X_{(\m\n)}$ denotes the symmetric traceless part of the tensor $X_{\m \n}$.
The linearized five-form equations are
\bea
E_{ \mu \nu \rho \sigma \tau}: \qquad
&&
5 \pa_{[ \mu} c_{\nu \rho \sigma \tau ]} = \frac{1}{4!} \eta_{\mu \nu \rho \sigma \tau}^{\; \; \; \; \; abcde}
\pa_{a} c_{bcde} + 2 \left(H^{\sigma}_{\sigma} - \frac{8}{3} h^{c}_{c}\right) \eta_{\mu \nu \rho \sigma \tau}; \\
E_{\mu \nu \rho \sigma a}: \qquad
&& \pa_a c_{\mu \nu \rho \sigma} +4 \pa_{[ \mu} c_{\nu \rho \sigma] a} = \frac{1}{4!}
\eta_{\mu \nu \rho \sigma a }^{\; \; \; \; \;  \tau bcde} (\pa_{\tau} c_{bcde} + 4 \pa_{b} c_{cde \tau})
+ 4 \eta_{\mu \nu \rho \sigma}^{\; \; \; \; \tau} h_{a \tau}; \nn \\
E_{\mu \nu \rho ab }: \qquad
&& 3 \pa_{[\mu} c_{\nu \rho] a b} +2 \pa_{[a} c_{b] \mu \nu \rho} = \frac{1}{12}
\eta_{\mu \nu \rho ab }^{\; \; \; \; \;  \sigma \tau cde} (3 \pa_{c} c_{d e \sigma \tau}
+ 2 \pa_{\sigma} c_{\tau c d e} ).  \nn
\eea
The linearized equations for fluctuations of the other type IIB fields decouple from these equations.

Projecting the traceless part of the $(\mu\nu)$ Einstein equation, $E_{(\mu \nu)}$, onto the scalar harmonics one obtains
\be
\left [ \frac{1}{2} ( \Box_x + \Box_y) H^I{}_{(\mu \nu)}
+ \frac{1}{2} \nabla_{(\mu} \nabla_{\nu)} H_{\rho}^{(I) \rho} - \nabla_{(\mu} \nabla^{\rho} H^{I}_{\nu) \rho}
-  R^o_{(\m | \rho \sigma| \n)} h^{(I) \rho \sigma}\right ] Y^{I} = 0, \label{spin2}
\ee
whilst the trace part of the $(\mu\nu)$ equation, $\frac{1}{5} E^{\mu}_{\mu}$, gives
\bea
&& \left [ - \frac{1}{5} \nabla^{\rho} \nabla^{\sigma} H^{I}_{\rho \sigma} + \frac{1}{10} (2 \Box_x + \Box_y)
H^{I}_{\lambda \lambda} - \frac{1}{6} (\Box_x + \Box_y) \pi^I \right . \label{spintrace} \\
&& \left . \qquad + \frac{1}{12} \eta^{\mu \nu \rho \sigma \tau} \pa_{\mu} b^{I}_{\nu \rho \sigma \tau}
+ \frac{16}{5} {H^{I}_{\sigma}}^{\sigma} - \frac{16}{3} \pi^I
\right ] Y^{I} = 0. \nn
\eea
There are five remaining equations which can be projected onto the scalar harmonics, three from the Einstein equations
and two from the five-form equations. These are
\bea
E_{\mu a}: \qquad
&& \left [ - \frac{1}{2} \nabla^{\rho} H_{\rho \mu}^{I} + \frac{1}{2} \nabla_{\mu} {H_{\rho}^{\rho}}^{I}
- \frac{4}{15} \nabla_{\mu} \pi^I + \nabla_{\mu} b^{I} + \frac{1}{24} \eta_{\mu}^{\; \nu \rho \sigma \tau}
b^{I}_{\nu \rho \sigma \tau} \right ] \nabla_{a} Y^{I} = 0; \nn  \\
E_{ (a b)}: \qquad
&& \left [ \frac{1}{2} {H_{\sigma}^{\sigma}}^{I} - \frac{8}{15} \pi^{I} \right ] \nabla_{(a} \nabla_{b)} Y^{I} = 0; \label{first}   \\
E^{a}_{a}: \qquad
&&
\left [ \left( \frac{1}{10} \Box_x - \frac{1}{150} \Box_y - \frac{16}{5}\right)
\pi^{I} + \frac{1}{10} \Box_y {H_{\sigma}^{\sigma}}^{I} + 2 \Box_{y} b^{I} \right ] Y^{I} = 0; \nn \\
E_{\mu \nu \rho \sigma \tau}: \qquad
&&
\left [ 5 \pa_{[\mu} b^{I}_{\nu \rho \sigma \tau ]} - \eta_{\mu \nu \rho \sigma \tau} \left(2 {H_{\sigma}^{\sigma}}^{I}
- \frac{16}{3} \pi^{I} + \Box_y b^{I}\right) \right ] Y^{I} = 0; \nn \\
E_{\mu \nu \rho \sigma a}: \qquad
&&
\left [ b^{I}_{\mu \nu \rho \sigma} + \eta_{\mu \nu \rho \sigma}^{\; \; \; \; \tau} \nabla_{\tau} b^{I}
\right ] \nabla_a Y^{I} = 0. \nn
\eea
The equations for $(H_{(\mu \nu)}^{I}, H_{\sigma \sigma}^{I}, b_{\mu \nu \rho \sigma}^I, \pi^I, b^I)$ are hence coupled,
but one can immediately eliminate $(H_{\sigma \sigma}^I, b^{I}_{\mu \nu \rho \sigma})$ using the second and fifth equations
above. Then one obtains coupled equations for $(b^I,\pi^I)$:
\bea
&&
\left [ (\Box_x + \Box_y - 32) \pi^{I} + 20 \Box_{y} b^{I} \right ] Y^{I} = 0; \\
&&
\left [ (\Box_x + \Box_y) b^{I} - \frac{16}{5} \pi^{I} \right ] Y^{I} = 0. \nn
\eea
Let us denote the eigenvalues of the scalar harmonics as $\Box_y Y^I = - \Lambda^{I} Y^{I}$; the mass eigenvalues
are then
\be
m_{s^I}^2 = (\Lambda^I + 16) - 8 \sqrt{\Lambda^I +4}; \qquad
m_{t^I}^2 = (\Lambda^I + 16) + 8 \sqrt{\Lambda^I + 4}, \label{s-field}
\ee
where the combinations $(s^I,t^I)$ are given by
\bea
s^{I} &=& \pi^I - \left(5 + \frac{5}{2} \sqrt{\Lambda^I + 4}\right) b^I; \\
t^I &=& \pi^{I} - \left(5 - \frac{5}{2} \sqrt{\Lambda^I+ 4}\right) b^{I}, \nn
\eea
with the masses being defined $(\Box_x - m_{s^{I}}^2) s^{I} = (\Box_x - m_{t^{I}}^2) t^{I} = 0$.

The modes $(s^I,t^I)$ are dual to scalar operators $({\cal O}_{s^I}, {\cal O}_{t^I})$ respectively. These operators
will acquire no sources or expectation values (at the linearized level) provided that both $\pi^I$ and $b^I$ are set to zero.
Setting these fluctuations to zero however enforces constraints on $H_{\mu \nu}^{I}$. In particular, from the first
two equations in \eqref{first} one finds that
\be
(H^{\sigma}_{\sigma})^{I} = \nabla^{\rho} H_{\rho \sigma}^{I} = 0,
\ee
so that the spin two field is both traceless and transverse. These constraints allow \eqref{spin2} to be simplified, resulting in
\be
\Box_x H^{I}_{(\mu \nu)} -  2 R^o_{(\m | \rho \sigma| \n)} H^{(I) (\rho \sigma)} - \Lambda^{I} H^{I}_{(\mu \nu)} = 0.
\ee
This is indeed precisely the equation \eqref{uu2eq} with $m^2 = \Lambda^I$.
Substituting the Riemann tensor in the case of Anti-de Sitter space gives
\be
(\Box_x + 2) H^{I}_{(\mu \nu)} = \Lambda^{I} H^{I}_{\mu \nu},
\ee
with the operator on the left hand side being the linearized Einstein operator.

Thus, to summarize, we see that the spectrum necessarily contains transverse traceless spin two fields whose mass is given in terms of the
eigenvalue of the associated scalar harmonic on the compact Sasaki-Einstein space. The presence of such fields does not depend on the details
of the Sasaki-Einstein space.

\bigskip

Now let us turn to the vector harmonic sector. The linearized field equations projected onto the vector harmonics give
\bea
E_{a \mu}: \qquad
&& \left [ (\Delta_{M} + \Delta_y) B_{\mu}^{I_v} + \frac{1}{3} \eta_{\mu}^{\nu \rho \sigma \tau} \pa_{\nu} b^{I_v}_{\rho \sigma \tau} +
2 \Delta_y b_{\mu}^{I_v} \right ] Y_{a}^{I_v} = 0; \label{vect} \\
E_{(ab)}: \qquad
&& \left [ \nabla^{\mu} B_{\mu}^{I_v} \right ] \nabla_{(a} Y_{b)}^{I_v} = 0; \nn \\
E_{\mu \nu \rho \sigma a}: \qquad
&& \left [ 4 \pa_{ [\mu} b^{I_v}_{\nu \rho \sigma]} + \eta_{\mu \nu \rho \sigma}^{\; \; \; \; \tau} \left(\Delta_y b_{\tau}^{I_v}
- 4 B_{\tau}^{I_v}\right) \right ] Y_{a}^{I_v} = 0; \nn \\
E_{\mu \nu \rho a b}: \qquad
&& \left [ b_{\mu \nu \rho}^{I_v} + \eta_{\mu \nu \rho}^{\; \; \; \sigma \tau} \nabla_{\sigma} b_{\tau}^{I_v} \right ]
\nabla_{[a} Y_{b]}^{I_v} = 0, \nn
\eea
where $\Delta_M$ denotes the Maxwell operator. The fields $b_{\mu \nu \rho}^{I_v}$ are therefore non-dynamical, as they
can be eliminated using the last
equation and the remaining equations reduce to
\bea
\left [ (\Delta_{M} + \Delta_y - 8) B_{\mu}^{I_v} + 4 \Delta_y b_{\mu}^{I_v} \right ] Y_{a}^{I_v} &=& 0;  \\
\left [ (\Delta_M + \Delta_y) b_{\mu}^{I_v} - 4 B_{\mu}^{I_v} \right ] Y_{a}^{I_v} &=& 0. \nn
\eea
These equations can be diagonalized to give two branches:
\bea
a_{\mu}^{I_v} &=& B_{\mu}^{I_v} - \left(1 + \sqrt{1 + \Lambda_{I_v}}\right) b^{I_v}_{\mu}; \qquad m_{a}^2 = (4 + \Lambda_{I_v})
- 4 \sqrt{ 1 + \Lambda_{I_v}}; \\
c_{\mu}^{I_v} &=& B_{\mu}^{I_v} - \left(1 - \sqrt{1 + \Lambda_{I_v}}\right) b^{I_v}_{\mu}; \qquad m_{c}^2 = (4 + \Lambda_{I_v})
+ 4 \sqrt{ 1 + \Lambda_{I_v}}, \nn
\eea
where $\Delta_y Y^{I_v}_{a} = - \Lambda_{I_v} Y^{I_v}_{a}$ and $m^2$ denotes the mass squared of the diagonal combination.
There is a special case when the vector harmonic corresponds to a Killing vector of the compact space, so that
$\Lambda_{I_v} = 8$, since then
\bea
a^{I_v}_{\mu} &=& B^{I_v}_{\mu} - 4 b_{\mu}^{I_v}; \qquad  m_{a}^2 =0, \\
c^{I_v}_{\mu} &=& B^{I_v}_{\mu} + 2 b_{\mu}^{I_v}; \qquad m_{c}^2 = 24, \nn
\eea
and $a^{I_v}_{\mu}$ corresponds to gauge fields in five dimensions. Thus we recover the well-known fact that gauge fields in the lower dimensional
theory are in one-to-one correspondence with the Killing vectors of the compact space.

The field equations associated with the tensor harmonics can also be projected out, resulting in diagonal equations for
the scalars $\phi^{I_t}$ and the antisymmetric tensors $b^{I_a}_{\mu \nu}$. From the
remaining type IIB fields in ten dimensions one obtains
a Kaluza-Klein spectrum of scalars, massive vector fields and antisymmetric tensors in five dimensions. As argued above, these modes are not relevant
in what follows and will not be discussed further here;  the details of this analysis in the AdS case may be found in \cite{Kim:1985ez},

\bigskip

Let us connect the above discussion to the action \eqref{action-Herzog}. We have derived diagonal field equations for the linearized
fluctuations about a background which is the product of a negative curvature Einstein manifold and a Sasaki-Einstein manifold.
Suppose we switch off all such fluctuations, apart from the spin two field $H_{\mu \nu}^{I}$
associated with one specific scalar harmonic $Y^{I}$ and one gauge field $a^{I_v}_{\mu}$ associated
with a Killing vector of the Sasaki-Einstein. (The Sasaki-Einstein always admits at least one Killing vector, the Reeb vector.) The action
for these fluctuations at the quadratic level follows from the reduction of the type IIB action over the Sasaki-Einstein manifold and
can be computed following the analysis in \cite{Arutyunov:1998hf} for the case of $S^5$.

In particular, the quadratic action for the gauge field $a$ is given by
\be
S[a] = - \frac{V_{SE}} {2 \kappa_{10}^2} \int d^{5} x \sqrt{-g} \frac{1}{12} (f_{\mu \nu} f^{\mu \nu}),
\ee
with $V_{SE}$ the volume of the compact manifold and $1/2 \kappa_{10}^2$ the ten-dimensional coupling, which
are related to the effective coupling in five dimensions given in \eqref{action-Herzog} as
\be
\frac{1}{2 \kappa^2} = \frac{V_{SE}}{2 \kappa_{10}^2}.
\ee
Here we
have assumed that the Killing vectors $K_{a}^{I_v}$ are normalized so that
\be \label{vec-norm}
\int_{SE} K^{(I_v) a} K_{a}^{I_v} = V_{SE}.
\ee
Defining $A_{\mu} = \frac{1}{\sqrt{3}} a_{\mu}$ one obtains a canonically normalized gauge field in five dimensions. For the corresponding $k=1$ vector $c_{\mu}$ one obtains
\be
S[a] = - \frac{V_{SE}} {2 \kappa_{10}^2} \int d^{5} x \sqrt{-g} \frac{1}{12} (f_{\mu \nu} (c) f^{\mu \nu} (c) + 48 c_{\mu} c^{\mu}), \label{c-field}
\ee
and a canonically normalised massive vector field is obtained by setting $C_{\mu} = \frac{1}{\sqrt{3}} c_{\mu}$.

The quadratic action for the spin two field $H_{\mu \nu} = H_{\mu \nu}^{I} Y^I$ is
\be
S[H] = \frac{1}{2 \kappa_{10}^2} \int d^{10}x \sqrt{-G} \left [ - \frac{1}{4} \nabla^{m} H_{\mu \nu} \nabla_{m} H^{\mu \nu} +
\frac{1}{2} R_{\mu \nu \rho \lambda} H^{\mu \rho} H^{ \nu \lambda} - \frac{1}{4} R_{\mu \nu} H^{\mu \lambda} H^{\nu}_{\lambda} \right ],
\ee
together with the constraints $H_{\mu}^{(I)\mu} = \nabla_{\mu} H^{(I) \mu \nu} = 0$. We consider complex scalar harmonics such that
\be
\int_{SE} Y^{I} (Y^J)^{\ast} = c_{I} V_{SE} \delta^{IJ},
\ee
with eigenvalue $\Box Y^{I} = - \Lambda^{I} Y^{I}$.
Then the action can be reduced over the compact manifold to give
\bea
S[H] &=&  \frac{V_{SE}}{2 \kappa_{10}^2} \int d^{5}x \sqrt{-g} c_I
\left [ - \frac{1}{2} \nabla^{\rho} H^I_{\mu \nu} \nabla_{\rho} (H^{(I) \mu \nu})^{\ast} - \frac{1}{2} m_{I}^2 H^{I}_{\mu \nu}
(H^{(I) \mu \nu})^{\ast} \right . \\
&& \qquad  \left .
+ R_{\mu \nu \rho \lambda} H^{(I) \mu \rho} (H^{(I)\nu \lambda})^{\ast}
+ {\rm h.c.} \right ], \nn
\eea
where $m_{I}^2 = \Lambda^I$. The reason for considering complex scalar harmonics is that they can have definite charge under a $U(1)$
symmetry of the compact space. An illustrative example would be the following. Consider the five-sphere
expressed as
\be
ds^2 = d \theta^2 + \sin^2 \theta d\Omega_3^2 + \cos^2 \theta d \varphi^2. \label{s5metric}
\ee
An example of a complex degree two harmonic satisfying $\Box Y^{(2)} = - 8 Y^{(2)}$ is
\be
Y^{(2)} = \cos^2 \theta e^{2 i \varphi}, \label{deg2}
\ee
which carries definite charge of two under the Killing vector $\partial_{\varphi}$. The corresponding real harmonics
\be
\cos^2 \theta \sin ({2  \varphi}), \qquad
\cos^2 \theta \cos ({2  \varphi}),
\ee
do not have definite charge under the symmetry.
By rescaling the spin two field in the case of a complex harmonic as
\be
\phi_{\mu \nu} = \frac{1}{\sqrt{2}} \sqrt{c_I} H^{I}_{\mu \nu},
\ee
the total action
\be
S = S[A_\mu] + S[\phi_{\mu \nu}]
\ee
gives precisely \eqref{action-Herzog} in the free field limit, with the constraints imposed.

\bigskip

The possible masses of the spin two field and its charge are related to the spectrum of scalar harmonics on
the compact space. In the case of $S^5$, the scalar harmonics lie in the $(0,k,0)$ representation of $SO(6)$
and $m_I^2 = \Lambda^I = k (k+4)$, corresponding to a conformal dimension $\Delta = (4 + k)$ with $k \ge 1$ for the spin two field. We will discuss the corresponding range for the $U(1)$ charge $q$ below. The corresponding ${\cal N} = 4$ SYM operators are also known
\be
{\cal O}_{ij} = {\rm Tr} \left (T_{ij} X^{(I_1} \cdots X^{I_k)} \right ),
\ee
where $T_{ij}$ is the stress energy tensor, $X^{I}$ are the $6$ scalar fields and the symmetric traceless product
corresponds to the specific $(0,k,0)$ spherical harmonic.

In the case of $T^{1,1}$ the scalar harmonics are labeled by $(j,l,r)$ of the $(SU(2) \times SU(2) \times U(1))$ isometry group
and the spin two mass is given by \cite{Gubser:1998vd,Ceresole:1999zs,Ceresole:1999ht}
\be
m^2 = 6 \left ( j (j+1) + l (l+1) - \frac{r^2}{8} \right ), \qquad
\frac{r}{2} = j_3 = - l_3. \label{mass-form}
\ee
The lowest dimension spin two operators are obtained from harmonics in which $l = j = r/2$ for which
\be
\Delta = 4 + \frac{3r}{2},
\ee
with $r$ a positive integer (and zero for the stress energy tensor). Such fields are in the same multiplet as the lowest dimension scalar operators, which have dimension $\Delta = 3r/2$. The effective $U(1)$ charge $q$ of the spin two field however depends on how the gauge field $U(1)$ is embedded into the full isometry group; we will discuss $q$ further at the end of this section.

\subsection{Higher order corrections}

Next we turn to the question of the interaction terms in \eqref{action-Herzog}.
Some quadratic contributions
to the field equations around $AdS_5 \times S^5$ have been calculated in order to compute holographically
three-point functions in ${\cal N} =4$ SYM,
whilst very few cubic contributions to the field equations have ever been computed, even around the $AdS_5 \times S^5$ background. Cubic terms involving scalar fields were computed first in \cite{Lee:1998bxa} and subsequently additional cubic terms (and associated three point functions) were obtained in \cite{Arutyunov:1999en}. In the context of Kaluza-Klein holography, cubic couplings were discussed in
\cite{Skenderis:2006uy, Skenderis:2006di, Skenderis:2007yb}.
Here we will build on earlier computations to extract the cubic terms of interest in \eqref{action-Herzog}.

The corrections to the ten-dimensional field equations up to second order in fluctuations
can be expressed as
\be
E_{mn} = T_{mn}^{(2)}, \qquad
E_{m_1 \ldots m_5} = T^{(2)}_{m_1 \ldots m_5}
\ee
where the quadratic corrections are given by \cite{Lee:1998bxa,Arutyunov:1999en}
\bea
&&T^{(2)}_{m_1 \ldots m_5} = -\frac{1}{2} h^l_l f^*_{m_1\ldots m_5}
+5 h^k_{[m_1} f^*_{m_2\ldots m_5]k} \la{cor1} \\
&&-\frac{5}{2} h^l_l  h^k_{[m_1} F^o_{m_2\ldots m_5]k}
+\left(\frac{1}{8} (h^l_l)^2 +\frac{1}{4} h^{ml} h_{ml}\right) F^o_{m_1 \ldots m_5}
+10 h^{k_1}_{[m_1} h^{k_2}_{m_2} F^o_{m_3m_4m_5]k_1k_2} \nn \\
&&T_{mn}^{(2)} = -R_{mn}^{(2)} \la{cor2} \\
&&+ \frac{1}{24} h^{kl} h_l^s F^o_{m k m_1m_2m_3} F^o_{ns}{}^{m_1m_2m_3}
+\frac{1 }{16} h^{k_1s_1} h^{k_2s_2}
F^o_{mk_1k_2m_1m_2} F^o_{ns_1s_2}{}^{m_1m_2} \nn \\
&&-\frac{1}{24} h^{ks} (f_{mkm_1m_2m_3} F^o_{ns}{}^{m_1m_2m_3}+
f_{nkm_1m_2m_3} F^o_{ms}{}^{m_1m_2m_3})
+\frac{1}{96} f_{mm_1\ldots m_4} f_{n}{}^{m_1\ldots m_4} \nn
\eea
where
\be
R_{mn}^{(2)}= - \nabla_k(h_l^k h^l_{mn}) + \frac{1}{2} \nabla_n(h_{kl} \nabla_m h^{kl})
+\frac{1}{2} h_{mn}^k \nabla_k h^l_l - h_{mk}^l h_{nl}^k,
\ee
and $h^{m}_{nl}$ was defined in \eqref{Def}.

We now consider the ten-dimensional field equations up to second order in the fluctuations of the spin two field and the gauge field, which
correspond to the cubic terms in the effective action. Let us first note that no term cubic in the spin two field can occur in the action, using charge conservation: the spin two field carries a definite $U(1)$ charge and there is no cubic combination which is neutral. Similarly the gauge field is Abelian and therefore there is no cubic
interaction term involving only this field. Therefore the cubic terms of interest involve a gauge field, a spin two and its complex conjugate. Looking at \eqref{action-Herzog} one notes that most of these terms are fixed entirely by gauge invariance, with the parameter $g$ being the only free parameter.

We can extract this cubic coupling from the quadratic corrections to the gauge field equation, due
to the spin two field $H_{(\mu \nu)}$ and its complex conjugate. From \eqref{vect}, we see that we need to compute
the corrections $T^{(2)}_{a \mu}$, $T^{(2)}_{\mu \nu \rho \sigma a}$ and $T^{(2)}_{\mu \nu \rho a b}$.
However, there are no quadratic corrections to the latter two equations from
the spin two field. Using the fact that the spin two field is transverse
and traceless to leading order, we first note that
\be
R_{a \mu}^{(2)} = - \nabla_{\rho} (H_{\nu}^{\rho} h^{\nu}_{a \mu})
+ \frac{1}{2} \nabla_{\mu} (H_{\nu \rho} \nabla_a H^{\nu \rho}) - h_{a K}^L h_{ \mu L}^K,
\ee
with
\bea
h^{L}_{a K} &=& \frac{1}{2} ( \nabla_a h^{L}_{K}) \delta^{L \nu} \delta_{K \mu}; \\
h^{a}_{\mu L} &=& - \frac{1}{2} \nabla^{a} h_{\mu L} \delta_{L \nu}; \nn \\
h^{\rho}_{\mu \nu} &=& \frac{1}{2} (\nabla_{\mu} H^{\rho}_{\nu} + \nabla_{\nu} H^{\rho}_{\mu}
- \nabla^{\rho} H_{\mu \nu}). \nn
\eea
Therefore
\bea
R_{a \mu}^{(2)} &=& - \frac{1}{2} \nabla_{\rho} (H_{\nu}^{\rho} \nabla_a H^{\nu}_{\mu})
+ \frac{1}{2} \nabla_{\mu} (H_{\nu \rho} \nabla_a H^{\nu \rho})
- \frac{1}{2} \nabla_{a} H^{\nu}_{\rho} (\nabla_{\mu} H^{\rho}_{\nu} + \nabla_{\nu} H^{\rho}_{\mu}
- \nabla^{\rho} H_{\mu \nu}); \nn \\
&=& - \frac{1}{2} H^{\rho \nu} \nabla_{\rho} \nabla_{a} H_{\mu \nu} - \frac{1}{4} \nabla_a H^{\rho \nu} \nabla_{\mu} H_{\nu \rho}.
\eea
This correction gives the only contribution to $T^{(2)}_{a \mu}$: $T^{(2)}_{a \mu} = - R^{(2)}_{a \mu}$ and thus
\bea
(\Delta_{M} - 16) B_{\mu}^{I_v} -32 b_{\mu}^{I_v} &=& - 2 \left [R_{a\mu}^{(2)} \right ]_{K_a^{I_{v}}};  \label{vec-c} \\
(\Delta_M  - 8) b_{\mu}^{I_v} - 4 B_{\mu}^{I_v} &=& 0. \nn
\eea
where in the first line the correction is projected onto the Killing vector $K_{a}^{I_v}$. Note that
the divergence identities $\nabla^{\mu} B_{\mu}^{I_v} = \nabla^{\mu} b _{\mu}^{I_v} = 0$ are also uncorrected.
Let us define the cubic overlap between a Killing vector and the scalar harmonics as
\be
\int_{SE} K^{(I_v) a} Y^{I} \nabla_a (Y^I)^{\ast} = i n C_{I_v I I^{\ast}} V_{SE},
\ee
with $C_{I_v I I^{\ast}}$ real and $n$ the $U(1)$ charge of the harmonic. One can understand the origin of this expression as follows.
Introduce local coordinates for the $U(1)$ such that the Killing vector is $K = {\cal K} \pa_{\phi}$; the normalization constant ${\cal K}$ is then
such that the Killing vector is normalized as in \eqref{vec-norm}. For example, in the case of $S^5$, if one chooses
the coordinate $\varphi$ in \eqref{s5metric} as the circle direction then the corresponding normalized Killing vector is $K = \sqrt{3} \pa_{\varphi}$.
In general, the scalar harmonic can be expressed as
$Y^{I} = e^{in \phi} Y^{I}_n (\theta^a)$, with $\theta^a$ the other coordinates on the compact manifold, excluding the $U(1)$ circle and hence
\be
\int_{SE} K^{(I_v) a} Y^{I} \nabla_a (Y^I)^{\ast} =  i n {\cal K} c_{I} V_{SE},
\ee
where $c_{I}$ denotes the normalization of the scalar harmonics. The resulting correction to the gauge field equation in five dimensions is
therefore
\be
\Delta_M A_{\mu} = \frac{ i n {\cal K}}{\sqrt{3}}
(\phi^{\ast})^{\rho \nu} \left ( \nabla_{\mu} \phi_{\rho \nu} - 2 \nabla_{\rho} \phi_{\mu \nu} \right ) + {\rm h.c.}.
\ee
Comparing with \eqref{uueq} this gives
\be
q = \frac{n {\cal K}}{\sqrt{3}}; \qquad g = 1.
\ee
We can also determine the $U(1)$ charge $q$ and $g$ in five dimensions using the corrections to the spin two equation computed in
appendix \ref{calc-spin2}. This involves computing the gauge field strength
corrections to the spin two equation at quadratic order, involving one spin two field and a vector
field. The calculation is somewhat more involved since we need to consider the corrections to all seven equations given in \eqref{spin2}, \eqref{spintrace} and \eqref{first} and is carried out in appendix \ref{calc-spin2}.

\bigskip

Next we turn to the question of which other fields are sourced by the spin two and gauge field. From the structure of the equations given above, it is clear that
terms quadratic in spin two fields also occur in the equations for other Kaluza-Klein fields: the quadratic terms can be projected onto different harmonics, and hence different fields.
For example, from \eqref{vec-c} one obtains
\be
(\Delta_M - 24) C_{\mu} = \frac{ i n {\cal K}}{\sqrt{3}}
(\phi^{\ast})^{\rho \nu} \left ( \nabla_{\mu} \phi_{\rho \nu} - 2 \nabla_{\rho} \phi_{\mu \nu} \right ) + {\rm h.c.}.
\ee
This implies that the spin two field induces the massive vector field associated with the Killing vector. More generally, any cubic interaction which is not explicitly forbidden by symmetry is likely to arise. However, the only correction at this order to the gauge field equation is that given above - terms involving $\phi^2$ or $(\phi^{\ast})^2$ are charged and do not contribute to this equation. Close to the transition temperature, when the condensate is of order $\epsilon$, the backreaction of such terms on all other Kaluza-Klein fields would be of order $\epsilon^2$.

The only correction to the spin two equation at this order is discussed in appendix \ref{calc-spin2}. Terms quadratic in spin two fields can only induce corrections to equations for spin two fields with charge $2q$, $0$ or $-2q$. Therefore such terms induce spin two fields with other charges, rather than correct the equation for a spin two field of charge $|q|$. Terms involving one gauge field and one spin two field generically source other fields; however, with the specific ansatz relevant for the condensed solution, these source terms are shown to vanish in appendix \ref{calc-spin2}.

Terms quadratic in gauge fields also generically source other fields: for example, the scalar $s$ fields of dimension two are part of the consistent truncation to gauged supergravity and they are indeed sourced by the gauge fields. In other words, the backreaction of the gauge field even in the ordinary phase sources other fields. By choosing the gauge field appropriately, we can restrict to the case in which the backreacted ordinary phase is the Reissner-Nordstrom brane, and in this case
such scalar fields vanish for the static gauge fields of interest. So at finite chemical potential, close to the transition temperature, the easiest way to take into account the (finite) effects of the backreaction of the gauge field is to work with the uplifted charged black brane which we consider in the next section.
One may also wish to consider a different choice of $U(1)$ Killing vector, such that the ordinary phase is an R charged black hole, in which scalar fields are turned on. We will discuss this possibility later in section \ref{backr}.

\bigskip

Finally, it is useful to consider the action for  charged scalar fields arising in the compactification. For the fields obtained in \eqref{s-field} the effective action has been computed and for complex fields of definite charge
\bea
S &=& \frac{1}{2 \kappa^2} \int d^5 x \sqrt{-g} \left ( - (D_{\mu} S)(D^{\mu} S^{\ast})+ m_s^2  |S|^2 \right ) \\
&& + \frac{1}{2 \kappa^2}   \int d^5 x \sqrt{-g} \left ( - (D_{\mu} T)(D^{\mu} T^{\ast}) + m_T^2  |T|^2 \right ),  \nn
\eea
where as before the effective charge $q$ appearing in the covariant derivative $\nabla_{\mu}$ is
\be
q = \frac{n {\cal K}}{\sqrt{3} }
\ee
for a field associated with a scalar harmonic of integral charge $n$.

\subsection{Summary and interpretation}

In this section we have shown that the only terms which contribute to the coupled spin two/gauge field equations are those given in \eqref{reduced}, namely
\bea
(\Box - m^2) \phi_{\mu \nu} &=& 2 R_{\mu \lambda \rho \nu} \phi^{\rho \lambda};  \\
D_{\mu} F^{\mu \nu} &=& i q \phi^{\ast}_{\alpha \beta} (D^{\nu} \phi^{\alpha \beta}) + {\rm h.c.}, \nn
\eea
provided that one (a) imposes
\be
D_{\mu} A_{\nu} \phi^{\mu \nu} = 0 = F_{\mu \rho} \phi^{\rho}_{\nu},
\ee
which is implicitly satisfied by the specific static ansatz
\be
A = V(z) dt; \qquad
\phi_{xy} = \frac{1}{2 z^2} h(z)
\ee
and (b) one works with small chemical potential $\mu/T \ll 1$ and close to the transition temperature. The corrections to the equations which have been dropped are suppressed by factors of $\epsilon^2$ and $(\mu/T)^2$, and the backreaction on the metric and other Kaluza-Klein fields is suppressed by the same factors.

If however the mass and charge parameters are such that condensation does not occur at $\mu/T \ll 1$, then it is necessary to drop the restriction to small chemical potential and work with the backreacted black brane solution. From the analysis above (and indeed from the previous literature), we see that one can indeed not increase $q$ arbitrarily at a given value of $m^2$ such that the backreaction of the gauge field can be neglected. In other words, condensation for spin two fields cannot occur at $\mu/T \ll 1$.

More precisely, for the $S^5$, the effective charge $q$ is bounded by the degree of the harmonic $k$ as
\be
|q| \le k,
\ee
with equality in the following case. Recalling that the Cartan of $S^5$ is $U(1)^3$, let the scalar harmonic have definite integral $U(1)$ charges $[n_1,n_2,n_3]$ with respect to the three $U(1)$ factors; group theory implies that $|n_1| + |n_2| + |n_3| \le k$. Then the harmonic given in \eqref{deg2} is a degree two harmonic with charges $[2,0,0]$ for which $q = k = 2$ with respect to the first $U(1)$ factor. Degree $k$ harmonics with charge $[k,0,0]$ have $q = k$ with respect to the first $U(1)$ factor. In the context of Reissner-Nordstrom black branes, the relevant harmonics are, as we will see in the next section, those of degree $k$ which have equal $U(1)$ charges $[m,m,m]$ where $m$ is an integer such that $k \ge 3 |m|$. Such harmonics give rise to an effective charge $q = n/\sqrt{3}$ with respect to the appropriately normalised diagonal $U(1)$.

Therefore we find that, for the $S^5$, the charged spin two and scalar fields satisfy
\be
\Delta_{\phi} = (4 + k), k \ge 1;
\qquad
\Delta_S = k,  k \ge 2;
\qquad
\Delta_T= 8 + k, k \ge 0,
\ee
but in each case $|q| \le k$. Recall that the critical temperature \eqref{critT} behaves as
\be
\frac{T_{c}}{\mu} = \frac{1}{ \pi \alpha_c} \frac{|q|}{\Delta},
\ee
with $1< \alpha_c <  2.4$. However, for all of the Kaluza-Klein fields
\be
\frac{q}{\Delta} \le 1,
\ee
with the maximum value obtained only for the maximally charged scalar $S$ fields. Therefore
\be
\frac{T_c}{\mu} \le \frac{1}{ \pi \alpha_c} \le \frac{1}{\pi},
\ee
and therefore the critical temperature can never be reached within the probe approximation for the gauge field. In the next section we hence consider spin two fields within the rotating D3-brane background corresponding to the uplift of the charged black hole in five dimensions.

Even in the backreacted solution, we noted in \eqref{mc-rel} that a condensate only forms at finite temperature if the mass to charge ratio is bounded: for $D = 4$, $m^2/q^2 < 2$ at large $m$. This condition gives a strong restriction on which Kaluza-Klein modes can condense since $q^2 < m^2$.

\section{Uplifted solutions} \label{uplift}

Close to the transition temperature it is useful to consider the condensate as a linear perturbation around a charged AdS black brane. In this section we will uplift this description to a ten-dimensional solution, making use of the nonlinear reduction formulas given in \cite{Cvetic:1999xp}. Note that one does not need the complete massless reduction ansatz for type IIB, which is still unknown, as it suffices to consider truncations involving only the metric/five form in which the gauge fields lie within a Cartan subalgebra of the full gauge group.

The ten-dimensional metric can be expressed in the form
\be
ds^2 = \left( g_{\mu \nu} + \frac{1}{3} A_{\mu} A_{\nu}\right) dx^{\mu} dx^{\nu} + ds^2 (S^5) + \frac{2}{\sqrt{3}}A_{\mu} K_a dx^{\mu}  dx^a,
\ee
where $K$ is a normalised Killing vector in the $S^5$ such that $K^a K_a  = 1$ (using the $S^5$ metric) and
\be
K = \pa_{\phi_1} + \pa_{\phi_2} + \pa_{\phi^3},
\ee
in coordinates for the $S^5$ such that
\be
ds^2 (S^5) = g^{o}_{ab} dx^a dx^b =  \sum_{i=1}^3 (d \mu_i^2 + \mu_i^2 d \phi_i^2),
\ee
with $\mu_1 = \sin \theta$, $\mu_2 = \cos \theta \sin \psi$ and $\mu_3 = \cos \theta \cos \psi$.

The five form field strength is $F_5 + \ast F_5$ where
\bea
F_5 &=& 4 \epsilon_5 + \frac{1}{2 \sqrt{3}} \sum_i d (\mu_i^2) \wedge \left( d \phi_i +  \frac{A}{\sqrt{3}} \right) \wedge \ast_5 F, \\
&=& 4 \epsilon_5 + \frac{1}{2 \sqrt{3} } (d K \wedge \ast_5 F) + \frac{1}{6} \sum_i d  (\mu_i^2) \wedge A \wedge \ast_5 F, \nn
\eea
where $\epsilon_5$ is the volume form on the manifold defined by $g_{\mu \nu}$
and $\ast$ is the Hodge dual on the ten-dimensional space, with $\ast_5$ being the Hodge dual on the five-dimensional space defined by the metric $g_{\mu \nu}$. With this reduction ansatz the ten dimensional equations of motion are satisfied provided the metric and gauge potential satisfy equations of motion following from the lower dimensional action \cite{Cvetic:1999xp}
\be
S = \frac{1}{2 \kappa^2} \int d^5 x \sqrt{-g} \left ( R + 12 - \frac{1}{4} F^{\mu\nu} F_{\mu \nu} + \frac{1}{12 \sqrt{3}}
\epsilon^{\mu \nu \rho \sigma \lambda} F_{\mu \nu} F_{\rho \sigma} A_{\lambda} \right ). \label{eym}
\ee
A particular solution of the lower dimensional action, which does not involve the Chern-Simons term, is the charged brane solution \eqref{rn}, which uplifts to the decoupling limit of the rotating D3-brane solution (with correlated angular momenta). For what follows it is useful to note that $F_5$ can in this case be obtained from the potential
\be
C_4 = - \frac{1}{z^4} dt \wedge d^3x + \frac{1}{2 \sqrt{3}} K \wedge \ast_5 F,
\ee
and thus $F_5$ only has components $F_{tzxyw}$ and $F_{xywab}$, where the brane spatial coordinates are $(x,y,w)$.

\subsection{Linearized perturbations}

In this background solution let us now switch on a generic metric perturbation $h_{mn}$ and a five form perturbation $f_{mnpqr}$. Following \eqref{lina} the perturbation of the ten-dimensional Ricci tensor to linear order is
\bea
R^{(1)}_{mn} &=& - \frac{1}{2} \Box h_{mn} - \frac{1}{2} \nabla_{m} \nabla_{n} h^{p}_{p} + R^o_{mpqn} h^{pq} \\
&& + \frac{1}{2} (\nabla_m \nabla^p h_{pn} + \nabla_n \nabla^p h_{pm} + R^{o p}_{m} h_{np} + R^{o p}_{n} h_{mp} ). \nn
\eea
where $R^o$ denotes the background curvature and $\Box = \nabla_m \nabla^m$ is the ten-dimensional operator. From \eqref{linb} the linearised Einstein equation is
\bea
&& R^{(1)}_{mn}
+ \frac{1}{24} h^{kl} F^o_{mk m_1 m_2 m_3} F^o_{nl}{}^{m_1 m_2 m_3} \\
&& \qquad - \frac{1}{96}(f_{m m_1 m_2 m_3 m_4} F^o_n{}^{m_1m_2m_3m_4}
+ f_{nm_1m_2m_3m_4} F^o_m{}^{m_1m_2m_3m_4}) = 0. \nn
\eea
In general, the linearised spectrum will be significantly more complicated than in the case where the metric is diagonal. If however we restrict to a metric perturbation $h_{xy}(x^{\mu},x^a)$, its linearised equation decouples. This follows from the $(xy)$ component of the linearised Einstein equation above, which give
\be
R^{(1)}_{xy} + \frac{1}{24} h^{kl} F^o_{x k m_1 m_2 m_3} F^o_{y l}{}^{m_1 m_2 m_3} = 0,
\ee
i.e. $h_{xy}$ decouples from the four form fluctuations. Moreover, the rest of the linearised equations are consistently solved by a transverse $h_{xy}$, with all other metric fluctuations and four form fluctuations vanishing. In other words, just as in the previous section, $h_{xy}$ does not couple to and source any other fluctuation.

Now let us restrict to a static fluctuation which is isotropic along the brane directions, i.e. $h_{xy}(z,x^a)$. The linearised equation of motion then reduces to
\bea
- \frac{1}{2} \Box h_{xy} + R^{o}_{xyxy} h^{xy} + R^{o x}_{x} h_{xy} + \frac{1}{24} h^{xy}
F^{o}_{yx m_1 m_2 m_3} F^{o}_{xy}{}^{m_1 m_2 m_3} = 0,
\eea
where we use $R^{o x}_{x} = R^{o y}_{y}$. This equation needs to be evaluated using the background fields given above. Using the background equation of motion this equation can immediately be simplified to give
\be
- \frac{1}{2} \Box h_{xy} + R^{o}_{xyxy} h^{xy}  = 0.
\ee
We will now show that this equation is equivalent to \eqref{Eff1}.
Note that
\be
R^{o}_{xy xy} (g_{mn}) = R^{o}_{xyxy} (g_{\mu \nu}) = - \frac{1}{z^4}
\ee
and in addition
\be
{\rm det} (g_{mn}) = {\rm det} (g_{\mu \nu}) {\rm det} (g^o_{ab}) = \frac{1}{z^{10} }{\rm det} (g^o_{ab}),
\ee
where we have used \eqref{rn}. Let the metric perturbation be expressed in terms of a spherical scalar harmonic of definite degree $k$ and definite
$U(1)$ charge $n$ as
\be
h_{xy} = \frac{1}{2} \sqrt{c_I} \phi_{xy} (z) Y^I (x^a),
\ee
with
\be
\Box Y^I = - k (k+4) Y^I; \qquad K \cdot Y^I = i n Y^I,
\ee
and $c^I$ denotes the normalization of the harmonic. As discussed at the end of the previous section,
the three $U(1)$ charges are equal and integral, i.e. $[m,m,m]$, so that $n = 3m$ with $|n| \le k$.

The equation of motion then reduces to
\be
\left ( (\nabla_{\mu} + i q A_{\mu})(\nabla^{\mu} - i q A^{\mu}) - m^2 \right ) \phi_{xy} = 2 R^{o}_{xyxy} \phi^{xy}, \label{fc}
\ee
where $\nabla_{\mu}$ is the covariant derivative associated with $g_{\mu \nu}$ and
\be
m^2 = k (k + 4);
\qquad q = \frac{n}{\sqrt{3}},  \qquad q \le \frac{k}{\sqrt{3}}.
\ee
This is indeed equivalent to \eqref{Eff1}. The condensate phase is only reached at finite temperature for small enough masses and large enough charges.

\subsection{Generalizations}

One can immediately generalise the previous analysis to Sasaki-Einstein reductions to five dimensions as follows. The metric ansatz is
\be
ds^2 = g_{\mu \nu} dx^{\mu} dx^{\nu} + ds^2 (B_{KE}) + \left(\eta + \frac{A}{\sqrt{3}}\right)^2,
\ee
where $\eta$ denotes the $U(1)$ fibration over the base Kahler-Einstein space $B_{KE}$ and $A$ is a gauge field on the five-dimensional spacetime with metric $g_{\mu \nu}$. Then the metric
\be
ds^2(B_{KE}) + \eta^2
\ee
is a Sasaki-Einstein metric, provided that the curvature on the base manifold is normalised as $R_{ij} = 6 g_{ij}$, with $\eta = d \varphi + P$ being the Reeb one form on the Sasaki-Einstein and $\omega = \frac{1}{2} dP$ being the Kahler form. The base manifold for $S^5$ is $CP^2$ with the Fubini-Study metric. The ansatz for the five form is $F_5 + \ast F_5$ where
\be
F_5 = 4 \epsilon_5 + \frac{1}{\sqrt{3}} \omega \wedge \ast_5 F,
\ee
where $\omega$ is the Kahler form on the base space, $\epsilon_5$ is the volume form on the five-dimensional spacetime defined by the $g_{\mu \nu}$, with $\ast_5$ being the five-dimensional Hodge dual.  The ten-dimensional equations of motion are then reproduced by the five-dimensional equations of motion following from the action \eqref{eym}.

The reduction from eleven dimensions to four dimensions on a Sasaki-Einstein is very similar. The metric ansatz in this case is
\be
ds^2 = g_{\mu \nu} dx^{\mu} dx^{\nu} + ds^2 (B_{KE}) + \left(\eta + \frac{A}{2}\right)^2,
\ee
where $\eta$ denotes the $U(1)$ fibration over the base Kahler-Einstein space $B_{KE}$ (with six dimensions) and $A$ is a gauge field on the four-dimensional spacetime with metric $g_{\mu \nu}$. Then the metric
\be
ds^2(B_{KE}) + \eta^2
\ee
is a Sasaki-Einstein metric, provided that the curvature on the base manifold is normalised as $R_{ij} = 8 g_{ij}$, with $\eta = d \varphi + P$ being the Reeb one form on the Sasaki-Einstein and $\omega = \frac{1}{2} dP$ being the Kahler form. The base manifold for $S^7$ is $CP^3$ with the Fubini-Study metric. The eleven dimensional Einstein equation
\be
R_{mn} = \frac{1}{12} \left(F_{mpqr} F_{n} {}^{pqr} - \frac{1}{12} F^{2} g_{mn}\right)
\ee
sets the normalisation of the four form $F_4$ for which the reduction ansatz can be written as
\be
F_4 = 2 \epsilon_4 - \omega \wedge \ast_4 F,
\ee
where $\omega$ is the Kahler form on the base space, $\epsilon_4$ is the volume form on the four-dimensional spacetime defined by the $g_{\mu \nu}$, with $\ast_4$ being the four-dimensional Hodge dual.  The eleven-dimensional equations of motion are then reproduced by the four-dimensional equations of motion following from the action
\be
S = \frac{1}{2 \kappa^2} \int d^4x \sqrt{-g} \left ( R + 6 - \frac{1}{4} F^2 \right ),
\ee
provided that one also restricts to solutions which satisfy $F \wedge F = 0$.

For both eleven and ten dimensional reductions, a specific solution for $g_{\mu \nu}$ and $A_{\mu}$ is the charged black brane solution given in \eqref{rn}. Following the same analysis as in the previous section,
the following linearised metric perturbations decouple
\be
g_{mn} \rightarrow g_{mn} + \frac{1}{2} \sqrt{c_I} \phi_{xy}(z) Y^I(x^a), \label{met2}
\ee
where $Y^I(x^a)$ is as before a scalar harmonic on the Sasaki-Einstein. Such metric perturbations satisfy the equation \eqref{fc}, with specific mass and charge parameters, the mass for $T^{1,1}$ being given in \eqref{mass-form}. With the metric parameterisation given above, the charge parameter $q$ is proportional to the R charge.

\subsection{Backreaction} \label{backr}

We now turn to the question of backreaction i.e. given the linearised perturbation can one generalise it to find a  non-linear solution. The difficulty in finding a non-linear solution arises from the fact that the metric perturbation in \eqref{met2} depends not just on the radial coordinate of the black brane but also on the coordinates of the Sasaki-Einstein. In general a charged scalar harmonic will depend explicitly on all five coordinates of the Sasaki-Einstein. For example, in the case of $T^{1,1}$ one can see this using the metric
\bea
ds^2 &=& \frac{2}{9} \left(d \phi + \cos \beta_1 d \alpha_1\right)^2 + \frac{1}{6} \left(d \beta_1^2 + \sin^2 \beta_1 d \alpha_1^2\right); \\
&& + \frac{2}{9} \left(d \phi - \cos \beta_2 d \alpha_2\right)^2 + \frac{1}{6} \left(d \beta_2^2 + \sin^2 \beta_2 d \alpha^2_2\right). \nn
\eea
The scalar harmonics can be written in a separable form as \cite{Gubser:1998vd}
\be
Y(x^a) = f _{j}(\beta_1) f_{l} (\beta_2) e^{i j_3 \alpha_1 + l_3 \alpha_2 + r \phi},
\ee
where $(j_3,l_3)$ are associated with the $SU(2)$ quantum numbers $(j,l)$ respectively and the specific form of the  functions $f_j(\beta)$ is not needed here. Recalling that
$r/2 = j_3 = - l_3$ we note that all harmonics with non-zero R charge depend explicitly on all five coordinates. The analysis for the more general case of $T^{p,q}$ is very similar.

To simplify the backreaction it might be interesting to look at the uplift of an R charged black branes with scalar fields, rather than Reissner-Nordstrom branes. Using again the reduction formulae for $S^5$ given in
\cite{Cvetic:1999xp}, let us consider a ten-dimensional metric of the form
\be
ds^2 = \Delta^{1/2} ds_5^2 + \Delta^{-1/2} \left( \Delta X^{-1/2} d\theta^2 + X^{-1} \cos^2 \theta (d \varphi + A)^2 + X^{1/2} \sin^2 \theta d \Omega_3^2\right)
\ee
where
\be
\Delta = X \cos^2 \theta + X^{-1/2} \sin^2 \theta
\ee
and $X$ is a function of the non-compact coordinates on the five-dimensional manifold only. When $X=1$ and the gauge field $A$ is zero, the metric is a diagonal product on the five-dimensional metric with the $S^5$ in coordinates \eqref{s5metric}. With the corresponding ansatz for the five form, see \cite{Cvetic:1999xp}, the equations of motion are equivalent to those obtained from the five dimensional action
\be
S = \frac{1}{2\kappa^2} \int d^5 x \sqrt{-g} \left ( R - \frac{1}{2} d \chi^2 +4 \left(e^{\sqrt{2/3} \chi} + 2 e^{ \frac{\chi}{\sqrt{6}}}\right) - \frac{1}{4} e^{2 \sqrt{2/3} \chi} F^2 \right ) ,
\ee
where $X = e^{-\sqrt{2/3} \chi}$. This action admits the well-known R charged black brane solution
\bea
ds^2 &=& - H^{-2/3} f dt^2 + H^{1/3} \left(f^{-1} dr^2 + r^2 dx \cdot dx\right); \\
X &=& e^{- \sqrt{2/3} \chi} = H^{-2/3}; \qquad
A = \frac{1}{\sinh \beta} \left( \frac{r_h^4 }{\mu_m} - H^{-1}\right) dt, \nn
\eea
with
\be
f = r^2 H - \frac{\mu_m}{r^2}; \qquad
H = 1 + \frac{\mu_m \sinh^2 \beta}{r^2},
\ee
with $r_h$ being the horizon location. (To express this metric in a form comparable with that of \eqref{rn} we should use a radial coordinate defined as $1/z_2 = r^2 H^{1/3}$.) Now $\chi$ is one of the dimension two $s$ scalar fields of the spherical reduction. It is not charged under the gauge field retained in the consistent truncation, but since it is associated with a rank two harmonic on the sphere it is charged with respect to other $U(1)$ generators in the $SO(6)$ R symmetry group and the dimension two operator dual to this field acquires an expectation value.

Now suppose one considers a small metric fluctuation in ten dimensions of the form
\be
h_{xy}(r, \theta,\varphi)
\ee
which is charged under the Killing vector $\partial_{\varphi}$. From the five dimensional perspective this will give rise to a massive charged spin two field. From the ten dimensional perspective the backreaction is easier than in the previous cases, since the fluctuation depends on only two coordinates on the sphere. Moreover condensation of the spin two would necessarily compete with that of the scalar field: implicitly there are two order parameters present, which may be interesting for modelling the pseudogap region (see conclusions).

\section{Conclusions and outlook} \label{conc}

Charged spin two fields are generic in Sasaki-Einstein reductions and can give rise to condensed d-wave superfluid phases. Massive spin two fields can never be retained as part of a consistent truncation but can be described using a probe approximation, either using effective lower dimensional equations or working directly with the ten/eleven dimensional supergravity solution. While the condensed solution exists, we were only able to show that it was thermodynamically favoured within the probe approximation of the bottom up model. Computation of the backreaction of the metric fluctuation in the higher picture would allow us to analyse the thermodynamics and the stability from the top down perspective.

While the focus of this paper was on d-wave phases, 
our analysis highlights the more general fact that Kaluza-Klein modes which are not included in consistent truncations can potentially give rise to novel phases in the dual theory. The rule of thumb is that the lowest dimension operators control the phase structure but while some such operators are retained in consistent truncations to gauged supergravity one needs to take into account that other operators of comparable dimension may lie outside these consistent truncations.

Within our top down approach the masses and charges of spin two fields are not independent and for any given dimension condensed phases only exist at finite temperature if the operator charge is large enough. The phenomenology of the condensed phases may be explored by analysing the behaviour of probe fermions, for which the masses, charges and couplings are also determined by the Kaluza-Klein reduction. The fermion spectral functions within the spin two condensed phases will be discussed in our subsequent work \cite{Dwave2}.

As well as the spin two fields discussed here, the top down models of course contain towers of other scalar fields, massive vectors etc. The richness of this spectrum could lead to a rich landscape of phases of the dual theory, some of which may also not be visible within consistent truncations. It would be interesting to explore
what phases can be realised and to address stability, although the latter is likely to be very difficult given the large number of Kaluza-Klein fields. In general one would expect that a number of different phases compete with each other.

As an example of competing phases we highlighted in section \ref{backr} the case of the spin two field competing with a charged scalar field; this would be a relatively simple but potentially instructive case to analyse. One of the most mysterious features of high $T_c$ superconductors is the pseudogap behaviour in the underdoped regime, just above the critical temperature. The physics underlying the pseudogap is not clear, with two of the main theories being the existence of preformed Cooper pairs (local rather than long range order) and an exotic order parameter competing with superconductivity. The latter scenario is precisely what is modelled by including coupled Kaluza-Klein fields and it would be interesting to explore this scenario further.

\section*{Acknowledgments}

This work is part of the research program of the Stichting voor Fundamenteel Onderzoek der Materie (FOM), which is financially supported by the Nederlandse Organisatie voor Wetenschappelijk Onderzoek (NWO). MT acknowledges support from a grant of the John Templeton Foundation. The opinions expressed in this publication are those of the authors and do not necessarily reflect the views of the John Templeton Foundation. KK acknowledges support via an NWO Vici grant of Kostas Skenderis and is grateful for the support of an GIST Global University Project. We would like to thank Kostas Skenderis and Paul McFadden for collaboration during earlier phases of this work.

\appendix

\section{Corrections to spin two equation}
\label{calc-spin2}

The corrections to the spin two equation at quadratic order, involving one spin two
field and a vector field, can be used to compute the gyromagnetic coupling.
This is somewhat involved, as we need to consider the corrections to
all seven equations given in \eqref{spin2}, \eqref{spintrace} and \eqref{first}. These corrections are given by
\bea
T^{(2)}_{\mu \nu} &=& - R^{(2)}_{\mu \nu}; \\
&=& \frac{1}{2} h^{a \rho} \left ( \nabla_a \nabla_{\mu} h_{\rho \nu} + \nabla_a \nabla_{\nu} h_{\rho \mu}
- 2 \nabla_{\rho} \nabla_a h_{\mu \nu} \right ) \nn \\
&& + \frac{1}{2} \left ( \nabla^{\rho} h_{\mu a} \nabla^a h_{\nu \rho} + \nabla^{\rho} h_{\nu a} \nabla^a h_{\mu \rho} \right ), \nn
\eea
where we again use the fact that $\nabla^{\mu} h_{\rho \mu} = \nabla^{a} h_{a \rho} = \nabla^{\rho} h_{a \rho} = h^{\mu}_{\mu} = 0$ at
the linearized order. The other corrections to the Einstein equations are
\bea
T^{(2)}_{ab} &=& - R^{(2)}_{ab} = \frac{1}{2} h^{\mu \rho} \nabla_{\mu} \left ( \nabla_{a} h_{\rho b} + \nabla_b h_{\rho a} \right ); \label{eab2} \\
T^{(2)}_{a \mu} &=& - R^{(2)}_{a \mu} - \frac{1}{24} h^{\nu \rho} f_{a \nu \sigma \tau \eta} {F^{o}}_{\mu \rho}^{\; \; \sigma \tau \eta};
\nn \\
&=& - \frac{1}{2} \left ( \nabla_{\nu} h^{\rho}_{\mu} - \nabla^{\rho} h_{\mu \nu} \right )
\left ( \nabla_{\rho} h^{\nu}_a - \nabla^{\nu} h_{a \rho} \right ) - h^{b \nu} \nabla_b \nabla_a h_{\mu \nu} \nn \\
&& - h^{\rho \nu} \nabla_{\rho} \left ( \nabla_{\mu} h^{\nu}_a - \nabla^{\nu} h_{\mu a} \right ), \nn
\eea
where the term involving $f_{a \nu \sigma \tau \eta}$ actually vanishes using the linearized field equations. For the five
form equations the correction $T^{(2)}_{\mu \nu \rho \sigma \tau}$ vanishes whilst
\bea
T^{(2)}_{\mu \nu \rho \sigma a} &=& 5 h^{\tau}_{[ \mu} f^{\ast}_{\nu \rho \sigma a ] \tau} + 10 h^{\tau}_{[\sigma} h^{\eta}_{a}
F^{o}_{\mu \nu \rho ] \tau \eta }; \\
&=& 40 h^{\tau}_{[\sigma} h^{\eta}_{a}
\eta_{\mu \nu \rho]  \tau \eta }. \nn
\eea
In the correction terms we use the leading form for the perturbations, namely
\bea
h_{\rho a} &=& B_{\rho} K_a = \frac{A_{\rho}}{\sqrt{3}} K_a;
\qquad h_{\mu \nu} = H^I_{\mu \nu} Y^I + (H^I_{\mu \nu} Y^I)^{\ast} \\
c_{\mu a b c} &=&  b_{\mu} \eta_{abc}{}^{de} \nabla_d K_e =
- \frac{A_{\mu}}{2 \sqrt{3}}  \eta_{abc}{}^{de} \nabla_d K_e. \nn
\eea
In explicitly evaluating the quadratic corrections to the spin two equation due to one spin two field
and a gauge field, a considerable simplification occurs: the projection of $T^{(2)}_{ab}$ onto scalar harmonics of the same degree as the spin two field vanishes. For the traceless part this follows from the identity
\be
\int_{SE} Y^{I,n} \nabla^{(a} \nabla^{b)} (Y^{J,n})^{\ast} \nabla_a K_{b}^{(I_v)} =
\frac{1}{2} \left ( \Lambda^{J} - \Lambda^I \right )  i n C_{I_v I J} V_{SE},
\ee
which vanishes when $\Lambda^{J} = \Lambda^I$. For the trace part the vanishing of the cubic overlap follows from
the transverse nature of the vector harmonic. This implies that the corrections to the second, third and fourth equations in
\eqref{first} vanish. These uncorrected equations can be used to eliminate $(\pi^I,b_{\mu\nu\rho\sigma}^I)$
in \eqref{spin2} and \eqref{spintrace}.  Projecting onto scalar harmonics, the correction to \eqref{spin2} gives
\bea
&& \frac{1}{2} ( \Box_x + \Box_y) H^I_{(\mu \nu)} + \frac{1}{2} \nabla_{(\mu} \nabla_{\nu)} H^I - \nabla_{(\mu} \nabla^{\rho} H^I_{\nu) \rho} - R^{o}_{ (\mu | \rho \sigma | \nu)} H^{I \rho \sigma} \label{c7} \\
&& \qquad = \frac{i n {\cal K} }{2 \sqrt{3}} \left [ A^{\rho} (\nabla_{(\mu} H^I_{ \nu) \rho } + \nabla_{(\nu} H^I_{\mu) \rho}
- 2 \nabla_{\rho} H^I_{( \mu \nu )} ) - 2 F_{\rho (\mu} H^I_{\nu) \rho}  \right ], \nn
\eea
From the corrections to \eqref{spintrace}, again projecting onto scalar harmonics one obtains
\be
\nabla^{\rho} \nabla^{\sigma} H^I_{\rho \sigma} + \frac{1}{4} \Box_{x} H^I + \frac{3}{4} \Box_y H^I + 24 H^I =  \frac{i n {\cal K} }{\sqrt{3}}
\nabla^{\rho} A^{\mu} H^I_{\rho \mu}. \label{c8}
\ee
To interpret the terms on the right hand side it is useful to consider the spin two equation of motion with a gyromagnetic ratio $g$,
\bea
(\Box - m^2) \phi_{\mu \nu} &=& D_{\mu} \phi_{\nu} + D_{\nu} \phi_{\mu} - D_{\< \mu} D_{\nu \>} \phi + 2 R_{\mu \rho \nu \lambda}
\phi^{\rho \lambda} \\
&& + g_{\mu \nu} \left [ (\Box - m^2 - D) \phi - D^{\rho} \phi_{\rho} \right ] + i g q  \left ( F_{\mu \rho} \phi^{\rho}_{\nu}
+ F_{\nu \rho} \phi^{\rho}_{\mu} \right ). \nn
\eea
Expanding this equation up to terms of first order in the gauge field gives
\bea
(\Box_x - m^2) \phi_{\mu \nu} &=& \nabla_{\mu} \phi_{\nu} + \nabla_{\nu} \phi_{\mu}
- \nabla_{\< \mu} \nabla_{\nu \>} \phi + 2 R_{\mu \rho \nu \lambda}
\phi^{\rho \lambda} \\
&& + g_{\mu \nu} \left [ (\Box_x - m^2 - D) \phi - \nabla^{\rho} \phi_{\rho} \right ] + i g q  \left ( F_{\mu \rho} \phi^{\rho}_{\nu}
+ F_{\nu \rho} \phi^{\rho}_{\mu} \right )  \nn \\
&& - i q (\nabla_{\mu} (A^{\rho} \phi_{\rho \nu}) + \nabla_{\nu} (A^{\rho} \phi_{\rho \mu} ))
+ i q g_{\mu \nu} \nabla^{\rho} A^{\sigma} \phi_{\rho \sigma}, \nn
\eea
where $\Box_{x} = \nabla^{\mu} \nabla_{\mu}$ and
in non-linear terms the leading order constraints $\nabla^{\mu} \phi_{\mu \nu} = \phi = \nabla^{\mu} A_{\mu} = 0$ have been used.

Comparing the traceless part of this equation with \eqref{c7} then gives
\be
q = \frac{n {\cal K}}{\sqrt{3}}, \qquad g = 1,
\ee
in agreement with the values found using the gauge field equation.

Expanding the constraints given in \eqref{const-1} and \eqref{const-2} to quadratic order gives
\bea
\frac{4}{3} ( m^2 + 3) \phi - (\nabla^2 \phi - \nabla^{\mu} \nabla^{\rho} \phi_{\mu \rho}) &=& i q \nabla^{\mu} A^{\rho}
\phi_{\rho \mu}; \label{c10} \\
\frac{4}{3} m^2 (m^2 + 5) \phi &=& - 2 i q (g-1) \nabla^{\mu} F^{\nu \rho} \nabla_{\nu} \phi_{\rho \mu}, \nn
\eea
where we use the leading order equation $F_{\mu} = 0$ to simplify the right hand side. Note that for $g=1$
the second constraint implies $\phi = 0$, and therefore this constraint remains unchanged. Substituting $\phi= 0$ into the other constraint and comparing with \eqref{c8} again gives the same value for $q$.

The divergence constraint gives
\bea
m^2 (\phi_{\mu} - \nabla_{\mu} \phi) &=& - i q m^2 A^{\rho} \phi_{\rho \mu} \\
&&  - i q (g \nabla_{\rho} F_{\mu \nu} \phi^{\nu \rho} + (2 -g) F^{\nu \rho} \nabla_{\rho} \phi_{\mu \nu}). \nn
\eea
Manipulating the corrections to the first and last equations in \eqref{first} gives
\bea
- \nabla^{\rho} H_{\rho \sigma} + \frac{1}{2} \nabla_{\sigma} H &=& \frac{1}{\sqrt{3} \Lambda^I} i n k
F^{\rho \nu} \nabla_{\rho} H_{\sigma \nu} \nn \\
&& + \frac{1}{\sqrt{3} \Lambda^{I}} i n k H^{\rho \nu} \nabla_{\rho} F_{\sigma \nu} \\
&& + \frac{1}{\sqrt{3} } i n k A^{\rho} H_{\rho \sigma}. \nn
\eea
which imposing the constraint $H = 0$, $\phi =0$ and recalling that $m^2 = \Lambda^I$
are consistent with the values obtained for $q$ and $g$.

Finally let us turn to the corrections to the other equations in this sector, namely the equations for
the spin one fluctuations $a_{\mu}$ and $c_{\mu}$; the scalar fields associated the symmetric tensor harmonics $\phi^{I_t}$ and the scalar fields associated with scalar harmonics $s^I$ and $t^I$.
Projecting the corrections involving one spin two field and one gauge field onto these equations we find
that these corrections actually vanish, when one uses the explicit ansatz for the spin two field and the gauge field. For example, projecting the first equation of \eqref{eab2} onto tensor harmonics gives the correction to the equations for $\phi^{I_t}$. Even if the harmonic projection is non-zero, the correction vanishes for the condensate solution since
\be
H^{I \mu \nu}  D_{\mu} A_{\nu} = 0.
\ee
The corrections to the other fluctuation equations vanish similarly.



\section{Fluctuation equations}
\label{schr-eq}

In this appendix we will rewrite the superfluid equations in a form which is useful for analysing the critical temperature. Let us start with the following equations of $\phi(r)$ and $\psi(r)$ in terms of a radial coordinate $r$.
\begin{align}
  & \phi''(r) + \frac{D-1}{r} \phi'(r) - \frac{\psi(r)^2}{f_o(r)}\phi(r) = 0 \label{phi} \,, \\
  & \psi''(r) + \left(\frac{f_o'(r)}{f_o(r)} + \frac{D-1}{r}\right)\psi'(r)
  + \left( \frac{\phi(r)^2}{f_o(r)^2} - \frac{m^2}{f_o(r)}\right)\psi(r) = 0 \,.
  \label{psi}
\end{align}
These equations are related to the equations in terms of $V(\zeta),h(\zeta)$: $r = 1/z$, $\phi(r) = \tilde{V}(\z)$, $\psi(r) = \tilde{h}(\z)$ and $f_o (r) := r^2(1-1/r^D)$, which includes $r^2$ factor compared to $f(\z) = 1-\z^D$.
Note that the last term in \eqref{phi} differs from the previous equations and \cite{Hartnoll:2008vx} by a factor of $2$. This is simply related to changing the normalization of $\psi$, $\psi \rightarrow \sqrt{2}\psi$. The normalization of $\psi$ in \eqref{phi} is consistent with that in our main draft.

The basic idea is to redefine the fields to remove the first derivative terms, which can be done by
\begin{equation} \label{rescaling}
  \phi(r) \equiv \varphi_0 \frac{1}{r^{\frac{D-1}{2}}} \varphi(r) \,, \qquad  \psi \equiv \chi_0 \frac{1}{r^{\frac{D-1}{2}}}\frac{1}{\sqrt{f_o}}\chi(r) \,,
\end{equation}
where $\varphi_0$ and $\chi_0$ are arbitrary constants. The equations \eqref{phi} and \eqref{psi} then become
\begin{align}
  & \varphi'' + M_\varphi^2(r) \varphi - \L_\varphi(r) \chi^2\varphi  = 0 \,, \\
  & \chi'' + M_\chi^2(r) \chi + \L_\chi(r) \varphi^2 \chi = 0 \,,
\end{align}
where
\begin{align}
  M_\varphi^2 &=  - \frac{(D-3)(D-1)}{4 r^2} \,, \\
  M_\chi^2 &= M_\varphi^2 - \frac{m^2}{f_o} - \frac{(D-1) f_o'}{2r f_o}
  + \frac{f_o'^2}{4 f_o^2} - \frac{f_o''}{2 f_o} \,, \\
  \L_\varphi &= \frac{\chi_0^2}{r^{D-1} f_o^2}  \,, \quad
  \L_\chi = \frac{\varphi_0^2}{r^{D-1} f_o^2} \,.
\end{align}
These equations can be obtained by extremizing the action
\begin{equation}
 S =  \int \dd r \left( \frac{1}{2} \varphi'^2 - \frac{1}{2} \chi'^2
  + \frac{1}{2} M_\chi^2 \chi^2 - \frac{1}{2} M_\varphi^2 \varphi^2 +
  \frac{1}{2}\L \varphi^2 \chi^2     \right) \,,
\end{equation}
where we choose $\varphi_0 = \chi_0$, so $\L_\varphi = \L_\chi =: \L$.

We can re-express these equations in the $\zeta$ coordinate
using rescaled fields
\begin{equation}
  V(\zeta) \equiv \calv_0 \zeta^{\frac{D-3}{2}} \calv(\zeta) \,, \qquad  h  \equiv \calh_0 \zeta^{\frac{D-1}{2}}\frac{1}{\sqrt{f}}\calh(\zeta) \,,
\end{equation}
so that the equations read
\begin{align}
  & \calv'' + M_{\calv}^2(\zeta) \calv - \L_{\calv}(\zeta) \calh^2 \calv  = 0 \,, \\
  & \calh'' + M_{\calh}^2(\zeta) \calh + \L_{\calh}(\zeta) \calv^2 \calh = 0 \;
\end{align}
where
\begin{align}
  M_{\calv}^2 &= - \frac{(D-3)(D-1)}{4 \zeta^2} \,, \\
  M_{\calh}^2 &= - \frac{D^2-1}{4 \zeta^2}- \frac{m^2}{\zeta^2f} - \frac{(D-1) f'}{2\zeta f}
  + \frac{f'^2}{4 f^2} - \frac{f''}{2 f}  \\
  &= \left( - \frac{m^2}{\zeta^2 f} + \frac{f^2 + D^2(-1+2\zeta^D)}{4\zeta^2f^2} \right) \,, \\
  \L_{\calv} &= \frac{\calh_0^2\zeta^{D-3}}{ f^2}  \,, \quad
  \L_{\calh} = \frac{\calv_0^2\zeta^{D-3}}{ f^2} \,.
\end{align}
These equations can be obtained by extremizing the action
\begin{equation}
 S =  \int \dd \zeta \left( \frac{1}{2} \calv'^2 - \frac{1}{2} \calh'^2
  + \frac{1}{2} M_{\calh}^2 \calh^2 - \frac{1}{2} M_{\calv}^2 \calv^2 +
  \frac{1}{2}\L \calv^2 \calh^2     \right) \,,
\end{equation}
where we choose $\calv_0 = \calh_0$, so $\L_{\calv} = \L_{\calh} =: \L$. We will use this form of the fluctuation equations below.

\section{Critical chemical potential (temperature)}
\label{appc}

In this appendix we look in detail at solution of the fluctuation equations
\begin{align}
  & \ddot{\tV} + \frac{3-D}{\zeta} \dot{\tV} - \frac{1}{\zeta^2 f} \tH^2 \tV = 0 \,,  \label{VecEq}\\
  & \ddot{\tH} +\left(\frac{\dot{f}}{f} - \frac{D-1}{\zeta} \right) \dot{\tH} +
    \left( \frac{\tV^2}{f^2} - \frac{\tm^2}{\zeta^2 f}  \right) \tH = 0  \,,
    \label{scalarEq}
\end{align}
where recall that
\begin{equation}
  \tV = q z_h V \,, \quad \tH = q h \,, \quad
  \zeta = \frac{z}{z_h} \,, \quad z_h =\frac{D}{4\pi T}  \,, \quad f=(1-\zeta^D).
\end{equation}
In the ordinary phase the relevant analytic solution is
\begin{equation}
  \tV = \tmu (1-\zeta^{D-2}) \,, \quad \tH=0 \,, \label{seed}
\end{equation}
which is a solution for any $\tmu$, with $\tmu$ defined in terms of the chemical potential $\mu$ in \eqref{tdmudef}. We now analyse the criteria for the existence of other solutions. From the last term of \eqref{scalarEq} we may interpret as an effective scalar mass the combination
\begin{equation}
  \tm_{\mathrm{eff}}^2 = \tm^2 - \frac{\zeta^2}{f}\tV^2, \label{meff}
\end{equation}
which is nothing but $m^2 + g^{tt} A_t^2$.
So, if $\tV$ is big enough, there is a possibility that the effective mass of scalar field is sufficiently negative to trigger a BF type instability. Therefore, it is natural to start with the form \eqref{seed}, plug it into \eqref{meff}, and  dial up $\tmu$ until we see the instability.
i.e. we assume the existence of  a perturbative solution to \eqref{VecEq} and \eqref{scalarEq} near the transition temperature of the form  \cite{Maeda:2008ir,Herzog:2010vz,Siopsis:2010uq}
\begin{align}
  \tV &= \tV_0 + \e^2 \tV_2 + \e^4\tV_4 + \cdots \,, \\
  \tH &= \e\tH_1 + \e^3\tH_3 + \e^5 \tH_5 + \cdots \,,
\end{align}
where, $\tV_0 = \tmu_c (1-\zeta^{D-2})$ and $\tmu_c$ is the critical chemical potential, which gives nonzero $\tH_1$. If $\tH_1 = 0$, then
all higher order terms vanish and we end up with the trivial solution \eqref{seed}.

An intuitive argument for the criteria for a condensate solution to exist was given in \cite{Horowitz:2010gk}:
the effective mass $m^2_\mathrm{eff}$ becomes sufficiently negative near the horizon to destabilize the field.
However, one has to be a little careful about this argument for the following reason.
At first sight, the effective mass does seem to become increasingly negative near the horizon because of the  $1/f$ term in $g^{tt}$ or \eqref{meff} which has a pole at the horizon but note that $A_t$ or $\tV$ has a zero near the horizon and thus there is a vanishing
contribution of $\tV$ to the effective mass near the horizon. There is no mass deformation near boundary either due to the $\zeta^2$ term in $g^{tt}$. Indeed, it turns out that a sufficiently negative mass actually arises in the intermediate regime, $\zeta \sim 0.5$.  One exception is the zero temperature case, where $g^{tt}$ develops a double pole and $g^{tt}\tV$ tends to a constant value near horizon. In this case, there is a relation between mass and charge for instability obtained by the BF mass violation near horizon \cite{Horowitz:2010gk}. 

\subsection{Method 1: Series expansion and matching}

Working perturbatively close to the transition temperature, we wish to find $\tmu_c$ giving non-zero $\tH_1$:
\begin{equation}
   \ddot{\tH}_1 +\left(\frac{\dot{f}}{f} - \frac{D-1}{\zeta} \right) \dot{\tH}_1 +
    \left( \frac{\tV_0^2}{f^2} - \frac{\tm^2}{\zeta^2 f}  \right)\tH_1  = 0  \,. \label{Eqh1}
\end{equation}
Let us solve \eqref{Eqh1} using series expansions near horizon and boundary, which are matched at some intermediate point.
The relevant series solutions are
\begin{equation}
\begin{split}
  \tH_B &= O_-\zeta^{\D_-}\left( 1+   \zeta^2 \frac{\tmu^2}{2(-2+D-2\D_-) } +\cdots\right) \\
  & + O_+\zeta^{\D_+} \left(1- \zeta^2 \frac{\tmu^2}{2(-2-D+2\D_+) + }   +\cdots\right) \,,
  \end{split}
\end{equation}
near the boundary and
\begin{equation}
\tH_H =   h_0\left(1 + \frac{\tm^2}{D}(1-\zeta) + \frac{2D\tm^2 + \tm^4 - (D-2)^2 \tmu^2}{4D^2}(1-\zeta)^2 +\cdots \right)
\end{equation}
near the horizon.

\begin{figure}[]
\centering
\subfigure[$D=3, O_+ = 2$]
  {\includegraphics[width=4.5cm]{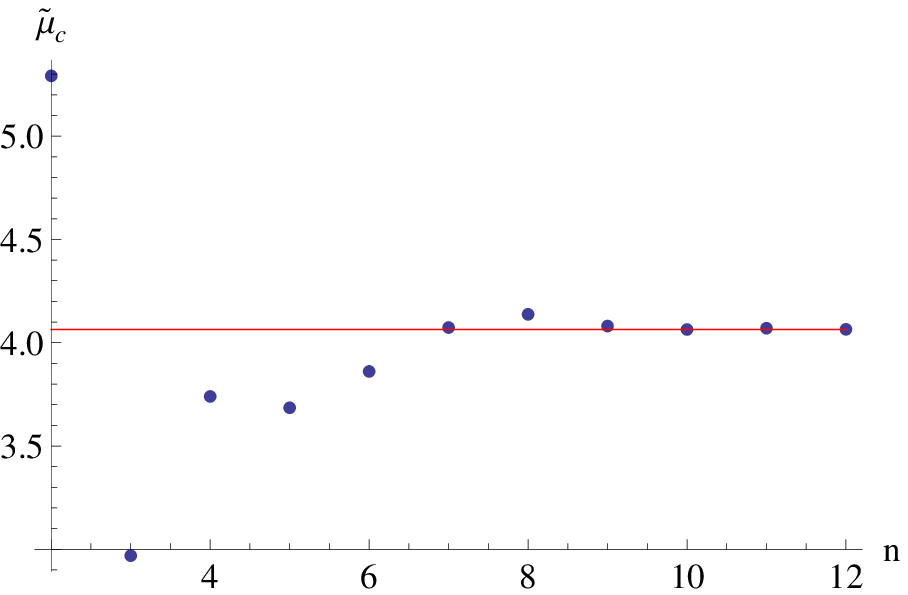}}
\subfigure[$D=3, O_- = 1$]
  {\includegraphics[width=4.5cm]{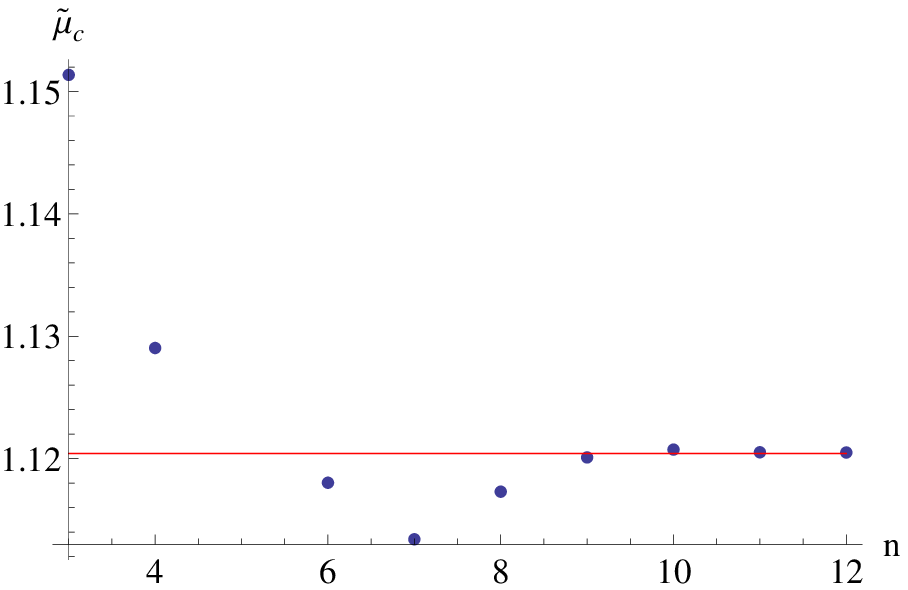}}
\subfigure[$D=4, O_+ = 2$]
  {\includegraphics[width=4.5cm]{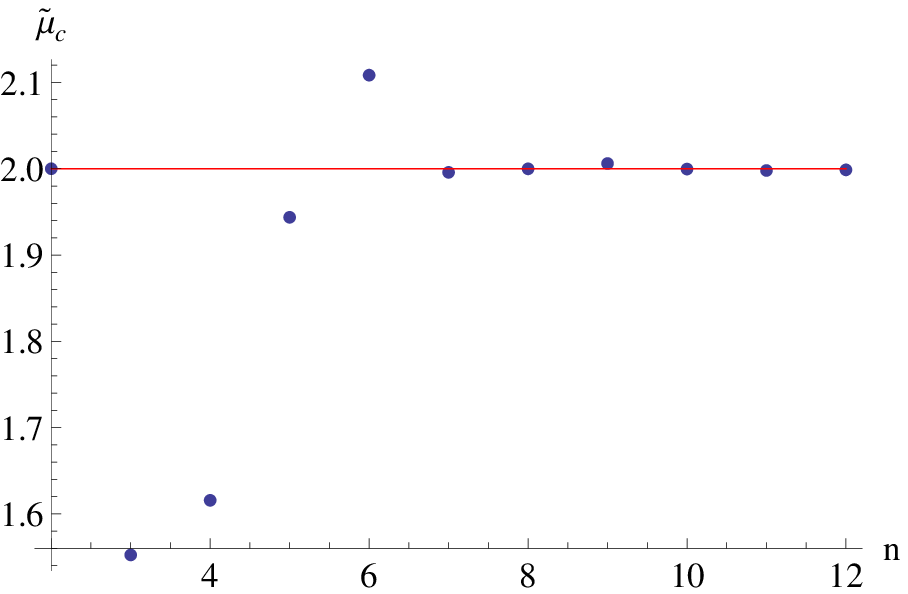}}
  \caption{ Critical chemical potential. $n$ is the power of terms $\zeta^n$ and $(1-\zeta)^n$ included. The red lines are precise values.   } \label{series}
\end{figure}

The matching condition is
\begin{equation}
  \tH_H(\zeta_m) = \tH_B(\zeta_m) \,, \quad  \tH_H'(\zeta_m) = \tH_B'(\zeta_m) \label{cond2}
\end{equation}
If $h_0 = 0$, then $O_- = O_+ = 0$ from \eqref{cond2}. So we assume $h_0 \ne 0$. We can choose either $O_- = 0$ or $O_+ = 0$. Then we are left with
three unknowns $O_+ (O_-), \tmu, \zeta_m$ and two constraints \eqref{cond2}.
i.e. we can have solutions $O_+(\zeta_m)$ and $\tmu_c(\zeta_m)$. In order to fix the ambiguity of $\zeta_m$, we impose another condition.
\begin{equation}
  \tH_H''(\zeta_m) = \tH_B''(\zeta_m)
\end{equation}
This method is only approximate with a few expanded terms.
However, with many terms retained, the method becomes very accurate, see figure \ref{series}.
See figure \ref{BF}(b) for a comparison to other methods.

\subsection{Method 2: Effective scalar in AdS$_2$ or Schrodinger equation}
\label{smallm-meth}

Next we exploit the heuristic idea of BF mass violation. For this purpose we need to define the background for the scalar. First, note that the scalar field $\phi(z)$ in AdS$_{d+1}$ reads
\begin{equation}
  \phi'' - \frac{d-1}{z}\phi' - \frac{m^2}{z^2} \phi = 0 \,.
  \label{AdSscalar}
\end{equation}
By redefining our field $\tH_1$, we can always change the coefficient of $\dot{\tH}_1$ of \eqref{Eqh1} to the form of the coefficient of $\phi'$ in \eqref{AdSscalar} with any $d$. We choose $d=1$, which will be interpreted
as an effective AdS$_2$ scalar equation or Schrodinger equation.

\begin{figure}[]
\centering
\subfigure[$D=3, m=10, \mu=0,10,20,30,37.15$ from top]
  {\includegraphics[width=6cm]{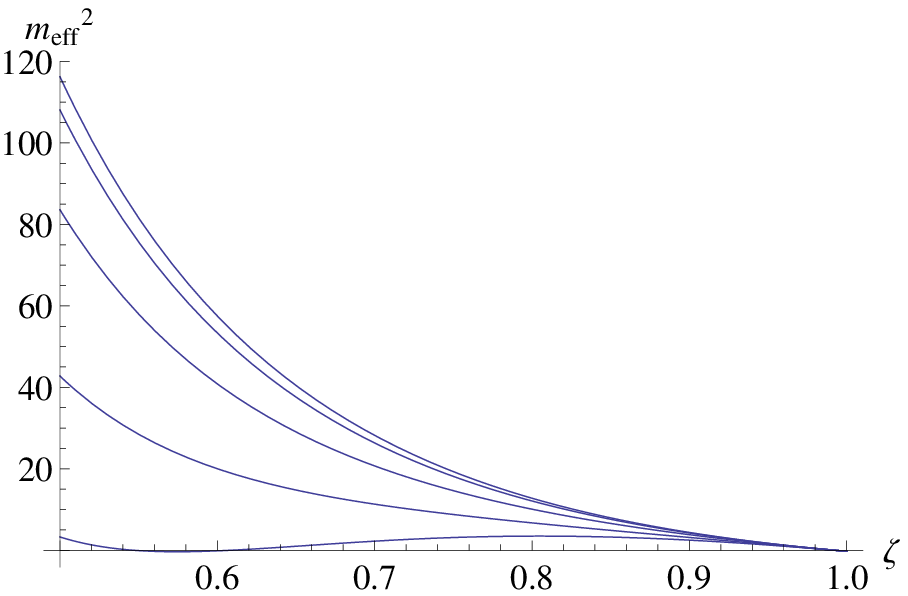}}
\subfigure[$D=3$, Line: semi-classical approximation, Red dots: BF bound analysis, Blue dots: numerical values by method 2.]
  {\includegraphics[width=6cm]{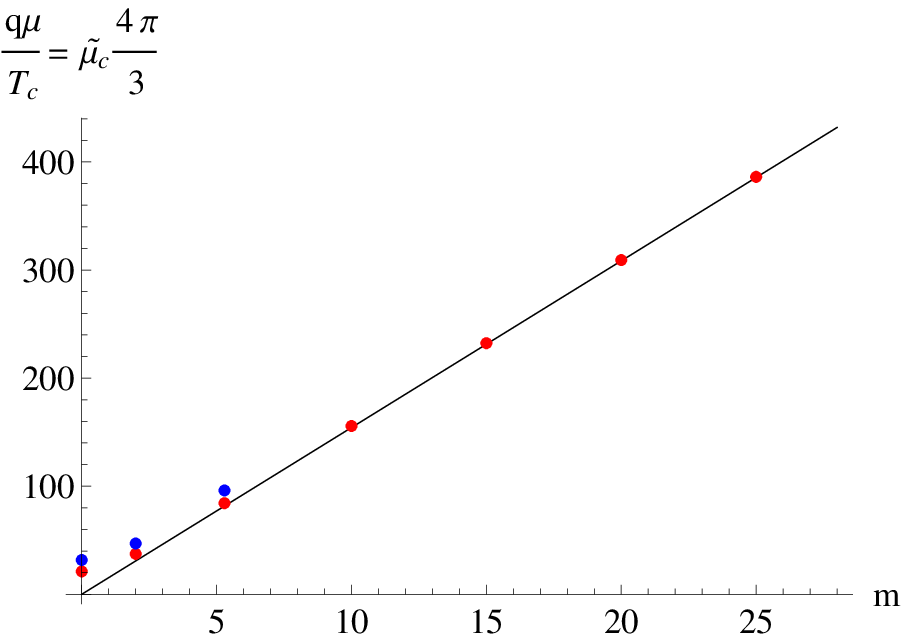}}
  \caption{ (a) Effective mass (b) critical chemical potential } \label{BF}
\end{figure}

In terms of the new fields $\calv$ and $\calh$ introduced in appendix \ref{schr-eq}
\begin{equation}
  \tV_0(\zeta) \equiv  \zeta^{\frac{D-3}{2}} \calv(\zeta) \,, \qquad  \tH_1  \equiv  \zeta^{\frac{D-1}{2}}\frac{1}{\sqrt{f}}\calh(\zeta) \,,
\end{equation}
the leading order and sub-leading order equations read
\begin{align}
  & \ddot{\calv} - \frac{(D-3)(D-1)}{4 \zeta^2} \calv  = 0 \,,  \label{calv}\\
  & \ddot{\calh} + \left( - \frac{\tm^2}{\zeta^2 f} + \frac{f^2 + D^2(-1+2\zeta^D)}{4\zeta^2f^2} \right)\calh + \frac{\zeta^{D-3}}{ f^2} \calv^2 \calh = 0 \,.
\end{align}
With $\calv$, which solves \eqref{calv} with a boundary condition $\tV_0(1)=0$,
\begin{equation}
  \calv = \tmu\zeta^{\frac{3-D}{2}} (1-\zeta^{D-2}) \,,
\end{equation}
the scalar equation becomes\footnote{It can be interpreted as
a Schrodinger equation with an inverse square potential. In this
case, the BF bound violation is interpreted as the existence of negative
energy bound state, which is allowed when $a < -1/4$ with a potential = $a/r^2$.}
\begin{equation}
  \ddot{\calh} - \calm \calh =0 \,,
\end{equation}
where
\begin{align}
  \calm &= - \frac{-4\tm^2 f + 4 \tmu^2 \zeta^2 (1-\zeta^{D-2})^2 +
  f^2 + D^2(-1+2\zeta^D)}{4\zeta^2f^2} \\
  &= \begin{cases}
  \left(\frac{D^2-1}{4}+\tm^2\right)\frac{1}{\zeta^2} - \mu^2 & \zeta \sim 0\\
  -\frac{1}{4}\frac{1}{(\zeta-1)^2} + \frac{D-D^2-4\tm^2}{4D(\zeta-1)}
  + \frac{1}{48}(D-1)(25+7D) + \frac{(3+D)\tm^2}{2D} - \frac{(D-2)^2 \tmu^2}{D^2}& \zeta \sim 1
\end{cases}
\end{align}
The BF bound of AdS$_2$ is $-1/4$. For a large $m^2$, the effective mass is well above the BF bound near the boundary, but saturates the BF bound near the horizon ($\zeta \ra 1-\zeta$), regardless of $D, m, \tmu$. Therefore, we may expect that the instability can happen by BF bound violation near
horizon. We define a $\zeta$ dependent mass by
\begin{equation}
  \tm_\mathrm{eff}^2(\zeta) = (1-\zeta)^2 \calm
\end{equation}
where $(1-\zeta)^2$ corresponds to $z^2$ in \eqref{AdSscalar}.
We find the $\tmu_c$ at a given $D$ and $\tm^2$ such that
\begin{equation}
  \tm_\mathrm{eff}^2(\zeta_c) = -1/4 \,, \qquad
  \dot{\tm}_\mathrm{eff}^2(\zeta_c) = 0 \,. \label{cond1}
\end{equation}
The first condition is nothing but the BF bound.  The second condition comes
from the smoothness of $\tm_\mathrm{eff}^2(\zeta)$. These two condition determines $\tmu_c$ and $\zeta_c$ uniquely. For example, see figure \ref{BF}(a). At $D=3$ and $\tm=10$, the curve with $\tmu_c = 37.15$ and $\zeta_c \sim 0.6$ satisfy \eqref{cond1}. By repeating for $\tm=0,2,5,10,15,20,25$ we can
identify corresponding $\tmu_c$s, which are plotted as red dots in figure  \ref{BF}(b).

Let us next consider the limit $\tm^2 \gg D^2$. The effective mass then
becomes
\begin{equation}
   \tm_{\mathrm{eff}}^2 \sim \left(\frac{\tm^2 f}{\zeta^2} - \tmu^2 (1-\zeta^{D-2})^2 \right) \frac{(1-\zeta)^2}{f^2} \,.
\end{equation}
For the purpose of the comparison to a semi-classical approximation, we
apply a weaker condition
\begin{equation}
  \tm_\mathrm{eff}^2(\zeta_c) = 0 \,, \qquad
  \dot{\tm}_\mathrm{eff}^2(\zeta_c) = 0 \,
\end{equation}
This condition is equivalent to
\begin{equation}
  V_\mathrm{eff}(\zeta_c) = 0 \,, \qquad
  \dot{V}_\mathrm{eff}(\zeta_c) = 0 \,
\end{equation}
with
\begin{equation}
   V_{\mathrm{eff}}  = \frac{\tm \sqrt{f}}{\zeta} - \tmu (1-\zeta^{D-2})  \,.
\end{equation}
Note that this equivalence is due to the regularity and positivity of $(1-\zeta)^2/f^2$ and the reduced condition $\tm^2_{\mathrm{eff}} =0$.
This agrees with the semi-classical analysis in the appendix C of \cite{Benini:2010pr}.
In $D=3$, $\tmu_c \sim 3.86 \tm$, which is the blue line in Fig. \ref{BF}(b).
The red dots are slightly bigger since they are obtained by a stronger conditions $\tm_\mathrm{eff}^2(\zeta_c) = -1/4$. Interestingly, it seems that the semi-classical approximation is quite good for small $\tm$, even though the condition $\tm^2 \gg D^2$ is used in the derivation.

\newpage

\section{Numerical analysis of condensate and comparison to analytic formula } \label{number}

The purpose of this section is two fold: 1) comparisons between analytic expressions and numerical results; 2) checks of the stability and reliability of our Mathematica code. For this purpose, we start with the original holographic superconductor model with parameters ($D=3, m^2 = -2, \D = 2 , \k =1, q=1$), for which considerable numerical data has been reported, enabling us to cross-check our numerical code with for example \cite{Hartnoll:2008vx}.

\subsection{Numerical condensate}

\begin{figure}[!h]
\centering
\subfigure[$\sqrt{\calo_2}/ T_c \sim (1-T)^{1/4}$]
  {\includegraphics[width=6cm]{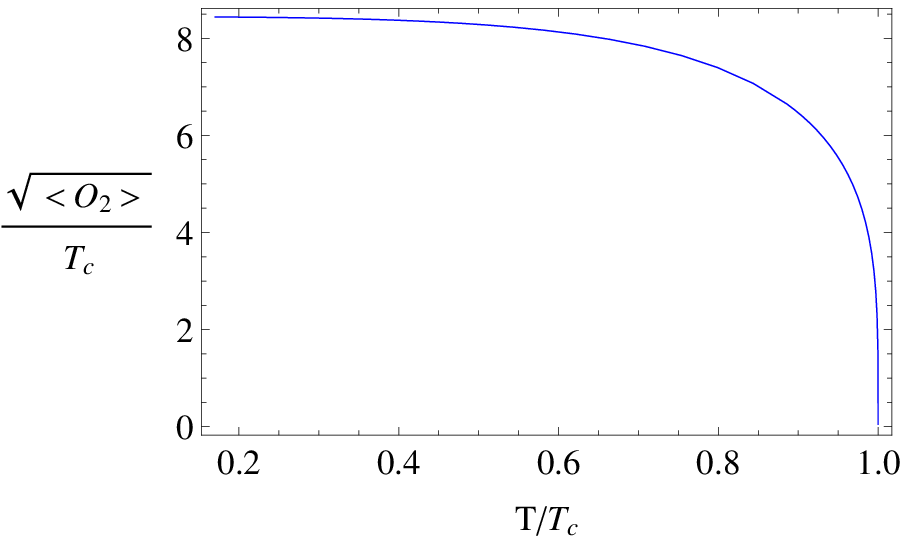}}
\subfigure[$ \calo_2/ T_c^2 \sim (1-T)^{1/2}$]
  {\includegraphics[width=6cm]{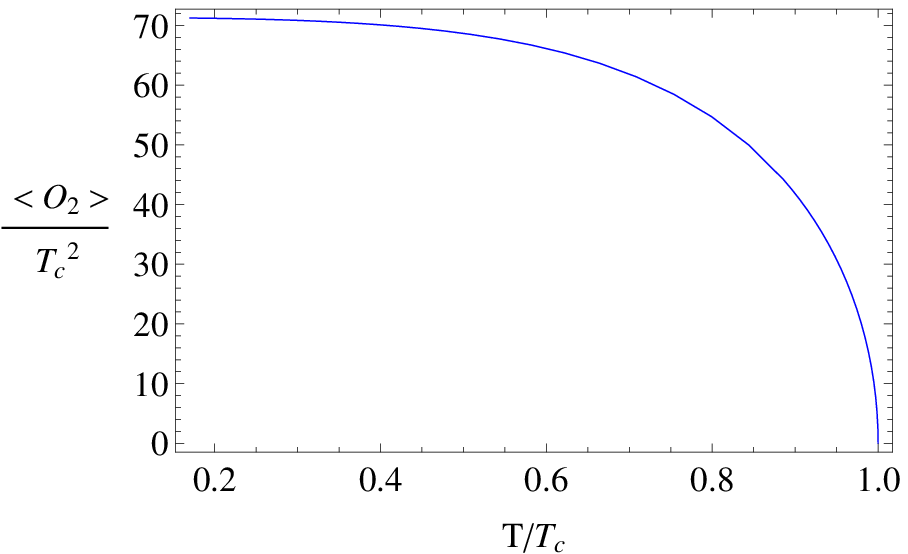}}
  \caption{Condensate vs Temperature: Figure (a) agrees with Fig.1 in \cite{Hartnoll:2008vx}. } \label{Fig:condN}
\end{figure}

\subsection{Comparison between numerics and analytic formula}

\begin{figure}[!h]
\centering
\subfigure[Red curves are by \eqref{condA}. $\zeta_m = 0.1,0.2,0.26,0.3,0.4,0.5$ from top to bottom.]
  {\includegraphics[width=6cm]{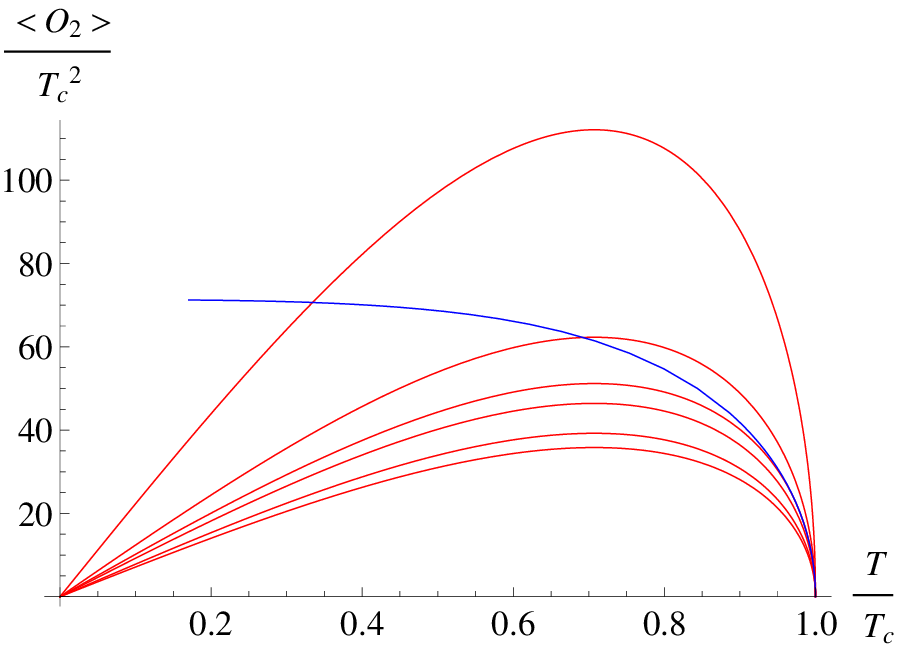}}
\subfigure[Zoom in of (a) near phase transition.]
  {\includegraphics[width=6cm]{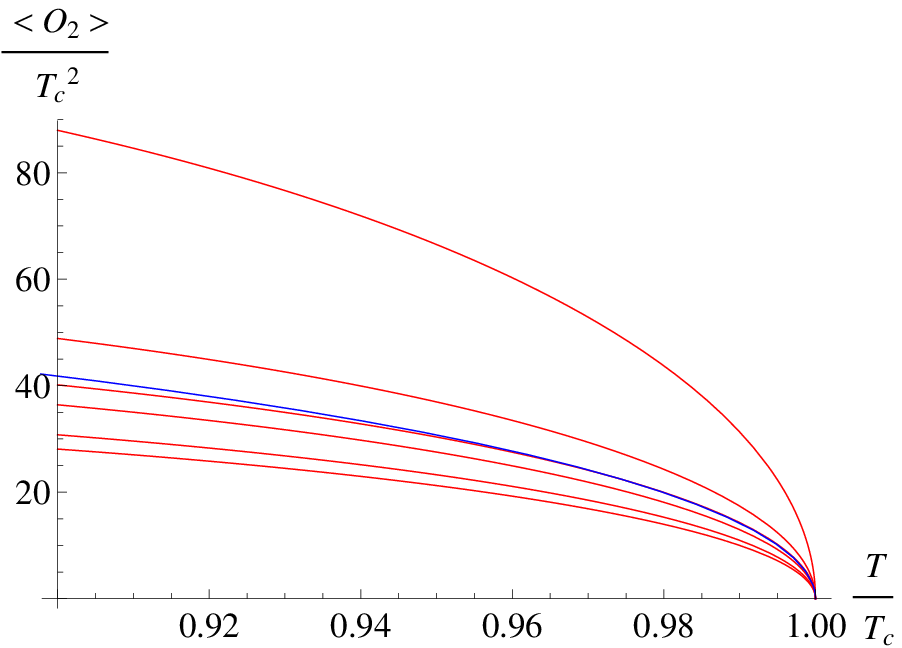}}
  \caption{ Comparison between numerics and analytic formula: the red curves are by \eqref{condA} and the blue one is from figure \ref{Fig:condN}(b). } \label{Fig:comp}
\end{figure}

We note the following:
\begin{itemize}
\item The analytic formula is supposed to be valid only near phase transition, so the deviations from the numerical result at small temperature are to be expected.
However, the analytic formula is quite accurate for a much wider regime than would have been expected.
\item The parameter $\zeta_m$ cannot be fixed in the analytic approximation. It reflects the fact that a priori the best matching point is unknown. Therefore, the $\zeta_m$ may be chosen differently depending on the purpose. The parameter was fixed as $\zeta_m=0.26$ since it gives the best looking fit to the numerics for the widest range.
\item $\zeta_m = 0.34$ was estimated in \cite{Gregory:2009fj} not by looking at the plots such as Fig \ref{Fig:comp}, but based on other numerical data, $T_c \sim 0.118 \rho^{1/2}$, a number borrowed from \cite{Hartnoll:2008vx}. However, from figure \ref{Fig:comp}, we note that $\zeta_m = 0.34$ does not yield a good fit. Since there may be some numerical uncertainties in the numerical methods between different research groups, this discrepancy is arguably not physically important.
One possibility is that $\zeta_m = 0.34$ may make the best fit only very near the phase transition. However, if we are interested in fermions moving in  finite condensate, it is more useful to focus on the region a little away from the phase transition point, which motivates the choice $\zeta_m=0.26$.
\end{itemize}

\subsection{Comparisons of the bulk fields}

\begin{figure}[!h]
\centering
  {\includegraphics[width=6cm]{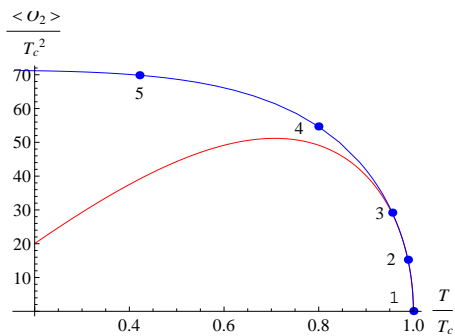}}
  \caption{ The points where the bulk fields are evaluated.
  The numbers $1,2,3,4,5$ correspond to figure \ref{Fig:compB}.
  Only the points $1,2,3$ are expected to show a good agreement between analytic expressions and numerical results. However, we present $4,5$ to show
how deviation occurs.($\zeta_m=0.26$ for the red.)
} \label{Fig:coms}
\end{figure}

Finally we compare the numerical bulk fields ($V,h$) to the following analytic expressions
\begin{equation} \label{VA}
\begin{split}
V(\zeta)&=
\begin{cases} \mu - \rho \zeta^{D-2} & \text{if } 0 \le \zeta \le \zeta_m,\\
              -v_1(1-\zeta) + \frac{1}{2}\left(-3+D-\frac{h_0^2}{D}\right)v_1(1-\zeta)^2& \text{if } \zeta_m \le \zeta \le 1 .
\end{cases}  \\
h(\zeta)&=
\begin{cases} h_+ \zeta^\D & \text{if } 0 \le \zeta \le \zeta_m,\\
              h_0 + \frac{h_0 m^2}{D}(1-\zeta) + \frac{h_0 (2Dm^2 + m^4 - v_1^2)}{4D^2}(1-\zeta)^2 & \text{if } \zeta_m \le \zeta \le 1 .
\end{cases}
\end{split}
\end{equation}
where $\mu,\rho,h_+$ are functions of $v_1,h_0$ for a given $D,\D,\zeta_m$.
We choose five points in figure \ref{Fig:coms}. For each point, there is $(v_1,h_0)$ obtained by numerical analysis, which will determine \eqref{VA}.
\begin{figure}[]
\centering
\subfigure[1]
  {\includegraphics[width=4.8cm]{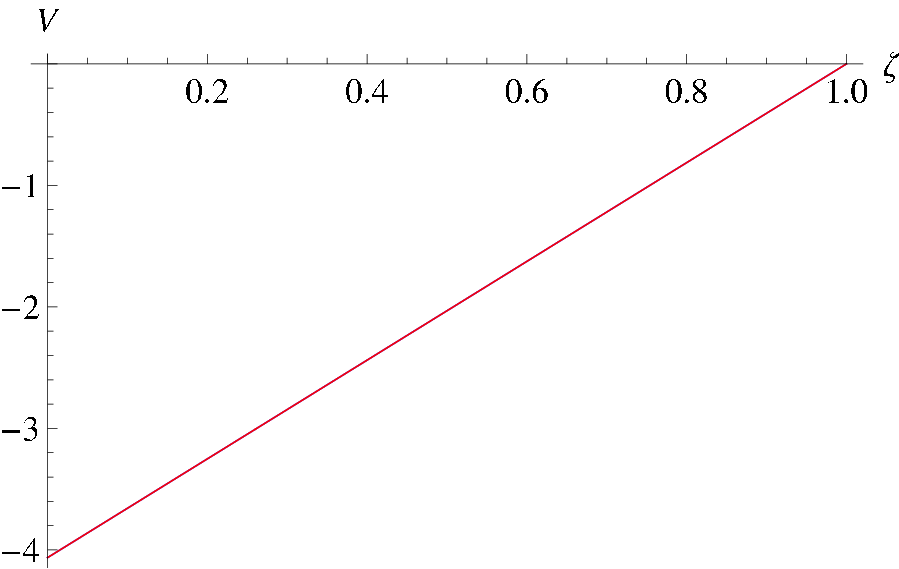}
   \includegraphics[width=4.8cm]{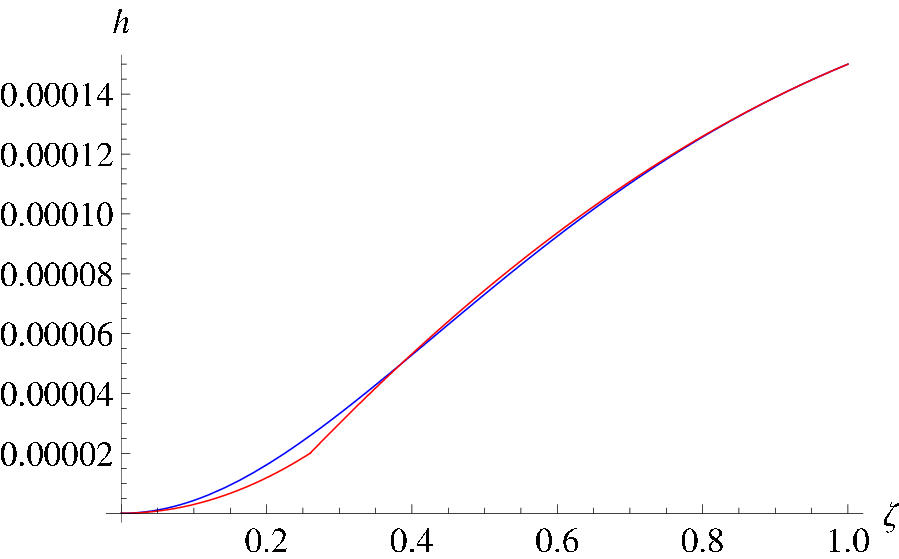}}
\subfigure[2]
  {\includegraphics[width=4.8cm]{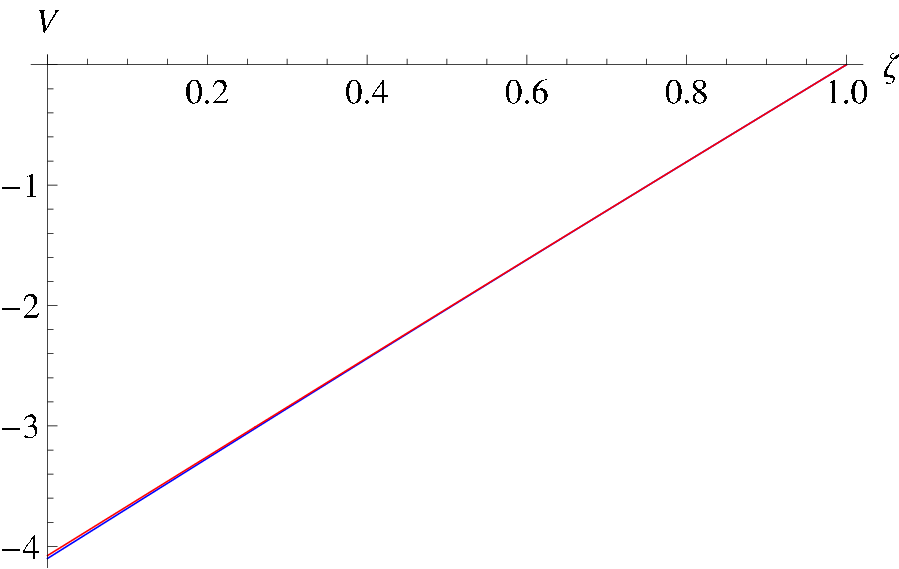}
   \includegraphics[width=4.8cm]{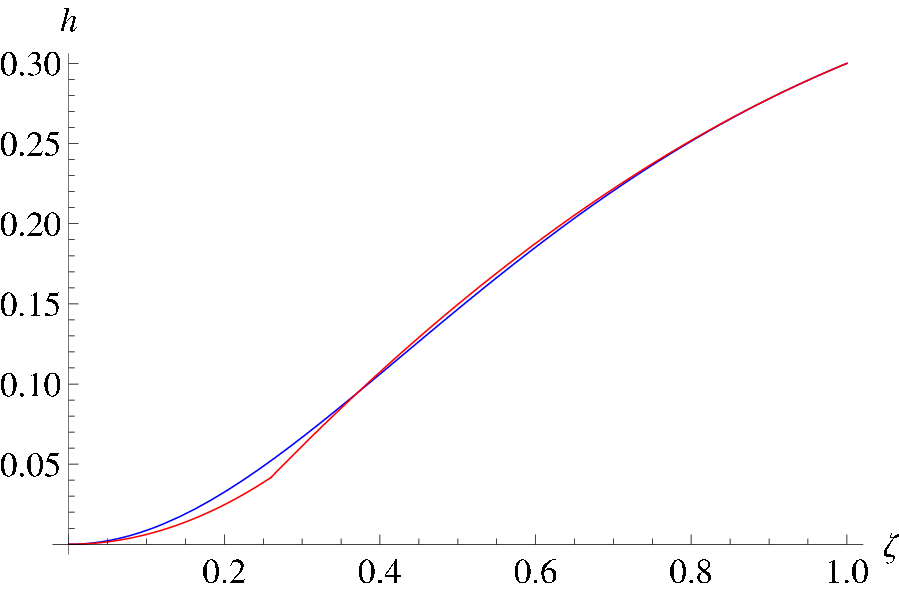}}
\subfigure[3]
  {\includegraphics[width=4.8cm]{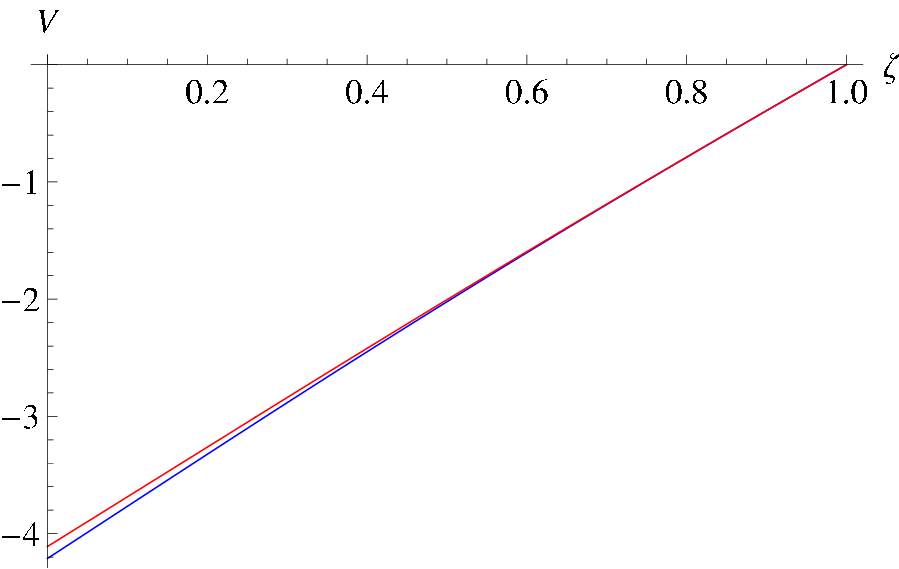}
   \includegraphics[width=4.8cm]{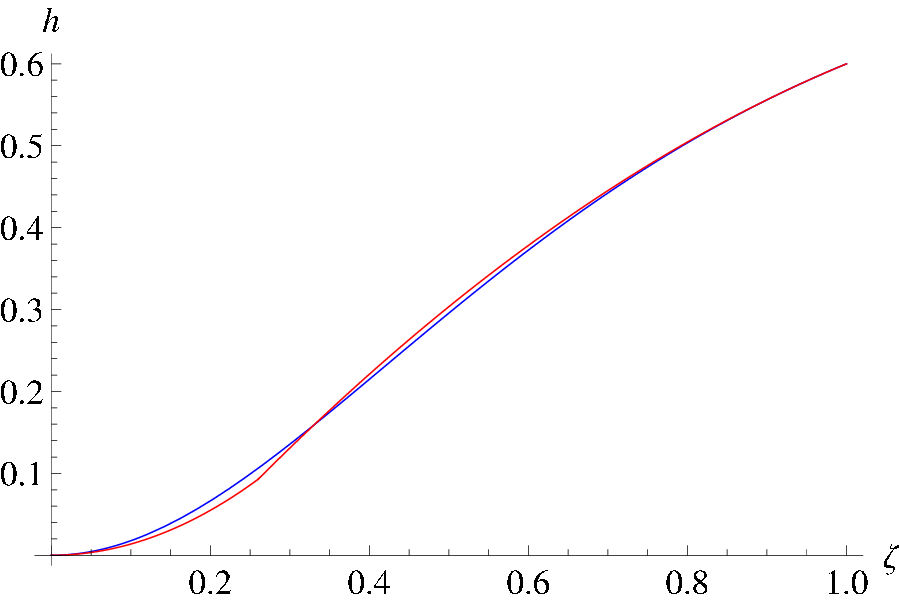}}
\subfigure[4]
  {\includegraphics[width=4.8cm]{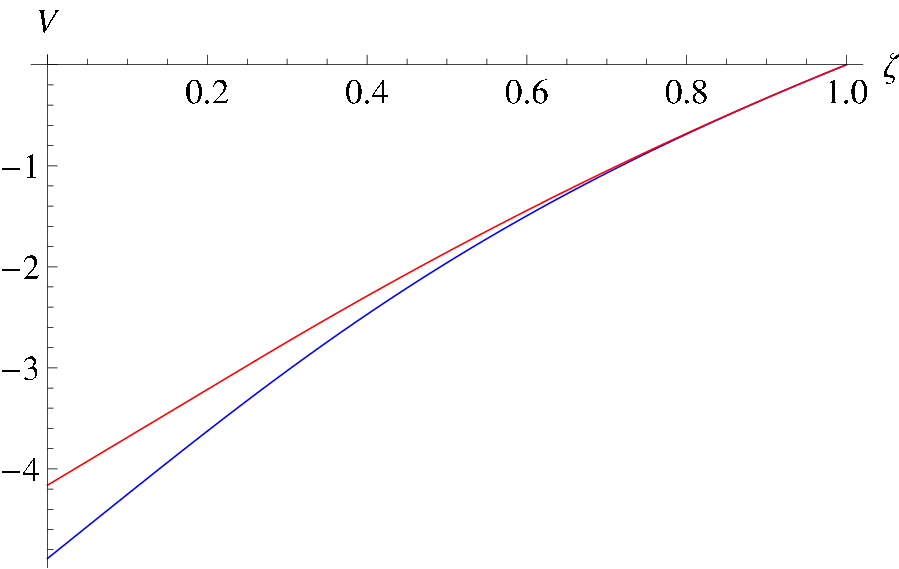}
   \includegraphics[width=4.8cm]{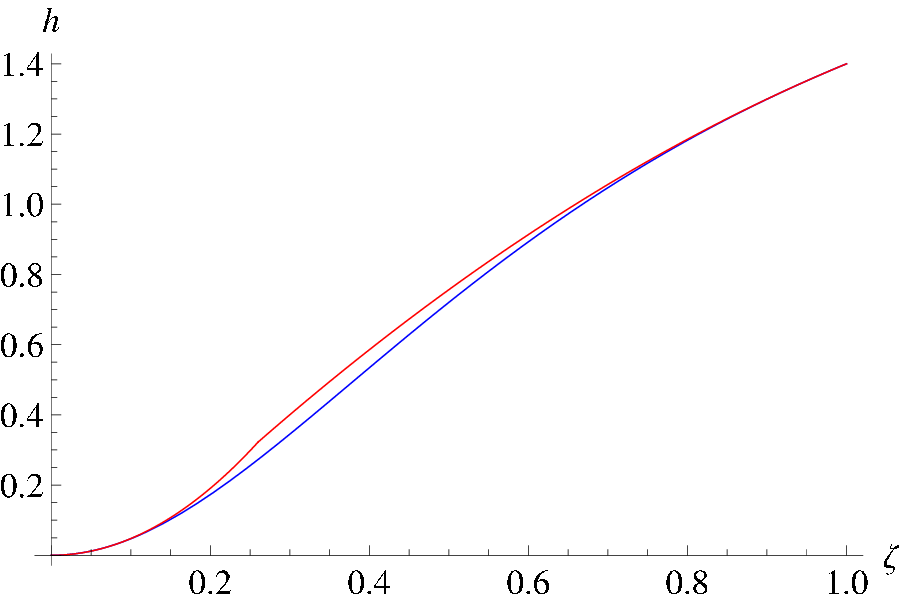}}
\subfigure[5]
  {\includegraphics[width=4.8cm]{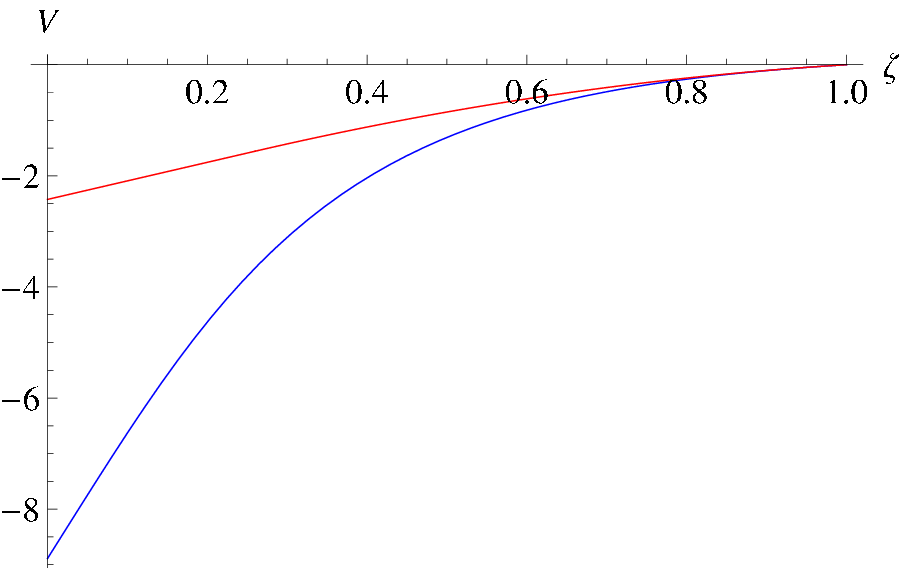}
   \includegraphics[width=4.8cm]{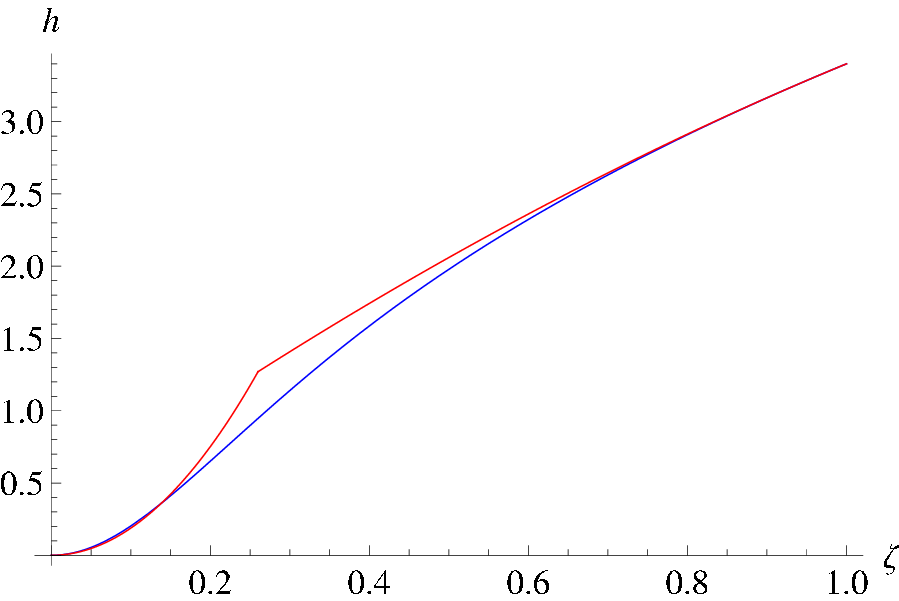}}
  \caption{ Comparison between numerical and analytic bulk fields: the red curves are by \eqref{VA} and the blue are numerical curves.
  The analytic curves are not smooth at the matching point. } \label{Fig:compB}
\end{figure}

\newpage


\begin{thebibliography}{10}

\bibitem{Hartnoll:2009sz}
S.~A. Hartnoll, {\it {Lectures on holographic methods for condensed matter
  physics}},  {\em Class.Quant.Grav.} {\bf 26} (2009) 224002,
  [\href{http://xxx.lanl.gov/abs/0903.3246}{{\tt arXiv:0903.3246}}].

\bibitem{Herzog:2009xv}
C.~P. Herzog, {\it {Lectures on Holographic Superfluidity and
  Superconductivity}},  {\em J.Phys.} {\bf A42} (2009) 343001,
  [\href{http://xxx.lanl.gov/abs/0904.1975}{{\tt arXiv:0904.1975}}].

\bibitem{McGreevy:2009xe}
J.~McGreevy, {\it {Holographic duality with a view toward many-body physics}},
  {\em Adv.High Energy Phys.} {\bf 2010} (2010) 723105,
  [\href{http://xxx.lanl.gov/abs/0909.0518}{{\tt arXiv:0909.0518}}].

\bibitem{Horowitz:2010gk}
G.~T. Horowitz, {\it {Introduction to Holographic Superconductors}},
  \href{http://xxx.lanl.gov/abs/1002.1722}{{\tt arXiv:1002.1722}}.

\bibitem{Gubser:2008px}
S.~S. Gubser, {\it {Breaking an Abelian gauge symmetry near a black hole
  horizon}},  {\em Phys.Rev.} {\bf D78} (2008) 065034,
  [\href{http://xxx.lanl.gov/abs/0801.2977}{{\tt arXiv:0801.2977}}].

\bibitem{Hartnoll:2008kx}
S.~A. Hartnoll, C.~P. Herzog, and G.~T. Horowitz, {\it {Holographic
  Superconductors}},  {\em JHEP} {\bf 12} (2008) 015,
  [\href{http://xxx.lanl.gov/abs/0810.1563}{{\tt arXiv:0810.1563}}].

\bibitem{Horowitz:2012ky}
G.~T. Horowitz, J.~E. Santos, and D.~Tong, {\it {Optical Conductivity with
  Holographic Lattices}},  {\em JHEP} {\bf 1207} (2012) 168,
  [\href{http://xxx.lanl.gov/abs/1204.0519}{{\tt arXiv:1204.0519}}].

\bibitem{Horowitz:2012gs}
G.~T. Horowitz, J.~E. Santos, and D.~Tong, {\it {Further Evidence for
  Lattice-Induced Scaling}},  {\em JHEP} {\bf 1211} (2012) 102,
  [\href{http://xxx.lanl.gov/abs/1209.1098}{{\tt arXiv:1209.1098}}].

\bibitem{Nakamura:2009tf}
S.~Nakamura, H.~Ooguri, and C.-S. Park, {\it {Gravity Dual of Spatially
  Modulated Phase}},  {\em Phys.Rev.} {\bf D81} (2010) 044018,
  [\href{http://xxx.lanl.gov/abs/0911.0679}{{\tt arXiv:0911.0679}}].

\bibitem{Ooguri:2010kt}
H.~Ooguri and C.-S. Park, {\it {Holographic End-Point of Spatially Modulated
  Phase Transition}},  {\em Phys.Rev.} {\bf D82} (2010) 126001,
  [\href{http://xxx.lanl.gov/abs/1007.3737}{{\tt arXiv:1007.3737}}].

\bibitem{Ooguri:2010xs}
H.~Ooguri and C.-S. Park, {\it {Spatially Modulated Phase in Holographic
  Quark-Gluon Plasma}},  {\em Phys.Rev.Lett.} {\bf 106} (2011) 061601,
  [\href{http://xxx.lanl.gov/abs/1011.4144}{{\tt arXiv:1011.4144}}].

\bibitem{Bergman:2011rf}
O.~Bergman, N.~Jokela, G.~Lifschytz, and M.~Lippert, {\it {Striped instability
  of a holographic Fermi-like liquid}},  {\em JHEP} {\bf 1110} (2011) 034,
  [\href{http://xxx.lanl.gov/abs/1106.3883}{{\tt arXiv:1106.3883}}].

\bibitem{Donos:2011qt}
A.~Donos, J.~P. Gauntlett, and C.~Pantelidou, {\it {Spatially modulated
  instabilities of magnetic black branes}},  {\em JHEP} {\bf 1201} (2012) 061,
  [\href{http://xxx.lanl.gov/abs/1109.0471}{{\tt arXiv:1109.0471}}].

\bibitem{Donos:2011bh}
A.~Donos and J.~P. Gauntlett, {\it {Holographic striped phases}},  {\em JHEP}
  {\bf 1108} (2011) 140, [\href{http://xxx.lanl.gov/abs/1106.2004}{{\tt
  arXiv:1106.2004}}].

\bibitem{Donos:2012gg}
A.~Donos and J.~P. Gauntlett, {\it {Helical superconducting black holes}},
  {\em Phys.Rev.Lett.} {\bf 108} (2012) 211601,
  [\href{http://xxx.lanl.gov/abs/1203.0533}{{\tt arXiv:1203.0533}}].

\bibitem{Donos:2012yu}
A.~Donos, J.~P. Gauntlett, J.~Sonner, and B.~Withers, {\it {Competing orders in
  M-theory: superfluids, stripes and metamagnetism}},  {\em JHEP} {\bf 1303}
  (2013) 108, [\href{http://xxx.lanl.gov/abs/1212.0871}{{\tt
  arXiv:1212.0871}}].

\bibitem{Chen:2010mk}
J.-W. Chen, Y.-J. Kao, D.~Maity, W.-Y. Wen, and C.-P. Yeh, {\it {Towards A
  Holographic Model of D-Wave Superconductors}},  {\em Phys.Rev.} {\bf D81}
  (2010) 106008, [\href{http://xxx.lanl.gov/abs/1003.2991}{{\tt
  arXiv:1003.2991}}].

\bibitem{Zeng:2010vp}
H.-B. Zeng, Z.-Y. Fan, and H.-S. Zong, {\it {d-wave Holographic Superconductor
  Vortex Lattice and Non-Abelian Holographic Superconductor Droplet}},  {\em
  Phys.Rev.} {\bf D82} (2010) 126008,
  [\href{http://xxx.lanl.gov/abs/1007.4151}{{\tt arXiv:1007.4151}}].

\bibitem{Chen:2011ny}
J.-W. Chen, Y.-S. Liu, and D.~Maity, {\it {$d+id$ Holographic
  Superconductors}},  {\em JHEP} {\bf 1105} (2011) 032,
  [\href{http://xxx.lanl.gov/abs/1103.1714}{{\tt arXiv:1103.1714}}].

\bibitem{Gao:2011aa}
D.~Gao, {\it {Vortex and droplet in holographic D-wave superconductors}},  {\em
  Phys.Lett.} {\bf A376} (2012) 1705--1709,
  [\href{http://xxx.lanl.gov/abs/1112.2422}{{\tt arXiv:1112.2422}}].

\bibitem{Ge:2012vp}
X.-H. Ge, S.~F. Tu, and B.~Wang, {\it {d-Wave holographic superconductors with
  backreaction in external magnetic fields}},  {\em JHEP} {\bf 1209} (2012)
  088, [\href{http://xxx.lanl.gov/abs/1209.4272}{{\tt arXiv:1209.4272}}].

\bibitem{Benini:2010qc}
F.~Benini, C.~P. Herzog, and A.~Yarom, {\it {Holographic Fermi arcs and a
  d-wave gap}},  \href{http://xxx.lanl.gov/abs/1006.0731}{{\tt
  arXiv:1006.0731}}.

\bibitem{Benini:2010pr}
F.~Benini, C.~P. Herzog, R.~Rahman, and A.~Yarom, {\it {Gauge gravity duality
  for d-wave superconductors: prospects and challenges}},  {\em JHEP} {\bf 11}
  (2010) 137, [\href{http://xxx.lanl.gov/abs/1007.1981}{{\tt
  arXiv:1007.1981}}].

\bibitem{Hartnett:2012np}
G.~S. Hartnett and G.~T. Horowitz, {\it {Geons and Spin-2 Condensates in the
  AdS Soliton}},  {\em JHEP} {\bf 1301} (2013) 010,
  [\href{http://xxx.lanl.gov/abs/1210.1606}{{\tt arXiv:1210.1606}}].

\bibitem{Maeda:2008ir}
K.~Maeda and T.~Okamura, {\it {Characteristic length of an AdS/CFT
  superconductor}},  {\em Phys.Rev.} {\bf D78} (2008) 106006,
  [\href{http://xxx.lanl.gov/abs/0809.3079}{{\tt arXiv:0809.3079}}].

\bibitem{Herzog:2010vz}
C.~P. Herzog, {\it {An Analytic Holographic Superconductor}},  {\em Phys.Rev.}
  {\bf D81} (2010) 126009, [\href{http://xxx.lanl.gov/abs/1003.3278}{{\tt
  arXiv:1003.3278}}].

\bibitem{Siopsis:2010uq}
G.~Siopsis and J.~Therrien, {\it {Analytic Calculation of Properties of
  Holographic Superconductors}},  {\em JHEP} {\bf 1005} (2010) 013,
  [\href{http://xxx.lanl.gov/abs/1003.4275}{{\tt arXiv:1003.4275}}].

\bibitem{Buchbinder:1999be}
I.~Buchbinder, V.~Krykhtin, and V.~Pershin, {\it {On consistent equations for
  massive spin two field coupled to gravity in string theory}},  {\em
  Phys.Lett.} {\bf B466} (1999) 216--226,
  [\href{http://xxx.lanl.gov/abs/hep-th/9908028}{{\tt hep-th/9908028}}].

\bibitem{Buchbinder:1999ar}
I.~Buchbinder, D.~Gitman, V.~Krykhtin, and V.~Pershin, {\it {Equations of
  motion for massive spin-2 field coupled to gravity}},  {\em Nucl.Phys.} {\bf
  B584} (2000) 615--640, [\href{http://xxx.lanl.gov/abs/hep-th/9910188}{{\tt
  hep-th/9910188}}].

\bibitem{Buchbinder:2000fy}
I.~Buchbinder, D.~Gitman, and V.~Pershin, {\it {Causality of massive spin-2
  field in external gravity}},  {\em Phys.Lett.} {\bf B492} (2000) 161--170,
  [\href{http://xxx.lanl.gov/abs/hep-th/0006144}{{\tt hep-th/0006144}}].

\bibitem{Buchbinder:2012iz}
I.~Buchbinder, T.~Snegirev, and Y.~Zinoviev, {\it {Cubic interaction vertex of
  higher-spin fields with external electromagnetic field}},  {\em Nucl.Phys.}
  {\bf B864} (2012) 694--721, [\href{http://xxx.lanl.gov/abs/1204.2341}{{\tt
  arXiv:1204.2341}}].

\bibitem{Deser:2001us}
S.~Deser and A.~Waldron, {\it {Partial masslessness of higher spins in (A)dS}},
   {\em Nucl.Phys.} {\bf B607} (2001) 577--604,
  [\href{http://xxx.lanl.gov/abs/hep-th/0103198}{{\tt hep-th/0103198}}].

\bibitem{Deser:2001dt}
S.~Deser and A.~Waldron, {\it {Inconsistencies of massive charged gravitating
  higher spins}},  {\em Nucl.Phys.} {\bf B631} (2002) 369--387,
  [\href{http://xxx.lanl.gov/abs/hep-th/0112182}{{\tt hep-th/0112182}}].

\bibitem{Porrati:2011uu}
M.~Porrati and R.~Rahman, {\it {Notes on a Cure for Higher-Spin Acausality}},
  {\em Phys.Rev.} {\bf D84} (2011) 045013,
  [\href{http://xxx.lanl.gov/abs/1103.6027}{{\tt arXiv:1103.6027}}].

\bibitem{Kulaxizi:2012xp}
M.~Kulaxizi and R.~Rahman, {\it {Holographic Constraints on a Vector Boson}},
  \href{http://xxx.lanl.gov/abs/1212.6265}{{\tt arXiv:1212.6265}}.

\bibitem{Gregory:2009fj}
R.~Gregory, S.~Kanno, and J.~Soda, {\it {Holographic Superconductors with
  Higher Curvature Corrections}},  {\em JHEP} {\bf 0910} (2009) 010,
  [\href{http://xxx.lanl.gov/abs/0907.3203}{{\tt arXiv:0907.3203}}].

\bibitem{Maldacena:2008wh}
J.~Maldacena, D.~Martelli, and Y.~Tachikawa, {\it {Comments on string theory
  backgrounds with non- relativistic conformal symmetry}},  {\em JHEP} {\bf 10}
  (2008) 072, [\href{http://xxx.lanl.gov/abs/0807.1100}{{\tt
  arXiv:0807.1100}}].

\bibitem{Cassani:2010uw}
D.~Cassani, G.~Dall'Agata, and A.~F. Faedo, {\it {Type IIB supergravity on
  squashed Sasaki-Einstein manifolds}},  {\em JHEP} {\bf 1005} (2010) 094,
  [\href{http://xxx.lanl.gov/abs/1003.4283}{{\tt arXiv:1003.4283}}].

\bibitem{Skenderis:2010vz}
K.~Skenderis, M.~Taylor, and D.~Tsimpis, {\it {A Consistent truncation of IIB
  supergravity on manifolds admitting a Sasaki-Einstein structure}},  {\em
  JHEP} {\bf 1006} (2010) 025, [\href{http://xxx.lanl.gov/abs/1003.5657}{{\tt
  arXiv:1003.5657}}].

\bibitem{Gauntlett:2010vu}
J.~P. Gauntlett and O.~Varela, {\it {Universal Kaluza-Klein reductions of type
  IIB to N=4 supergravity in five dimensions}},  {\em JHEP} {\bf 1006} (2010)
  081, [\href{http://xxx.lanl.gov/abs/1003.5642}{{\tt arXiv:1003.5642}}].

\bibitem{Liu:2010sa}
J.~T. Liu, P.~Szepietowski, and Z.~Zhao, {\it {Consistent massive truncations
  of IIB supergravity on Sasaki-Einstein manifolds}},  {\em Phys.Rev.} {\bf
  D81} (2010) 124028, [\href{http://xxx.lanl.gov/abs/1003.5374}{{\tt
  arXiv:1003.5374}}].

\bibitem{Cvetic:1999xp}
M.~Cvetic, M.~Duff, P.~Hoxha, J.~T. Liu, H.~Lu, {\em et.~al.}, {\it {Embedding
  AdS black holes in ten-dimensions and eleven-dimensions}},  {\em Nucl.Phys.}
  {\bf B558} (1999) 96--126,
  [\href{http://xxx.lanl.gov/abs/hep-th/9903214}{{\tt hep-th/9903214}}].

\bibitem{Gubser:1998bc}
S.~S. Gubser, I.~R. Klebanov, and A.~M. Polyakov, {\it {Gauge theory
  correlators from non-critical string theory}},  {\em Phys. Lett.} {\bf B428}
  (1998) 105--114, [\href{http://xxx.lanl.gov/abs/hep-th/9802109}{{\tt
  hep-th/9802109}}].

\bibitem{Witten:1998qj}
E.~Witten, {\it {Anti-de Sitter space and holography}},  {\em Adv. Theor. Math.
  Phys.} {\bf 2} (1998) 253--291,
  [\href{http://xxx.lanl.gov/abs/hep-th/9802150}{{\tt hep-th/9802150}}].

\bibitem{Skenderis:2006uy}
K.~Skenderis and M.~Taylor, {\it {Kaluza-Klein holography}},  {\em JHEP} {\bf
  05} (2006) 057, [\href{http://xxx.lanl.gov/abs/hep-th/0603016}{{\tt
  hep-th/0603016}}].

\bibitem{Kim:1985ez}
H.~J. Kim, L.~J. Romans, and P.~van Nieuwenhuizen, {\it {The Mass Spectrum of
  Chiral N=2 D=10 Supergravity on S**5}},  {\em Phys. Rev.} {\bf D32} (1985)
  389.

\bibitem{Arutyunov:1998hf}
G.~E. Arutyunov and S.~A. Frolov, {\it {Quadratic action for type IIB
  supergravity on AdS(5) x S(5)}},  {\em JHEP} {\bf 08} (1999) 024,
  [\href{http://xxx.lanl.gov/abs/hep-th/9811106}{{\tt hep-th/9811106}}].

\bibitem{Gubser:1998vd}
S.~S. Gubser, {\it {Einstein manifolds and conformal field theories}},  {\em
  Phys.Rev.} {\bf D59} (1999) 025006,
  [\href{http://xxx.lanl.gov/abs/hep-th/9807164}{{\tt hep-th/9807164}}].

\bibitem{Ceresole:1999zs}
A.~Ceresole, G.~Dall'Agata, R.~D'Auria, and S.~Ferrara, {\it {Spectrum of type
  IIB supergravity on AdS(5) x T(11): Predictions on N = 1 SCFT's}},  {\em
  Phys. Rev.} {\bf D61} (2000) 066001,
  [\href{http://xxx.lanl.gov/abs/hep-th/9905226}{{\tt hep-th/9905226}}].

\bibitem{Ceresole:1999ht}
A.~Ceresole, G.~Dall'Agata, and R.~D'Auria, {\it {KK spectroscopy of type IIB
  supergravity on AdS(5) x T(11)}},  {\em JHEP} {\bf 11} (1999) 009,
  [\href{http://xxx.lanl.gov/abs/hep-th/9907216}{{\tt hep-th/9907216}}].

\bibitem{Lee:1998bxa}
S.~Lee, S.~Minwalla, M.~Rangamani, and N.~Seiberg, {\it {Three-point functions
  of chiral operators in D = 4, N = 4 SYM at large N}},  {\em Adv. Theor. Math.
  Phys.} {\bf 2} (1998) 697--718,
  [\href{http://xxx.lanl.gov/abs/hep-th/9806074}{{\tt hep-th/9806074}}].

\bibitem{Arutyunov:1999en}
G.~Arutyunov and S.~Frolov, {\it {Some cubic couplings in type IIB supergravity
  on AdS(5) x S(5) and three-point functions in SYM(4) at large N}},  {\em
  Phys. Rev.} {\bf D61} (2000) 064009,
  [\href{http://xxx.lanl.gov/abs/hep-th/9907085}{{\tt hep-th/9907085}}].

\bibitem{Skenderis:2006di}
K.~Skenderis and M.~Taylor, {\it {Holographic Coulomb branch vevs}},  {\em
  JHEP} {\bf 0608} (2006) 001,
  [\href{http://xxx.lanl.gov/abs/hep-th/0604169}{{\tt hep-th/0604169}}].

\bibitem{Skenderis:2007yb}
K.~Skenderis and M.~Taylor, {\it {Anatomy of bubbling solutions}},  {\em JHEP}
  {\bf 0709} (2007) 019, [\href{http://xxx.lanl.gov/abs/0706.0216}{{\tt
  arXiv:0706.0216}}].

\bibitem{Dwave2}
K.-Y. Kim, K.~Skenderis, and M.~Taylor, {\it {Fermions in top down d-wave
  models}},  [\href{http://xxx.lanl.gov/abs/13xx.xxxx}{{\tt arXiv:13xx.xxxx}}].

\bibitem{Hartnoll:2008vx}
S.~A. Hartnoll, C.~P. Herzog, and G.~T. Horowitz, {\it {Building a Holographic
  Superconductor}},  {\em Phys.Rev.Lett.} {\bf 101} (2008) 031601,
  [\href{http://xxx.lanl.gov/abs/0803.3295}{{\tt arXiv:0803.3295}}].

\end{thebibliography}

\providecommand{\href}[2]{#2}\begingroup\raggedright\endgroup

\end{document}